%% file: diss.tex
\documentclass[11pt,twoside,a4paper,reqno]{article}
\usepackage{german}
\usepackage[centertags]{amsmath}
\allowdisplaybreaks[4]
\usepackage{amssymb}
\usepackage{amsthm}
\usepackage{latexsym}
\theoremstyle{plain}
\newtheorem{theorem}{Theorem}
\newtheorem{proposition}{Proposition}
\theoremstyle{definition}
\newtheorem{definition}{Definition}
\numberwithin{equation}{section}
\newcommand{\al}{\alpha}
\newcommand{\be}{\beta}
\newcommand{\ga}{\gamma}
\newcommand{\de}{\delta}
\newcommand{\la}{\lambda}
\newcommand{\si}{\sigma}
\newcommand{\ro}{\rho}

\newcommand{\ep}{\epsilon}
\newcommand{\vp}{\varphi}
\newcommand{\vep}{\varepsilon}
\newcommand{\om}{\omega}
\newcommand{\ti}[1]{\tilde{#1}}
\newcommand{\wti}[1]{\widetilde{#1}}
\newcommand{\pal}{{^{\scriptstyle{,}}\!\alpha}}
\newcommand{\pbe}{{^{\scriptstyle{,}}\!\beta}}
\newcommand{\pga}{{^{\scriptstyle{,}}\!\gamma}}

\newcommand{\pxi}{{^{\scriptstyle{,}}\!\xi}}
\newcommand{\pro}{{^{\scriptstyle{,}}\!\rho}}

\newcommand{\pep}{{^{\scriptstyle{,}}\!\varepsilon}}
\newcommand{\pmu}{{^{\scriptstyle{,}}\!\mu}}
\newcommand{\pnu}{{^{\scriptstyle{,}}\!\nu}}
\newcommand{\psig}{{^{\scriptstyle{,}}\!\sigma}}
\newcommand{\tot}{\dagger}
\newcommand{\abs}[1]{\mathsf{#1}}
\newcommand{\contract}[1]{%
\overset{\makebox%
          {\, \rule[-.27em]{.3pt}{.3em}\hrulefill%
           \rule[-.27em]{.3pt}{.3em}\!\, }}%
        {\rule{0pt}{0.75em}#1}}%
\title{\vspace*{-15mm} {\LARGE{
Gravity as the Spin-2 Quantum Gauge Theory}} \\[15mm]}
\author{{\bf Dissertation} \\[10mm]
{\bf zur} \\[2mm]
{\bf Erlangung der naturwissenschaftlichen Doktorw\"urde} \\[2mm]
{\bf (Dr. sc. nat.)} \\[2mm]
{\bf vorgelegt der} \\[2mm]
{\bf Mathematisch-naturwissenschaftlichen Fakult\"at} \\[2mm]
{\bf der} \\[10mm]
{\bf Universit\"at Z\"urich} \\[10mm]
{\bf von} \\[4mm]
{\bf Mark WELLMANN} \\[4mm]
{\bf aus}\\[4mm]
{\bf Elmshorn (Deutschland)} \\[10mm]
{\bf Begutachtet von} \\[7mm]
{\bf Prof. Dr. G\"unter SCHARF} \\[10mm]}
\date{Z\"urich 2001}  
\begin{document}
\pagestyle{empty}
\maketitle
\thispagestyle{empty}
\newpage
\begin{sf}
{\small
\noindent
Die vorliegende Arbeit wurde von der Mathematisch-naturwissenschaftlichen 
Fakult\"at
der Universit\"at Z\"urich auf Antrag von Prof. Dr. G\"unter Scharf  
und Prof. Dr. Norbert Straumann
als Dissertation angenommen.
}
\newpage
\vspace{15cm}
\newpage
\centerline{\Large Zusammenfassung}
\vspace{3 cm}
In dieser Arbeit wird die Theorie eines quantisierten Spin-2 Feldes behandelt. Dies geschieht im Rahmen der kausalen St\"orungstheorie nach Epstein und Glaser. Die Arbeit besteht aus zwei Teilen. Im ertsen Teil untersuchen wir die Eichstruktur eines masselosen selbstwechselwirkenden Spin-2 Feldes. Dabei nehmen wir einen reinen feldtheoretischen Standpunkt ein, d.h. es werden keine geometrischen Aspekte der allgemeinen Relativit\"atstheorie vorausgesetzt. Aus dem Prinzip der Operatoreichinvarianz werden in erster Ordnung St\"orungstheorie notwendige und hinreichende Bedingungen abgeleitet, die eine eichinvariante Theorie eines Spin-2 Feldes erf\"ullen muss. Dieses Prinzip besagt, dass die Eichvariation bzgl. der Eichladung $Q$ der Selbstkopplung des Spin-2 Feldes eine Divergenz im Sinne der Vektoranalysis sein muss. Es zeigt sich, dass die allgemeinste trilineare Selbstkopplung des Spin-2 Feldes sich von der Einstein-Hilbert Kopplung nur um Co-R\"ander und Divergenzen unterscheidet.

Im zweiten Teil dieser Arbeit (Kap.9) besch\"aftigen wir uns mit dem Langabstandsverhalten der Theorie eines Spin-2 Feldes welches an massive skalare Materie koppelt. Es wird der adiabatische Limes, bei dem die Abschaltung der Wechselwirkung im Unendlichen entfernt wird, in Strahlungskorrekturen zur Zweiteilchenstreuung untersucht. Wir berechnen den Wirkungsquerschnitt f\"ur Graviton Bremsstrahlung in dem von einem der streuenden Teilchen ein Graviton niedriger Energie emittiert wird. Es zeigt sich, dass der Wirkungsquerschnitt im adiabatischen Limes logarithmisch divergiert. Desweiteren werden das Infrarotverhalten der Graviton Selbstenergie sowie die Selbstenergie der massiven skalaren Materie untersucht. Die Graviton Selbstenergie ist endlich im adiabati\-schen Limes, w\"ahrend bei der Selbstenergie der Materie ebenfalls eine logarithmische Divergenz im Wirkungsquerschnitt entsteht.  
\newpage
\nonfrenchspacing
\centerline{\Large Summary}
\vspace{3 cm}
This work deals with the theory of a quantized spin-2 field in the framework of causal perturbation theory. It is divided into two parts. In the first part we analyze the gauge structure  of a massless self-interacting quantum tensor field. We look at this theory from a pure field theoretical point of view without assuming any geometrical aspect  from general relativity. To first order in the perturbation expansion of the $S$-matrix we derive necessary and sufficient conditions for such a theory to be gauge invariant, by which we mean that the gauge variation of the self-coupling with respect to the gauge charge operator $Q$ is a divergence in the sense of vector analysis. The most general trilinear self-coupling of the graviton field turns out to be the one derived from the Einstein-Hilbert action plus coboundaries and divergences.

In the second part of this work (sect.9) we consider a massive scalar field coupled to gravity. We are interested in the long range behaviour of this theory. Radiative corrections for two particle scattering are investigated in the adiabatic limit, where the cutoff of the interaction at infinity is removed. We compute the differential cross section for graviton bremsstrahlung in which one of the scattered particles emits a graviton of low energy. It is shown that such processes are logarithmically divergent in the adiabatic limit. Furthermore we show that the differential cross section for two particle scattering with a graviton self-energy insertion is finite in the adiabatic limit while for matter self-energy it is logarithmically divergent, too.   
\newpage
\end{sf}
\tableofcontents
\newpage
\mbox{}
\newpage
\pagestyle{plain}
\setcounter{page}{1}
\input{faust.tex}
\section{Introduction}
One of the most interesting unsolved problems in modern theoretical physics is to combine Einstein's general theory of relativity with the principles of quantum physics. It is experimentally tested to a high degree of accuracy that the classical description of matter breaks down on the atomic scale and that quantum corrections become important. For the theory of general relativity one expects that it is valid up to the scale of the Planck length, $l_P$ which is the unique combination with the dimension of length of the fundamental constants of nature, the speed of light, $c$, Planck's constant, $\hbar$ and Newton's gravitational constant, $G$, namely $l_P\equiv(G\hbar/c^3)^{1/2}\sim 10^{-33}cm$. Beyond this scale it is believed that effects of a full theory of quantum gravity come into play. Therefore one might ask the questions: How does gravity behave at microscopic scales? What does the spacetime look like at the order of the Planck length? To answer these questions a full theory of quantum gravity is needed. Although there are many different approaches to the problem, none of them succeeded in overcoming the fundamental problems arising. One of the main problems in quantum gravity is that we have to give a meaning to the quantized metric field. Since the metric defines the geometry of spacetime we are forced to explain what a quantized geometry should be. The postulate of microcausality becomes meaningless because there is no longer a fixed background on which concepts like timelike, spacelike or lightlike objects can be defined. 

In this work we deal with the theory of a covariantly quantized tensor field in four-dimensional Minkowski spacetime. The idea to describe quantum gravity in this way goes back to the work of Rosenfeld, Fierz and Pauli \cite{ro:zqdw,fie:rtktmbs,paufie:rftmbsief} in the thirties and it was further developed by Feynman and deWitt in the sixties \cite{fe:qtg,dew:qtog1,dew:qtog2,dew:qtog3}. In this approach the metric tensor field $g_{\mu\nu}$ is splitted into two parts, the constant background field $\eta_{\mu\nu}$ which describes the causal structure of Minkowski spacetime and a dynamical part $h_{\mu\nu}$ which will then be quantized in this spacetime. These quanta of the gravitational field, called gravitons in the following, have zero mass and spin 2. Altough pure quantum gravity is finite at one loop it was realized by Deser, van Nieuwenhuizen and Boulware \cite{devnieu:oldqemf,devnbo:unqg} as well as by t'Hooft and Veltman \cite{thvelt:nrqeds} in the mid seventies that a theory of a spin-2 field coupled to matter is non-renormalizable. Later in the eighties Goroff and Sagnotti \cite{gosa:ubeg} computed the divergences of pure quantum gravity at two loop level. After the discovery that such a spin-2 theory is non-renormalizable people lost their interest in it. Clearly such a theory has not much predictive power because of the proliferation of undetermined constants which appear in the perturbative calculations. Nevertheless it was shown recently by Grillo \cite{gr:diss} in the calculation of quantum corrections to the Newtonian potential that normalization terms only affect the potential at the origin. Therefore the long range part of the potential remains untouched. This will also be the case in higher orders, see \cite{sch:qgt}. In addition to that we think that one can learn something about the gauge structure of quantum gravity by considering this approach. The theory of a quantized tensor field will be treated here in the framework of causal perturbation theory which has the advantage that no ultraviolet divergences occur due to the mathematically correct treatment of the distributions. 
%
%
\section{Causal Construction of the $S$-Matrix}
Causal perturbation theory goes back to ideas of Bogoliubov \cite{bog:itqf} and was carefully developed by Epstein and Glaser \cite{eg:rlp} in the seventies. Later Scharf \cite{sch:qed} applied it successfully to QED. In this approach one considers the $S$-matrix as the central object. The idea is to fix the first order interaction and then to construct higher orders of the scattering-matrix $S$ by induction using free fields only. Since the free quantum fields as basic objects are operator-valued distributions the scattering-matrix will be constructed as an operator-valued functional on some test function space. The most important ingredient for the inductive construction is the causality requirement, which roughly states that the scattering-matrix of a sum of two test functions factorizes if the supports of the test functions can be separated in time. As test function space one considers for convenience the Schwartz space of functions of rapid decrease, since this space is invariant under Fourier transformation. Then all expressions are well defined tempered distributions. In doing this we cut off the interaction at large but finite distances which is unphysical in most cases and has to be removed at the end. This is the so called adiabatic limit where we take the limit that the test function goes to a constant. In this way we investigate the long range behaviour of the theory. 

The starting point for the construction is the $S$-matrix given as a formal power series in the coupling constant of the theory
\begin{equation}
  S(g)=\mathbf{1}+\sum_{n=1}^{\infty}\frac{1}{n!}\int d^4 x_1\ldots d^4x_n\,T_n(x_1,\ldots,x_n)g(x_1)\ldots g(x_n)
\end{equation}
where $g$ is a test function from the Schwartz space $\mathcal{S}(\mathbb{R}^4)$ of rapidly decreasing functions. The $T_n(x_1,\ldots,x_n)$ are time-ordered products of interaction Lagrangians. They are given in terms of Wick monomials of free field operators. These free fields are operator-valued distributions on Fock space \cite{wg:fod,sw:pct}. If an ordering of the arguments of $T_n$ would be given, such that $x_1^0>x_2^0>\ldots>x_n^0$ then we could write
\begin{equation}
  T_n(x_1,\ldots,x_n)=T_1(x_1)\ldots T_1(x_n)
\end{equation}
where $T_1$ defines the interaction to first order of perturbation theory. In general no such time-ordering is given and one has to construct $T_n$ carefully to every order. The naive construction of the $T_n$ with the help of the step-function is ill-defined
\begin{equation}
  T_n(x_1,\ldots,x_n)=\sum_{\si\in S_n}\theta(x_{\si(1)}^0-x_{\si(2)}^0)\ldots\theta(x_{\si(n-1)}^0-x_{\si(n)}^0)T_1(x_{\si(1)})\ldots T_1(x_{\si(n)})
\end{equation}
because this exression involves pointwise products of distributions which are not defined a priori. The use of this time-ordering would led to the well-known UV-divergences. The Epstein-Glaser method gives an inductive construction of the time-ordered products which are well defined and free of UV-divergences. In this method the $S$-matrix is constructed by causality. This means that if we take two test functions $g_1$ and $g_2$ for which we can find a separation of the supports in time by a spacelike surface, i.e. $\text{supp}(g_1)>\text{supp}(g_2)$ then the $S$-matrix should factorize according to
\begin{equation}
  S(g_1+g_2)=S(g_1)S(g_2).
\end{equation}
This means for the time-ordered products
\begin{equation}
  T_n(x_1,\ldots,x_n)=T_m(x_1,\ldots,x_m)T_{n-m}(x_{m+1},\ldots,x_n)
\end{equation}
if $\{x_1,\ldots,x_m\}>\{x_{m+1},\ldots,x_n\}$. This factorization property lies at the heart of the causal construction. Now we assume that all $T_m$ for $2\leq m\leq n-1$ are already constructed. One can show that they have the form
\begin{equation}
  T_m(x_1,\ldots,x_m)=\sum_{k}:O_{k}(x_1,\ldots,x_m):t_m^{(k)}(x_1-x_m,\ldots,x_{m-1}-x_m)
\end{equation}
where $:O_{k}(x_1,\ldots,x_m):$ is a product of normally ordered free field operators and $t_m^{(k)}(x_1-x_m,\ldots,x_{m-1}-x_m)$ is a translation invariant numerical distribution on $\mathbb{R}^4$. Then Epstein and Glaser have shown that $T_m$ is a well-defined operator-valued distribution. For the explicit construction of the $n$-th order we proceed as follows: First of all we construct auxiliary distributions $R_n^{\prime}(x_1,\ldots,x_n)$ and $A_n^{\prime}(x_1,\ldots,x_n)$ from all the known $T_m$ of lower order by carrying out all possible contractions between field operators appearing in $T_m$ and using Wick's theorem to obtain normally ordered expressions according to
\begin{align}
  R_n^{\prime}(x_1,\ldots,x_n) & = \sum_{P_2}T_{n-n_1}(Y,x_n)\ti{T}_{n_1}(X)\\
  A_n^{\prime}(x_1,\ldots,x_n) & = \sum_{P_2}\ti{T}_{n_1}(X)T_{n-n_1}(Y,x_n)
\end{align}
where $\ti{T}_m$ are appear in the expansion of $S(g)^{-1}$. The sum runs over all partitions of the set of points $\{x_1,\ldots,x_m\}$ into two subsets $X,Y$ so that $X$ is not empty. Then we introduce the difference 
\begin{equation}
  D_n(x_1,\ldots,x_n)=R_n^{\prime}(x_1,\ldots,x_n)-A_n^{\prime}(x_1,\ldots,x_n)
\end{equation} 
which has the following form
\begin{equation}
  D_n(x_1,\ldots,x_n)=\sum_k:O_k(x_1,\ldots,x_n):d_n^{(k)}(x_1-x_n,\ldots,x_{n-1}-x_n)  
\end{equation}
where $:O_k(x_1,\ldots,x_n):$ is a normally ordered product of free field operators and $d_n^{(k)}(x_1-x_n,\ldots,x_{n-1}-x_n)$ is a numerical distribution that is build out of products of positive and negative frequency parts of the Pauli-Jordan distribution. Such products are well-defined. Due to the translation invariance, $d_n^{(k)}$ depends only on $n-1$ independent coordinates. The most important property of the distribution $D_n$ is causality, i.e.
\begin{equation}
  \text{supp}\bigl(D_n(x_1,\ldots,x_n)\bigr)\subseteq\Gamma_{n-1}^+(x_n)\cup\Gamma_{n-1}^-(x_n)
\end{equation}
where
\begin{equation}
  \Gamma_{n-1}^{\pm}(x_n)=\bigl\{(x_1,\ldots,x_n)\in\mathbb{R}^{4n}|x_j\in(x_n+\overline{V^{\pm}}),\forall j=1,\ldots,n-1\bigr\}.
\end{equation}
The support properties are entirely encoded in the numerical part of $D_n$. The next step in the construction of $T_n$ is to split the distribution $D_n$ into retarded and advanced parts, $R_n$ and $A_n$, respectively such that
\begin{align*}
  \text{supp}\bigl(R_n(x_1,\ldots,x_n)\bigr) & \subseteq\Gamma_{n-1}^+(x_n) \\
  \text{supp}\bigl(A_n(x_1,\ldots,x_n)\bigr) & \subseteq\Gamma_{n-1}^-(x_n). 
\end{align*}
To achieve this we have to split the numerical part $d_n^{(k)}$. The critical point for this operation is the total diagonal, i.e.
\begin{equation}
  \Gamma_{n-1}^+(x_n)\cap\Gamma_{n-1}^-(x_n)=\Delta_n=\{x_1=\ldots=x_n\}.
\end{equation}
Due to the fact that $d_n^{(k)}$ is translation invariant we can shift the critical point to the origin. The behaviour of the distribution $d_n^{(k)}$ at the origin is measured by the singular order. The singular order can be defined for an arbitrary distribution $t\in\mathcal{S}^{\prime}(\mathbf{R}^d)$ \cite{stei:axft}.
\begin{definition}
  One says that the distribution $t$ has scaling degree $s$ at $x=0$, if
  \begin{equation*}
    s=\inf\{s^{\prime}\in\mathbb{R}|\lambda^{s^{\prime}}t(\lambda x)\stackrel{\lambda\searrow 0}{\longrightarrow}0\ \text{in the sense of distributions}\}. 
  \end{equation*}
  In spacetime dimension $d=4$ the singular order is then related to the scaling degree by $\om:=[s]-4$, where $[s]$ is the greatest integer less or equal to $s$. 
\end{definition}

In order to avoid UV-divergences one has to do the splitting operation with the correct singular order. It turns out that if the singular order $\omega<0$ then the splitting is trivial and the multiplication with the discontinuous step function can be done without the appearence of UV-divergences. On the other hand, if $\omega\geq 0$ then the splitting operation is non-trivial and moreover non-unique. In this case we obtain for the retarded part
\begin{equation}
  \begin{split}
    d_n^{(k)}(x_1-x_n,\ldots,x_{n-1}-x_n) & \ \longrightarrow r_n^{(k)}(x_1-x_n,\ldots,x_{n-1}-x_n)\\    
                                          & +\sum_{|a|=0}^{\omega}C_{a,k}D^a\delta^{(4(n-1))}(x_1-x_n,\ldots,x_{n-1}-x_n) \\
  \end{split}
\end{equation}
where $\omega\geq 0$ is the singular order of the distribution $d_n^{(k)}$ and $r_n^{(k)}$ is a special splitting solution. The splitting solution is therefore only determined up to a sum of normalization terms with finite coefficients $C_{a,k}$. They are not determined by the splitting procedure itself and there must be imposed further physical conditions like Lorentz covariance, gauge invariance, etc. to restrict them. The retarded part can be determined in momentum space by a dispersion integral \cite{sch:qed}. Finally $T_n$, including it's normalization terms is given by
\begin{equation}
  \begin{split}
    T_n(x_1,\ldots,x_n) & = \ R_n(x_1,\ldots,x_n)-R_n^{\prime}(x_1,\ldots,x_n) \\
                        & = \ \sum_k:O_{k}(x_1,\ldots,x_n):t_n^{(k)}(x_1-x_n,\ldots,x_{n-1}-x_n). \\
  \end{split}
\end{equation}
In this way we obtain a well-defined time-ordered $n$-point function which is renormalized without introducing any regularization presrciption. With this $T_n$ it is possible to compute UV-finite physical quantities, like amplitudes or cross-sections.   
%
%
\section{Gauge Invariance in Classical Linearized Gravity}
The general theory of relativity can be derived from the Einstein-Hilbert Lagrangian 
\begin{equation}
  L_{{\scriptscriptstyle EH}}=-\frac{2}{\kappa^2}\sqrt{-g}R
  \label{eq:Einstein-Hilbert-Lagrangefunktion}
\end{equation}
where $R=g^{\mu\nu}R_{\mu\nu}$ is the Ricci scalar, $g$ is the determinant of the matrix $(g^{\mu\nu})$ and $\kappa^2=32\pi G$ ($G$ is Newton's gravitational constant)\cite{wa:gr,we:gc}. It is convenient to work with Goldberg variables~\cite{go:clgr}
\begin{equation}
  \ti{g}^{\mu\nu}=\sqrt{-g}g^{\mu\nu}.
  \label{eq:Goldberg-Variablen}
\end{equation}
As was already mentioned in the introduction we will in this section consider the linearized theory of gravity, therefore we expand the inverse metric $\ti{g}^{\mu\nu}$ in an asymptotically flat geometry
\begin{equation}
  \ti{g}^{\mu\nu}=\eta^{\mu\nu}+\kappa h^{\mu\nu}.
  \label{eq:Definition-Graviton-Feld}
\end{equation}
Here $\eta^{\mu\nu}=\text{diag}(+1,-1,-1,-1)$ is the metric of Minkowski spacetime. All tensor indices will be raised and lowered with $\eta^{\mu\nu}$. The quantity $h^{\mu\nu}$ is a symmetric second rank tensor field, which describes gravitons after quantization. Formally (\ref{eq:Einstein-Hilbert-Lagrangefunktion}) becomes an infinite power series in $\kappa$
\begin{equation}
  L_{{\scriptscriptstyle EH}}=\sum_{j=0}^{\infty}\kappa^jL_{{\scriptscriptstyle EH}}^{(j)}.
  \label{eq:Entwicklung-Einstein-Hilbert-Lagrangefunktion}
\end{equation}
The lowest order term $L_{{\scriptscriptstyle EH}}^{(0)}$ is quadratic in $h^{\mu\nu}(x)$ and defines the free asymptotic fields. The linearized Euler-Lagrange equations of motion for $h^{\mu\nu}(x)$ are
\begin{equation}
  \Box h^{\mu\nu}(x)-\frac{1}{2}\eta^{\mu\nu}\Box h(x)-h^{\mu\ro,\nu}_{\pro}(x)-h^{\nu\ro,\mu}_{\pro}(x)=0
\end{equation}
where $h(x)=h^{\mu}_{\mu}(x)$. This equation is invariant w.r.t. gauge transformations of the form
\begin{equation}
  h^{\mu\nu}\longrightarrow \ti{h}^{\mu\nu}+u^{\mu,\nu}+u^{\nu,\mu}-\eta^{\mu\nu}u^{\ro}_{\pro}
\label{classicalgauge}
\end{equation}
where $u^{\mu}$ is a vector field which satisfies the wave equation
\begin{equation}
  \Box u^{\mu}(x)=0.
\end{equation}
These gauge transformation correspond to the general covariance in it's linearized form of the metric tensor $g_{\mu\nu}(x)$. The corresponding gauge condition, compatible with (\ref{classicalgauge}) is the Hilbert-gauge 
\begin{equation}
  h^{\mu\nu}_{\pmu}=0.
\end{equation}
Then the dynamical equation for the graviton field $h^{\mu\nu}$ reduces to the wave equation
\begin{equation}
  \Box h^{\mu\nu}(x)=0.
  \label{eq:Wellengleichung-Graviton-Feld}
\end{equation}
The first order term $L_{{\scriptscriptstyle EH}}^{(1)}$ gives the trilinear self-coupling of the gravitons
\begin{equation}
  L_{{\scriptscriptstyle EH}}^{(1)}=\frac{1}{2}\,h^{\ro\si}\Bigl(h^{\al\be}_{\pro}h^{\al\be}_{\psig}-\frac{1}{2}\,h^{}_{\pro}
                                h^{}_{\psig}+2\,h^{\al\ro}_{\pbe}h^{\be\si}_{\pal}+h^{}_{\pal}h^{\ro\si}_{\pal}
                                -2\,h^{\al\ro}_{\pbe}h^{\al\si}_{\pbe}\Bigr).
  \label{eq:klassische-Einstein-Hilbert-Lagrangefunktion}
\end{equation}

There exists many alternative derivations of this result (\ref{eq:klassische-Einstein-Hilbert-Lagrangefunktion}), starting from massless tensor fields and requiring consistency with gauge invariance in some sense \cite{ve:qtg,no:brstgf,op:ifs2ee}. In classical theory the work closest to our non-geometrical point of view is the one of Ogievetsky and Polubarinov~\cite{op:ifs2ee}. These authors analyze spin-$2$ theories by working with a generalized Hilbert-gauge condition to exclude the spin one part from the outset. They impose an invariance under infinitesimal gauge transformations of the form
\begin{equation}
  \delta h^{\mu\nu}=\partial^{\mu}u^{\nu}+\partial^{\nu}u^{\mu}+\eta^{\mu\nu}\partial_{\al}u^{\al}
  \label{eq:klassische-Eichinvarianz}
\end{equation}
and get Einstein's theory at the end. Instead Wyss~\cite{wy:ug} considers the coupling to matter. Then the self-coupling of the tensor-field (\ref{eq:klassische-Einstein-Hilbert-Lagrangefunktion}) is necessary for consistency. Wald~\cite{wa:stfgc} derives a divergence identity which is equivalent to an infinitesimal gauge invariance of the theory. Einstein's theory is the only non-trivial solution of this identity. In quantum theory the problem was studied by Boulware and Deser~\cite{bd:cgrqg}. These authors require Ward identities associated with the graviton propagator to implement gauge invariance. All authors get Einstein's theory as the unique classical limit if the theory is purely spin two without a spin one admixture.
%
%
\section{Perturbative Quantum Gauge Invariance}
In this work we will study the problem without any reference to the metric tensor by means of perturbative quantum gauge invariance. This method which was worked out for spin-1 non-abelian gauge theories (massless~\cite{as:ym} and massive~\cite{ds:et,as:pgiet2}) in last years proceeds as follows: First one defines infinitesimal gauge variations on free fields. In the case of tensor fields it looks like (\ref{eq:klassische-Eichinvarianz}) where $u^{\mu}(x)$, instead of being an arbitrary function, is now a Fermi field which satisfies the wave equation. $u^{\mu}(x)$ may be regarded as a free Fadeev-Popov ghost field. These ghost fields play a very important role in connection with gauge invariance. We write down the most general trilinear coupling $T_1$ between the graviton and ghost fields which is compatible with Lorentz covariance, power counting and certain basic properties (like zero ghost number). Next we impose first order gauge invariance which strongly restricts the form of $T_1$. Among the possible solutions we recover Einstein's theory $L_{{\scriptscriptstyle EH}}^{(1)}$. The general solution can be written as a linear combination of $L_{{\scriptscriptstyle EH}}^{(1)}$ and divergences as well as coboundaries. In the perturbative construction of the $S$-matrix we next have to calculate the time-ordered product $T\{T_1(x)T_1(y)\}=T_2(x,y)$ by means of causality~\cite{eg:rlp,sch:qed}. Then Schorn~\cite{sch:giqgca} has shown that second order gauge invariance gives further restrictions, in particular, in the case of gravity it requires quartic normalization terms of the form $L_{{\scriptscriptstyle EH}}^{(2)}$ in the above expansion (\ref{eq:Entwicklung-Einstein-Hilbert-Lagrangefunktion}). In this way the so-called proliferation of couplings can be overcome by perturbative gauge invariance.

Our fundamental free asymptotic fields are a symmetric tensor field of rank two $h^{\mu\nu}(x)$ and ghost and anti-ghost fields $u^{\mu}(x)$ and $\ti{u}^{\nu}(x)$. We consider these fields in the background of Minkowski spacetime. A symmetrical tensor field has ten degrees of freedom, which are more than the five independent components of a spin-$2$ field. The additional degrees of freedom can be eliminated by imposing two further conditions~\cite{op:ifs2ee}, namely
\begin{equation}
  h^{\mu\nu}(x)_{\pnu}=0\quad \text{and}\quad {h^{\mu}}_{\mu}(x)=0.
  \label{eq:Elimination von Spin 0 und Spin 1 Komponenten}
\end{equation}
They are disregarded in the construction of the gauge theory and must be considered later in the characterization of physical states~\cite{gr:cqg,gr:cqg2}.

Our tensor field $h^{\mu\nu}(x)$ will be quantized as a massless field~\cite{fe:qtg,gu:qegfla,gu:eotg} as follows
\begin{equation}
  \bigl[h^{\al\be}(x),h^{\mu\nu}(y)\bigr]=-ib^{\al\be\mu\nu}D_0(x-y)
  \label{eq:h-Quantisierung}
\end{equation}
where $D_0(x-y)$ is the massless Pauli-Jordan distribution and the tensor $b^{\al\be\mu\nu}$ is constructed from the Minkowski metric $\eta^{\mu\nu}$ in the following way
\begin{equation}
  b^{\al\be\mu\nu}=\frac{1}{2}\bigl(\eta^{\al\mu}\eta^{\be\nu}+\eta^{\al\nu}\eta^{\be\mu}-
                   \eta^{\al\be}\eta^{\mu\nu}\bigr).
  \label{eq:b-tensor}
\end{equation}
We can write down the Fourier-representation of the field $h^{\mu\nu}(x)$. It is given by
\begin{equation}
  h^{\al\be}(x)=(2\pi)^{-3/2}\int \frac{d^3\vec{k}}{\sqrt{2\omega(\vec{k})}}\bigl(a^{\al\be}(\vec{k})\exp(-ikx)+a^{\al\be}(\vec{k})^{\tot}\exp(+ikx)\bigr).
\label{h-field}
\end{equation}
Here $\omega(\vec{k})=|\vec{k}|$ and $a^{\al\be}(\vec{k})$, $a^{\al\be}(\vec{k})^{\tot}$ are annihilation and creation operators on a bosonic Fock-space. From (\ref{eq:h-Quantisierung}) we find that they have the following commutation relations
\begin{equation}
  \bigl[a^{\al\be}(\vec{k}),a^{\mu\nu}(\vec{k}^{\prime})^{\tot}\bigr]=b^{\al\be\mu\nu}\delta^{(3)}(\vec{k}-\vec{k}^{\prime}).
\label{h-com}
\end{equation}

In analogy to spin-$1$ theories one introduces a gauge charge operator by
\begin{equation}
  Q:=\int\limits_{x^{0}=t} h^{\al\be}(x)_{\pbe}\overset{\leftrightarrow}{\partial_0}u^{\al}d^3x.
  \label{eq:Q}
\end{equation}
For the construction of the physical subspace and in order to prove the unitarity of the $S$-matrix we want to have a nilpotent operator $Q$. Therefore we have to quantize the ghost fields with anticommutators
\begin{equation}
  \bigl\{u^{\mu}(x),\ti{u}^{\nu}(y)\bigr\}=i\eta^{\mu\nu}D_0(x-y)
  \label{eq:u-Quantisierung}
\end{equation}
and all other anti-commutators vanishing. All asymptotic fields fulfil the wave equation
\begin{equation}
  \begin{split}
    \Box\, h^{\mu\nu}(x)   & = 0 \\
    \Box\, u^{\al}(x)      & = 0 \\
    \Box\, \ti{u}^{\be}(x) & = 0 \\
  \end{split}
  \label{eq:Wellengleichung}
\end{equation}
The gauge charge $Q$ (\ref{eq:Q}) defines a gauge variation by
\begin{equation}
  d_QF:=QF-(-1)^{n_g(F)}FQ
  \label{eq:Eichvariation}
\end{equation}
where $n_g$ is the ghostnumber. This is the number of ghost fields minus the number of anti-ghost fields in the Wick monomial $F$. The operator $d_Q$ obeys the Leibniz rule
\begin{equation}
  d_Q(AB)=(d_QA)B+(-1)^{n_g(A)}Ad_QB
  \label{eq:Leibniz-Regel}
\end{equation}
where $A$ and $B$ are arbitrary operators.
We obtain the following gauge variations of the fundamental fields:
\begin{align}
  d_Qh^{\mu\nu}   & = -\frac{i}{2}\bigl(u^{\mu}_{\pnu}+u^{\nu}_{\pmu}-
                      \eta^{\mu\nu}u^{\al}_{\pal}\bigr) \label{eq:Eichvariation-h-Tensor} \\
  d_Qh            & = \ iu^{\mu}_{\pmu} \label{eq:Eichvariation-h-Spur} \\
  d_Q\ti{u}^{\mu} & = \ ih^{\mu\nu}_{\pnu} \label{eq:Eichvariation-Anti-Geist} \\
  d_Qu^{\mu}      & = \ 0 \label{eq:Eichvariation-Geist}
\end{align}
From (\ref{eq:Eichvariation-h-Tensor}) we immediately see
\begin{equation}
  d_Qh^{\mu\nu}_{\pmu}=0.
  \label{eq:Eichvariation-h-Tensor-Ableitungskontraktion}
\end{equation}
The result (\ref{eq:Eichvariation-h-Tensor}) agrees with the infinitesimal gauge transformations of the Goldberg variables, so that our quantization (\ref{eq:h-Quantisierung}) and choice of $Q$ corresponds to the classical framework described in section 3. The asymptotic fields will be used to construct the time-ordered products $T_n$ in the adiabatically switched $S$-matrix
\begin{equation}
  S(g)=\mathbf{1}+\sum_{n=1}^{\infty}\frac{1}{n!}\int d^4x_1\ldots d^4x_n\,T_n(x_1,\ldots,x_n)g(x_1)\ldots g(x_n)
  \label{eq:S-Matrix}
\end{equation}
where $g\in\mathcal{S}(\mathbb{R}^4)$ is a test function. The time-ordered products $T_n$ are operator-valued distributions and they can be expressed by normally ordered products of free fields, see section 2. It is very important that gauge invariance of the $S$-matrix can be directly formulated in terms of the $T_n$ \cite{sch:qgt}. First order gauge invariance means that $d_QT_1$ is a divergence in the sense of vector analysis, i.e.
\begin{equation}
  d_QT_1(x)=i\partial_{\mu}T_{1/1}^{\mu}(x).
  \label{eq:Eichinvarianz-1.-Ordnung}
\end{equation}
The definition of the $n$-th order gauge invariance then reads 
\begin{equation}
  d_QT_n=\bigl[Q,T_n\bigr]=i\sum_{l=1}^{n}\frac{\partial}{\partial x^{\mu}_l}T_{n/l}^{\mu}(x_1,\ldots,x_l,\ldots,x_n).
  \label{eq:Eichinvarianz-n.te-Ordnung}
\end{equation}
Here $T_{n/l}^{\mu}$ is the time ordered product with a gauge variated vertex $T_{1/1}^{\mu}(x_l)$ at position $x_l$ and ordinary vertices $T_1$ at the other arguments.  
%
%
\section{Structure of the Interaction}
Here we introduce the self-couplings of the quantum tensor field $h^{\mu\nu}(x)$. The simplest expression leading to a self-interacting spin-$2$ field theory is a trilinear coupling of the quantum fields $h^{\mu\nu}(x)$ and $h(x)\equiv {h^{\mu}}_{\mu}(x)$. We require Lorentz invariance and in addition to that two derivatives acting on the fields. This is for the following reasons: First of all, by inspection of all trilinear self-interaction terms without derivatives, it is easily seen that such a theory cannot be gauge invariant to first order of perturbation theory. Therefore an interaction without derivatives can be ruled out. Secondly, it is impossible to form a Lorentz-scalar from three rank-$2$ tensor fields with only one derivative. Last but not least the corresponding trilinear expression in the expansion of the Einstein-Hilbert Lagrangian contains two derivatives as well. Therefore we're able to reproduce the results from classical general relativity.

In the following all fields are free fields obeying the free field equations of motion. All products of two or more fields at the same spacetime point $x$ are viewed as normal products. Then the general ansatz for a combination of three field operators contains $12$ terms\footnote{We use the following convention regarding the indices. All vector and tensor indices are written as superscript, whereas all partial derivatives are written as subscript in the abbreviated form with a prime in front of the index, i.e.: $A(x)_{\pnu}=\partial A(x)/\partial x^{\nu}$. All indices will be raised and lowered by the Minkowski metric $\eta_{\mu\nu}$ and will be properly contracted like $A^{\mu}B^{\mu}:=\eta_{\mu\nu}A^{\mu}B^{\nu}$.}:
\begin{equation}
  \begin{split}
    T_1^h(x):= & \ a_1:h^{\mu\nu}(x)_{\pmu}h^{}(x)_{\pnu}h(x):+\,a_2:h^{\mu\nu}(x)h^{}(x)_{\pmu}
                 h^{}(x)_{\pnu}: \\
               & +a_3:h^{\al\be}(x)_{\pal}h^{\be\mu}(x)_{\pmu}h(x):+\,a_4:h^{\al\be}(x)_{\pal}h^{\be\mu}(x)
                 h^{}(x)_{\pmu}: \\
               & +a_5:h^{\al\be}(x)h^{\be\mu}(x)_{\pal}h^{}(x)_{\pmu}:+\,a_6:h^{\al\be}(x)_{\pmu}
                 h^{\be\mu}(x)_{\pal}h(x): \\
               & +a_7:h^{\mu\nu}(x)_{\pmu}h^{\al\be}(x)_{\pnu}h^{\al\be}(x):+\,a_8:h^{\mu\nu}(x)
                 h^{\al\be}(x)_{\pmu}h^{\al\be}(x)_{\pnu}: \\
               & +a_9:h^{\mu\nu}(x)_{\pal}h^{\nu\al}(x)_{\pbe}h^{\mu\be}(x):+\,a_{10}:h^{\mu\nu}(x)_{\pal}
                 h^{\nu\al}(x)h^{\mu\be}(x)_{\pbe}: \\
               & +a_{11}:h^{\mu\nu}(x)h^{\nu\al}(x)_{\pal}h^{\mu\be}(x)_{\pbe}:+\,a_{12}:h^{\mu\nu}(x)
                 h^{\nu\al}(x)_{\pbe}h^{\mu\be}(x)_{\pal}: \\
  \end{split}
  \label{eq:T1h}
\end{equation}
Here we have omited all terms which are divergences. These are terms with a contraction on the two derivatives, e.g. 
\begin{equation*}
  h^{}(x)_{\pal}h^{}(x)_{\pal}h(x)=1/2\,\partial_{\al}\bigl(h^{}(x)_{\pal}h(x)h(x)\bigr).
\end{equation*}
Furthermore all terms with two derivatives acting on the same field can be transformed into a divergence plus a term already contained in (\ref{eq:T1h}). These terms would modify our ansatz only in a redefinition of some parameters $a_i$ and can be omited without losing generality.

As in the cases of Yang-Mills theory~\cite{dhks:ccymt1,dhks:ccymt2} and Einstein gravity~\cite{sch:giqgca} we expect to get a gauge invariant first order coupling only if we couple the tensor field $h^{\mu\nu}$ also to ghost and anti-ghost fields. The most general expression with zero ghost-number is 
\begin{equation}
  \begin{split} 
    T_1^u(x):= & \ b_1:u^{\ro}(x)_{\pnu}\ti{u}^{\mu}(x)_{\pro}h^{\mu\nu}(x):+\,b_2:u^{\ro}(x)_{\pnu}
                 \ti{u}^{\mu}(x)h^{\mu\nu}(x)_{\pro}: \\
               & +b_3:u^{\ro}(x)\ti{u}^{\mu}(x)_{\pnu}h^{\mu\nu}(x)_{\pro}:+\,b_4:u^{\ro}(x)_{\pro}
                 \ti{u}^{\mu}(x)_{\pnu}h^{\mu\nu}(x): \\
               & +b_5:u^{\ro}(x)_{\pro}\ti{u}^{\mu}(x)h^{\mu\nu}(x)_{\pnu}:+\,b_6:u^{\ro}(x)
                 \ti{u}^{\mu}(x)_{\pro}h^{\mu\nu}(x)_{\pnu}: \\
               & +b_7:u^{\ro}(x)_{\pmu}\ti{u}^{\mu}(x)_{\pro}h(x):+\,b_8:u^{\ro}(x)_{\pmu}\ti{u}^{\mu}(x)
                 h^{}(x)_{\pro}: \\
               & +b_9:u^{\ro}(x)\ti{u}^{\mu}(x)_{\pmu}h^{}(x)_{\pro}:+\,b_{10}:u^{\ro}(x)_{\pro}
                 \ti{u}^{\mu}(x)_{\pmu}h(x): \\
               & +b_{11}:u^{\ro}(x)_{\pro}\ti{u}^{\mu}(x)h^{}(x)_{\pmu}:+\,b_{12}:u^{\ro}(x)
                 \ti{u}^{\mu}(x)_{\pro}h^{}(x)_{\pmu}: \\
               & +b_{13}:u^{\ro}(x)_{\pmu}\ti{u}^{\mu}(x)_{\pnu}h^{\ro\nu}(x):+\,b_{14}:u^{\ro}(x)_{\pmu}
                 \ti{u}^{\mu}(x)h^{\ro\nu}(x)_{\pnu}: \\
               & +b_{15}:u^{\ro}(x)\ti{u}^{\mu}(x)_{\pmu}h^{\ro\nu}(x)_{\pnu}:+\,b_{16}:u^{\ro}(x)_{\pnu}
                 \ti{u}^{\mu}(x)_{\pmu}h^{\ro\nu}(x): \\
               & +b_{17}:u^{\ro}(x)_{\pnu}\ti{u}^{\mu}(x)h^{\ro\nu}(x)_{\pmu}:+\,b_{18}:u^{\ro}(x)
                 \ti{u}^{\mu}(x)_{\pnu}h^{\ro\nu}(x)_{\pmu}: \\
               & +b_{19}:u^{\mu}(x)_{\pnu}\ti{u}^{\mu}(x)_{\pro}h^{\ro\nu}(x):+\,b_{20}:u^{\mu}(x)_{\pnu}
                 \ti{u}^{\mu}(x)h^{\ro\nu}(x)_{\pro}: \\
               & +b_{21}:u^{\mu}(x)\ti{u}^{\mu}(x)_{\pnu}h^{\ro\nu}(x)_{\pro}: \\
  \end{split}
  \label{eq:T1u}
\end{equation} 
We will suppress all arguments of the field operators as well as the double dots of normal ordering in subsequent expressions. The complete first order coupling is then given by:
\begin{equation}
  T_1(x) := T_1^h(x)+T_1^u(x).
  \label{eq:T1}
\end{equation}
In the following analysis we want to study in which way the parameters of the theory $a_1,\ldots, a_{12}$ and $b_1,\ldots, b_{21}$ will be restricted due to first order gauge invariance.
%
%
\input{divclass.tex}
%
%
\section{Relations from First Order Gauge Invariance}
Now we will work out the consequences of first order gauge invariance. This means that we compute the gauge variation of the first order coupling $T_1$ and require that the result can be written in the form $\partial_{\mu}\wti{T}_{1/1}^{\mu}$, where $\wti{T}_{1/1}^{\mu}$ is given by (\ref{eq:T1/1}) with the redundant terms eliminated according to the discussion above.  
%
%
\subsection{Type E Divergences}
In this subsection we consider the type $E$ divergences explicitly. From the comparison of these divergences with  $d_QT_1|_{Type E}$ we will get linear relations among the coupling parameters of $T_1$. We require the following equation to be satisfied
\begin{equation}
  d_Q T_1|_{Type E} = \partial_{\mu} \wti{T}_{1/1}^{\mu,E}.
  \label{eq:Eichinvarianz-TypE}
\end{equation}
Calculating the right side of this equation we get
\begin{equation}
  \begin{split}
    \partial_{\mu} \wti{T}_{1/1}^{\mu,E} = & \ d_{29} u^{\al}_{\pmu\psig} h^{\al\nu}_{\pnu} h^{\mu\si}+
                                             d_{30} u^{\al}_{\pmu} h^{\al\nu}_{\psig\pnu} h^{\mu\si}+
                                             d_{31} u^{\al}_{\pmu} h^{\al\nu}_{\pnu} h^{\mu\si}_{\psig}+
                                             d_{32} u^{\al}_{\pmu\psig} h^{\al\nu} h^{\mu\si}_{\pnu} \\
                                         & + d_{33} u^{\al}_{\pmu} h^{\al\nu}_{\psig} h^{\mu\si}_{\pnu}+
                                             d_{34} u^{\al}_{\pmu} h^{\al\nu} h^{\mu\si}_{\psig\pnu}+
                                             d_{35} u^{\al} h^{\al\nu}_{\pmu\psig} h^{\mu\si}_{\pnu}+
                                             d_{36} u^{\al} h^{\al\nu}_{\pmu} h^{\mu\si}_{\psig\pnu} \\
                                         & + d_{37} u^{\al}_{\pmu\pnu} h^{\al\nu}_{\psig} h^{\mu\si}+
                                             d_{38} u^{\al}_{\pnu} h^{\al\nu}_{\psig\pmu} h^{\mu\si}+
                                             d_{39} u^{\al}_{\pnu} h^{\al\nu}_{\pmu} h^{\mu\si}_{\psig}+
                                             d_{40} u^{\al}_{\psig\pnu} h^{\al\nu} h^{\mu\si}_{\pmu} \\
                                         & + d_{41} u^{\al}_{\pnu} h^{\al\nu} h^{\mu\si}_{\pmu\psig}+
                                             d_{42} u^{\al} h^{\al\nu}_{\pmu\pnu} h^{\mu\si}_{\psig}+
                                             d_{43} u^{\al} h^{\al\nu}_{\pnu} h^{\mu\si}_{\pmu\psig}+
                                             d_{44} u^{\al}_{\psig\pnu\pmu} h^{\al\nu}h^{\mu\si} \\
                                         & + d_{45} u^{\al} h^{\al\nu}_{\psig\pnu\pmu} h^{\mu\si}+
                                             d_{46} u^{\al} h^{\al\nu} h^{\mu\si}_{\pmu\pnu\psig}. \\
  \end{split}
  \label{eq:TypE-Monome}
\end{equation}
The new constants are defined as follows
\begin{equation}
  \begin{split}
    d_{29} := & \ c_{42}+c_{48},\quad d_{30} := c_{40}+c_{42}+c_{51},\quad d_{31} :=  c_{42}+c_{47}+c_{52} \\
    d_{32} := & \ c_{43}+c_{48},\quad d_{33} := c_{43}+c_{44}+c_{51},\quad d_{34} := c_{41}+c_{43}+c_{52} \\ 
    d_{35} := & \ c_{44}+c_{49},\quad d_{36} := c_{41}+c_{44}+c_{53},\quad d_{37} := c_{39}+c_{45}+c_{51} \\
    d_{38} := & \ c_{45}+c_{49},\quad d_{39} := c_{45}+c_{46}+c_{53},\quad d_{40} := c_{39}+c_{46}+c_{52} \\
    d_{41} := & \ c_{46}+c_{50},\quad d_{42} := c_{40}+c_{47}+c_{53},\quad d_{43} := c_{47}+c_{50} \\ 
    d_{44} := & \ c_{39}+c_{48},\quad d_{45} := c_{40}+c_{49},\quad d_{46} := c_{41}+c_{50}. \\
  \end{split}
  \label{eq:d-TypE}
\end{equation}
From equation~(\ref{eq:Eichinvarianz-TypE}) we see that
\begin{equation}
  \begin{split}
    d_{29} = & -\frac{i}{2}\,a_{10},\quad d_{30} = -ib_{19},\quad d_{31} = -i\bigl(a_{11}+b_{14}+b_{20}\bigr) \\
    d_{32} = & -\frac{i}{2}\,a_{9},\quad d_{33} = -ia_{12},\quad d_{34} = -ib_{13} \\
    d_{35} = & \ 0,\quad d_{36} = -ib_{18},\quad d_{37} = -i\Bigl(2a_{8}+\frac{1}{2}\,a_{9}\Bigr) \\
    d_{38} = & \ 0,\quad d_{39} = -i\bigl(a_{7}+b_{17}\bigr),\quad d_{40} = -i\Bigl(a_{7}+\frac{1}{2}\,a_{10}\Bigr) \\
    d_{41} = & -ib_{16},\quad d_{42} = -ib_{21},\quad d_{43} = -ib_{15} \\
    d_{44} = & \ 0,\quad d_{45} = 0,\quad d_{46} = 0. \\
  \end{split}
  \label{eq:Zuordnung-TypE}
\end{equation}
Finally we arrive at the divergence form if we invert the system~(\ref{eq:d-TypE}). This is done in appendix B. Let $M_E\in Mat(18\times 15,\mathbb{Z})$ be the coefficient matrix of (\ref{eq:d-TypE}). Then we can write this system of equations as
\begin{equation}
  M_E\cdot \mathbf{c}^E=\mathbf{d}^E
\label{eq:d-TypE-MatrixForm}
\end{equation}
where $\mathbf{c}^E\in\mathbb{C}^{15}$ and $\mathbf{d}^E\in\mathbb{C}^{18}$ are the column vectors with components $(c_{39},\ldots,c_{53})$ and $(d_{29},\ldots,d_{46})$ respectively. Now we observe two things:
\begin{enumerate}
  \item  For a solution to exist it is necessary to fulfil 
    \begin{equation}
      \begin{split}
        d_{32} & +d_{36}-d_{33}-d_{46}-d_{44}-d_{39}+d_{37}+d_{41}= 0 \\
        d_{32} & +d_{36}-d_{33}-2d_{44}+d_{40}-d_{39}+d_{37}-d_{31}+d_{29}+d_{43}-d_{46}= 0 \\
        d_{32} & -d_{33}-2d_{44}+d_{40}-d_{39}+d_{37}+d_{35}-d_{31}+d_{29}+d_{42}-d_{45}= 0 \\
        d_{36} & +2d_{32}-d_{33}-2d_{44}+d_{40}-d_{39}+d_{37}-d_{34}= 0 \\
        d_{42} & +2d_{29}-d_{30}+d_{40}-2d_{44}-d_{39}+d_{37}-d_{31}= 0 \\
        d_{38} & +d_{40}-d_{39}-d_{44}-d_{31}+d_{29}+d_{42}-d_{45}= 0. \\
      \end{split}
      \label{eq:TypE-d-Relationen}
    \end{equation}
  \item $rank(M_E)=12$.
\end{enumerate}
 
From 2. we get the information that the representation of $d_QT_1|_{Type E}$ as a divergence is not unique. But the important results are the equations (\ref{eq:TypE-d-Relationen}), because we obtain relations among the coupling parameters if we use (\ref{eq:Zuordnung-TypE}):  
\begin{align}
  a_{7} & -2a_{8}-a_{9}+a_{12}-b_{16}+b_{17}-b_{18}= 0 \label{eq:TypE-Relation1} \\
  -2a_{8} & -a_{9}-a_{10}+a_{11}+a_{12}+b_{14}-b_{15}+b_{17}-b_{18}+b_{20}= 0 \label{eq:TypE-Relation2} \\
  -2a_{8} & -a_{9}-a_{10}+a_{11}+a_{12}+b_{14}+b_{17}+b_{20}-b_{21}= 0 \label{eq:TypE-Relation3} \\
  -2a_{8} & -\frac{3}{2}a_{9}-\frac{1}{2}a_{10}+a_{12}+b_{13}+b_{17}-b_{18}= 0 \label{eq:TypE-Relation4} \\
  -2a_{8} & -\frac{1}{2}a_{9}-\frac{3}{2}a_{10}+a_{11}+b_{14}+b_{17}+b_{19}+b_{20}-b_{21}= 0 \label{eq:TypE-Relation5} \\
  -a_{10} & +a_{11}+b_{14}+b_{17}+b_{20}-b_{21}= 0. \label{eq:TypE-Relation6}
\end{align}
These equations are direct consequences of first order gauge invariance.
%
%
\subsection{Divergences of Type $F,G,J$}
In analogy to the case of Lorentz type $E$ we obtain linear relations among the coupling parameters from the types $F,G$ and $J$. One finds the following $9$ linear independent relations
\begin{align}
  -a_4 & -a_5-a_9-a_{10}-b_2+b_3-b_4= 0 \label{eq:Rela7} \\
  -2a_3 & -2a_6-a_9-a_{10}-a_{11}-a_{12}+b_3-2b_4+2b_5-b_6= 0 \label{eq:Rela8} \\
  a_5 & -a_6= 0 \label{eq:Rela9} \\
  -a_4 & +a_5-2a_6-\frac{1}{2}a_9-\frac{1}{2}a_{10}-b_1-b_4= 0 \label{eq:Rela10} \\
  -2a_2 & -a_5+a_6-a_8+b_8+b_{11}= 0 \label{eq:Rela11} \\
 -2a_1 & +2a_2-a_4+2a_5-3a_6-a_7+a_8-b_8-2b_{10}+b_{11}-2b_{12}= 0 \label{eq:Rela12} \\
  a_5 & -a_6-b_9-b_{12}= 0 \label{eq:Rela13} \\
  -\frac{1}{2}a_4 & +\frac{3}{2}a_5-2a_6+b_7-b_8-b_{12}= 0 \label{eq:Rela14} \\
  -2a_2 & +a_5-a_6-a_8-b_8-b_{11}= 0. \label{eq:Rela15}
\end{align}
Together with the six relations from type $E$ (\ref{eq:TypE-Relation1}--\ref{eq:TypE-Relation6}) we get $15$ linear independent equations which restrict the admissible theories. By construction these equations are necessary for a spin-$2$ theory to be gauge invariant.
%
%
\subsection{Nilpotency of $Q$}
The gauge charge operator is by definition nilpotent $(Q^2=0)$. As a consequence the application of twice the gauge variation to every expression must vanish, i.e.
\begin{equation}
  (d_Q)^2T_1(x)=0.
  \label{eq:Q2}
\end{equation}
If we now use the gauge invariance of $T_1$ to first order, we get additional constraints for the $Q$-vertex $T_{1/1}^{\mu}$, namely
\begin{equation}
  d_Q\left(\partial_{\mu}^x T_{1/1}^{\mu}(x)\right)=0.
  \label{eq:Q2-Bedingung}
\end{equation}
This equation gives us restrictions on the parameters of $T_{1/1}^{\mu}$. After a lengthy calculation one arrives at exactly $63$ linear independent coefficients. The remaining ones can be expressed as linear combinations of them. One might think that these linear dependences may produce further necessary conditions beside the fifteen above. But it turns out that this is not the case. As was mentioned above it is shown in appendix B that the arbitrariness in the Lorentz types $E,F,G$ and $J$ can be shifted into a form with vanishing divergence. In view of this one can say that the $Q$-vertex is unique modulo redundant terms. In order to deal with more simple algebraic expressions it is convenient to work with the full set of $98$ different parameters since otherwise we must work with long linear combinations of the independent parameters because the Lorentz types get mixed. We have convinced ourself that the relation (\ref{eq:Q2-Bedingung}) is always satisfied as soon as gauge invariance to first order holds.
%
%
\section{Gauge Invariant Spin-$2$ Theories}
The preceding section has shown what kind of restrictions we obtain if we require the theory to be gauge invariant. The $15$ equations (\ref{eq:TypE-Relation1}--\ref{eq:Rela15}) we have found for the $33$ parameters $a_1,\ldots,a_{12}$ as well as $b_1,\ldots,b_{21}$ play a central role. Now we can look at an arbitrary solution to this set of equations. The corresponding $T_1$ is then gauge invariant to first order for the following reason. We have to write the gauge variation of this $T_1$ as a divergence in the sense of vector analysis. Because of the generality of our ansatz for the $Q$-vertex every term in $d_QT_1$ can be uniquely identified with a $d_j$-monomial in $\wti{T}_{1/1}^{\mu}$. With the help of the equations from appendix A we can then find a unique divergence for the types $A,B,C,D,H$ and $K$. For the other types we can also find a divergence which is unique modulo redundant terms.

Summing up we have proven the following proposition
\begin{proposition}
Let $T_1$ and $\wti{T}_{1/1}^{\mu}$ be given as above, furthermore let $f$ be the following mapping
\begin{align*}
  f : \text{(Vectorspace of Wick-monomials)} & \longrightarrow (\text{Vectorspace of coefficients}\  a_i,\,b_j) \\
          a_1h^{\mu\nu}_{\pmu}h^{}_{\pnu}h+\ldots+
          b_{21}u^{\mu}\ti{u}^{\mu}_{\pnu}h^{\ro\nu}_{\pro} & \longmapsto (a_1,\ldots,a_{12},b_1,\ldots,b_{21}).
\end{align*}
Let $V\in\mathbb{R}^{33}$ be the space of solutions to (\ref{eq:TypE-Relation1}--\ref{eq:Rela15}). $V$ is an $18$-dimensional subspace of $\mathbb{R}^{33}$, which is characterized through the following injective linear mapping $L:\mathbb{R}^{18}\longrightarrow \mathbb{R}^{33}$ :
\begin{equation*}
  \begin{split}
   &  \bigl(a_6,a_{12},b_3,b_4,b_5,b_6,b_7,b_{10},b_{11},b_{12},b_{13},b_{14},b_{16},b_{17},b_{18},b_{19},
    b_{20},b_{21}\bigr)\longmapsto \\
   & \Bigl(-b_7-b_{10}-\frac{1}{2}\bigl(b_{16}-b_{17}+b_{18}\bigr), \frac{1}{4}\bigl(b_{13}+b_{17}-b_{18}
     -b_{19}\bigr),-a_6+\frac{1}{2}b_3-b_4+b_5 \\
   & -\frac{1}{2}\bigl(b_6+3\,b_{13}-b_{14}\bigr)-b_{17}+\frac{1}{2}\bigl(3\,b_{18}-b_{19}+b_{20}-b_{21}\bigr),
     -a_6+2\,\bigl(b_7+b_{11}-b_{12}\bigr), \\
   & a_6,a_6,b_{16}-b_{17}+b_{18},\frac{1}{2}\bigl(-b_{13}-b_{17}+b_{18}+b_{19}\bigr),a_{12}+b_{13}+b_{17}
     -b_{18}-b_{19},-a_{12} \\
   & +b_{13}+b_{17}-b_{18}+b_{19},-a_{12}+b_{13}-b_{14}-b_{18}+b_{19}-b_{20}+b_{21},a_{12},-b_4-2\,\bigl(b_7 \\
   & +b_{11}-b_{12}\bigr)-b_{13}-b_{17}+b_{18},b_3-b_4-2\,\bigl(b_7+b_{11}-b_{12}+b_{13}+b_{17}-b_{18}\bigr), \\
   & b_3,b_4,b_5,b_6,b_7,-b_{11},-b_{12},b_{10},b_{11},b_{12},b_{13},b_{14},-b_{18}+b_{21},b_{16},b_{17},
     b_{18},b_{19},b_{20},b_{21}\Bigr) \\
  \end{split}
\end{equation*}
Then we have the two equivalent statements:
\begin{align*}
  (A1)\quad  & d_Q T_1(x)=\partial_{\mu}^x \wti{T}_{1/1}^{\mu}(x)\quad 
       \text{and}\quad d_Q\left(\partial_{\mu}^x \wti{T}_{1/1}^{\mu}(x)\right)=0 \\
 (A2)\quad  & f(T_1)\in V=\text{im}(L)
\end{align*}
where $\text{im}(L)$ means the image of the linear mapping $L$.
\end{proposition} 

This proposition determines all gauge invariant spin-$2$ theories up to first order of perturbation theory. Among them there is the trilinear coupling in the expansion of the Einstein-Hilbert Lagrangian. It is given by $L(0,1,-1,-1,\underbrace{0,\ldots,0}_{11\,\text{times}},1,0,0)$ which can be written explicitly 
\begin{equation}
  T_1^{h,EH}=\kappa\Bigl[-\frac{1}{4}\,h^{\mu\nu}h^{}_{\pmu}h^{}_{\pnu}+\frac{1}{2}\,h^{\mu\nu}h^{\al\be}_{\pmu}h^{\al\be}_{\pnu}
             +h^{\mu\nu}h^{\nu\al}_{\pbe}h^{\mu\be}_{\pal}\Bigr].
  \label{eq:Einstein-Hilbert-Kopplung}
\end{equation}
The ghost coupling turns out to be the one first suggested by Kugo and Ojima~\cite{ko:scpsmuimqgt}, namely
\begin{equation}
  T_1^{u,KO}=\kappa\bigl[u^{\ro}_{\pnu}\ti{u}^{\mu}_{\pro}h^{\mu\nu}-u^{\ro}\ti{u}^{\mu}_{\pnu}h^{\mu\nu}_{\pro}-u^{\ro}_{\pro}
             \ti{u}^{\mu}_{\pnu}h^{\mu\nu}+u^{\mu}_{\pnu}\ti{u}^{\mu}_{\pro}h^{\ro\nu}\bigr].
  \label{eq:Kugo-Ojima-Kopplung}
\end{equation}
From the viewpoint of gauge properties of a quantized tensor field we have obtained a set of $18$ linear independent gauge theories. We claim that the most general gauge invariant theory which is given by a linear combination of these 18 linear independent ones is physically equivalent (in the sense explained below) to the trilinear coupling of Einstein-Hilbert~(\ref{eq:Einstein-Hilbert-Kopplung}) plus the ghost coupling of Kugo-Ojima~(\ref{eq:Kugo-Ojima-Kopplung}) up to first order of perturbation theory. 

Let $P_{phys}$ be the projection from the whole Fock-space $\mathcal{F}$ onto the physical subspace $\mathcal{F}_{phys}$, which can be expressed in terms of the kernel and the range of the gauge charge operator $Q$ by 
\begin{equation}
  \mathcal{F}_{phys}=\text{ker}\,Q/\text{ran}\,Q
  \label{eq:physikalischer-Unterraum}
\end{equation}
(see e.g.~\cite{kr:cptmvbt,gr:cqg2}). Then two $S$-matrices $S,S^{\prime}$ describe the same physics if all matrix elements between physical states agree in the adiabatic limit $g\rightarrow 1$, i.e.
\begin{equation}
  \lim_{g\rightarrow 1}(\phi,P_{phys}S(g)P_{phys}\psi)=\lim_{g\rightarrow 1}(\phi,P_{phys}S^{\prime}(g)P_{phys}\psi),\ \forall \phi,\psi\in \mathcal{F}.
  \label{eq:physikalische-Aequivalenz}
\end{equation}
For theories with massless fields the existence of the adiabatic limit is a problem. To avoid this we work with a perturbative version of (\ref{eq:physikalische-Aequivalenz}):
\begin{equation}
  P_{phys}T_nP_{phys}-P_{phys}T^{\prime}_nP_{phys}=\text{divergences}.
  \label{eq:physikalische-Aequivalenz:stoerungstheoretische-Form}
\end{equation}
Obviously (\ref{eq:physikalische-Aequivalenz:stoerungstheoretische-Form}) for all $n$ implies (\ref{eq:physikalische-Aequivalenz}) if the adiabatic limit exists. Specializing to first order $n=1$ we see that two couplings $T_1$ and $T^{\prime}_1$ which differ by a divergence are physically equivalent to first order. Furthermore, if they differ by a coboundary, i.e. a term 
\begin{equation}
  T_1^{cb}=d_QX
  \label{eq:Corand-Theorien}
\end{equation}
where $X$ has ghostnumber $n_g(X)=-1$, they are also equivalent because of the equation
\begin{equation}
  P_{phys}(d_QX)P_{phys}=P_{phys}QXP_{phys}+P_{phys}XQP_{phys}=0
  \label{eq:projizierte-Coraender}
\end{equation}
since by inspection of (\ref{eq:physikalischer-Unterraum}) we have  
\begin{equation}
   QP_{phys}=0=P_{phys}Q.
  \label{eq:QPPQ}
\end{equation}

Let us return to the space of solutions $V$ from proposition 1. Every vector in $V$ corresponds through the mapping $f^{-1}$ to a gauge invariant theory to first order of perturbation theory. As was mentioned earlier the trilinear coupling of Einstein-Hilbert lies in this space. Now we look at the other theories beside the Einstein-Hilbert coupling. For this purpose we choose a suitable basis in $V$. It turns out that a basis can be choosen so that all theories beside the classical Einstein-Hilbert coupling consists of divergences and coboundaries only, i.e. we have the following theorem: 
\begin{theorem}
Up to first order of perturbation theory the most general gauge invariant trilinear self-coupling of a quantized tensor field $h^{\mu\nu}(x)$ is physically equivalent (in the sense described above) to the one obtained from the expansion of the Einstein-Hilbert Lagrangian (given by~(\ref{eq:Einstein-Hilbert-Kopplung}) without the two divergence terms, see~\cite{sch:giqgca}).
\end{theorem}

The proof of this theorem is as follows: With the notation of proposition 1 we choose a basis $(v_1,\ldots,v_{17},v_{{\scriptscriptstyle EH}})$ in $V$ which displays the vector 
\begin{equation}
  v_{{\scriptscriptstyle EH}}:=L(0,1,-1,-1,\underbrace{0,\ldots,0}_{11\,\text{times}},1,0,0)\in V
\end{equation}
 corresponding to the Einstein-Hilbert coupling with Kugo-Ojima ghost coupling explicitly. We can choose the remaining basis vectors $v_1,\ldots,v_{17}$ in such a way that they have the following property:
\begin{equation}
   f^{-1}(v_i)=\sum d_QX+\text{divergences},\quad \forall i=1,\ldots,17
  \label{eq:Corand-Divergenz-Summe}
\end{equation}
where $X$ is of the form 
\begin{equation}
  X\sim\partial\,\mid \ti{u}hh\quad \text{or}\quad X\sim\partial\,\mid u\ti{u}\ti{u}.
  \label{eq:Coraender}
\end{equation}
We consider the following set of vectors $\{v_i|\,i=1,\ldots,17\}\in V$:
\begin{align}
  v_1    & = (0,0,-1,-1,1,1,\underbrace{0,\ldots,0}_{27\,\text{times}}) \label{eq:v1} \\
  v_2    & = (\underbrace{0,\ldots0}_{8\,\text{times}},1,-1,-1,1,\underbrace{0,\ldots,0}_{21\,\text{times}}) \label{eq:v2} \\
  v_3    & = (\underbrace{0,\ldots,0}_{13\,\text{times}},1,1,0,0,1,\underbrace{0,\ldots,0}_{15\,\text{times}}) \label{eq:v3} \\
  v_4    & = (\underbrace{0,\ldots,0}_{12\,\text{times}},-1,-1,0,1,1,\underbrace{0,\ldots,0}_{16\,\text{times}}) \label{eq:v4} \\
  v_5    & = (\underbrace{0,\ldots,0}_{16\,\text{times}},1,2,\underbrace{0,\ldots,0}_{15\,\text{times}}) \label{eq:v5} \\
  v_6    & = (0,0,1,\underbrace{0,\ldots,0}_{13\,\text{times}},1,\underbrace{0,\ldots,0}_{16\,\text{times}}) \label{eq:v6} \\
  v_7    & = (\underbrace{0,\ldots,0}_{18\,\text{times}},1,1,0,-1,-1,\underbrace{0,\ldots,0}_{10\,\text{times}}) \label{eq:v7} \\
  v_8    & = (0,0,0,2,\underbrace{0,\ldots,0}_{8\,\text{times}},-2,-2,0,0,0,0,1,0,0,-1,\underbrace{0,\ldots,0}_{11\,\text{times}}) \label{eq:v8} \\
  v_9    & = (-1,\underbrace{0,\ldots,0}_{20\,\text{times}},1,\underbrace{0,\ldots,0}_{11\,\text{times}}) \label{eq:v9} \\
  v_{10} & = (\underbrace{0,\ldots,0}_{18\,\text{times}},1,0,-1,-1,0,1,\underbrace{0,\ldots,0}_{9\,\text{times}}) \label{eq:v10} \\
  v_{11} & = (0,0,-3/2,0,0,0,0,0,1,1,1,1,0,-2,-1,-1,\underbrace{0,\ldots,0}_{8\,\text{times}},1,0,0,0,0,0,1,0,0) \label{eq:v11} \\
  v_{12} & = (\underbrace{0,\ldots,0}_{25\,\text{times}},1,1,0,0,0,0,0,1) \label{eq:v12} \\
  v_{13} & = (\underbrace{0,\ldots,0}_{25\,\text{times}},1,0,0,0,0,0,-1,0) \label{eq:v13} \\
  v_{14} & = (0,0,1/2,0,0,0,0,0,-1,1,0,-1,\underbrace{0,\ldots,0}_{19\,\text{times}},1,0) \label{eq:v14} \\
  v_{15} & = (\underbrace{0,\ldots,0}_{24\,\text{times}},-1,0,1,1,0,-1,0,0,0) \label{eq:v15} \\
  v_{16} & = (-1/2,0,0,0,0,0,1,\underbrace{0,\ldots,0}_{20\,\text{times}},1,0,0,0,0,0) \label{eq:v16} \\
  v_{17} & = (\underbrace{0,\ldots,0}_{26\,\text{times}},1,0,-1,-1,0,1,0) \label{eq:v17}. 
\end{align}
It's easy to see that these vectors together with $v_{{\scriptscriptstyle EH}}$ form a basis of $V$. What remains to be done is to show that they indeed have the property (\ref{eq:Corand-Divergenz-Summe}). After a lenghty calculation we have found: 
\begin{align}
  \begin{split}
    f^{-1}(v_1) = & -h^{\al\be}_{\pal}h^{\be\mu}_{\pmu}h-h^{\al\be}_{\pal}h^{\be\mu}h^{}_{\pmu}+h^{\al\be}h^{\be\mu}_{\pal}h^{}_{\pmu}+
                    h^{\al\be}_{\pmu}h^{\be\mu}_{\pal}h \\
                = & \  \partial_{\al}\bigl(h^{\al\be}_{\pmu}h^{\be\mu}h-h^{\al\be}h^{\be\mu}_{\pmu}h\bigr) \\
  \end{split} \\
  \begin{split}
    f^{-1}(v_2) = & \ h^{\mu\nu}_{\pal}h^{\nu\al}_{\pbe}h^{\mu\be}-h^{\mu\nu}_{\pal}h^{\nu\al}h^{\mu\be}_{\pbe}-h^{\mu\nu}h^{\nu\al}_{\pal}h^{\mu\be}_{\pbe}+
                    h^{\mu\nu}h^{\nu\al}_{\pbe}h^{\mu\be}_{\pal} \\
                = & \  \partial_{\al}\bigl(h^{\mu\nu}h^{\nu\al}_{\pbe}h^{\mu\be}-h^{\mu\nu}h^{\nu\al}h^{\mu\be}_{\pbe}\bigr) \\
  \end{split} \\
  \begin{split}
    f^{-1}(v_3) = & \ u^{\ro}_{\pnu}\ti{u}^{\mu}h^{\mu\nu}_{\pro}+u^{\ro}\ti{u}^{\mu}_{\pnu}h^{\mu\nu}_{\pro}+u^{\ro}\ti{u}^{\mu}_{\pro}h^{\mu\nu}_{\pnu} \\
                = & \ i\,d_Q\bigl(u^{\mu}\ti{u}^{\nu}_{\pmu}\ti{u}^{\nu}\bigr)+\partial_{\nu}\bigl(u^{\ro}\ti{u}^{\mu}h^{\mu\nu}_{\pro}\bigr) \\
  \end{split} \\
  \begin{split}
    f^{-1}(v_4) = & -u^{\ro}_{\pnu}\ti{u}^{\mu}_{\pro}h^{\mu\nu}-u^{\ro}_{\pnu}\ti{u}^{\mu}h^{\mu\nu}_{\pro}+u^{\ro}_{\pro}\ti{u}^{\mu}_{\pnu}h^{\mu\nu}+
                    u^{\ro}_{\pro}\ti{u}^{\mu}h^{\mu\nu}_{\pnu} \\
                = & \ \partial_{\nu}\bigl(u^{\ro}_{\pro}\ti{u}^{\mu}h^{\mu\nu}-u^{\nu}_{\pro}\ti{u}^{\mu}h^{\mu\ro}\bigr) \\
  \end{split} \\
  \begin{split}
    f^{-1}(v_5) = & \ u^{\ro}_{\pro}\ti{u}^{\mu}h^{\mu\nu}_{\pnu}+2\,u^{\ro}\ti{u}^{\mu}_{\pro}h^{\mu\nu}_{\pnu} \\
                = & \  i\,d_Q\bigl(u^{\mu}\ti{u}^{\nu}_{\pmu}\ti{u}^{\nu}\bigr)+\partial_{\nu}\bigl(u^{\ro}_{\pro}\ti{u}^{\mu}h^{\mu\nu}+u^{\ro}\ti{u}^{\mu}_{\pro}
                    h^{\mu\nu} \\
                  & +u^{\ro}\ti{u}^{\mu}h^{\mu\nu}_{\pro}-u^{\nu}_{\pro}\ti{u}^{\mu}h^{\mu\ro}-u^{\nu}\ti{u}^{\mu}_{\pro}h^{\mu\ro}\bigr) \\
  \end{split} \\
  \begin{split}
    f^{-1}(v_6) = & \ h^{\al\be}_{\pal}h^{\be\mu}_{\pmu}h+u^{\ro}_{\pro}\ti{u}^{\mu}h^{\mu\nu}_{\pnu} \\
                = & -i\,d_Q\bigl(\ti{u}^{\mu}h^{\mu\nu}_{\pnu}h\bigr) \\
  \end{split} \\
  \begin{split}
    f^{-1}(v_7) = & \ u^{\ro}_{\pmu}\ti{u}^{\mu}_{\pro}h+u^{\ro}_{\pmu}\ti{u}^{\mu}h^{}_{\pro}-u^{\ro}_{\pro}\ti{u}^{\mu}_{\pmu}h-
                    u^{\ro}_{\pro}\ti{u}^{\mu}h^{}_{\pmu} \\
                = & \ i\,d_Q\Bigl(\frac{1}{2}\,\ti{u}^{\mu}_{\pmu}hh+\ti{u}^{\mu}h^{}_{\pmu}h\Bigr)+\partial_{\mu}\Bigl(\frac{1}{2}\,h^{\mu\nu}_{\pnu}hh+
                    u^{\mu}_{\pro}\ti{u}^{\ro}h\Bigr) \\
  \end{split} \\
  \begin{split}
    f^{-1}(v_8) = & \ 2\,h^{\al\be}_{\pal}h^{\be\mu}h^{}_{\pmu}-2\,u^{\ro}_{\pnu}\ti{u}^{\mu}_{\pro}h^{\mu\nu}-2\,u^{\ro}_{\pnu}\ti{u}^{\mu}h^{\mu\nu}_{\pro}
                    +u^{\ro}_{\pmu}\ti{u}^{\mu}_{\pro}h-u^{\ro}_{\pro}\ti{u}^{\mu}_{\pmu}h \\
                = & \ i\,d_Q\Bigl(\frac{1}{4}\,\ti{u}^{\mu}_{\pmu}hh+\frac{1}{2}\,\ti{u}^{\mu}h^{}_{\pmu}h+\ti{u}^{\mu}_{\pnu}h^{\mu\nu}h
                    +\ti{u}^{\mu}h^{\mu\nu}_{\pnu}h \\
                  & -\ti{u}^{\mu}h^{\mu\nu}h^{}_{\pnu}\Bigr)+\partial_{\mu}\Bigl(\frac{1}{4}\,h^{\mu\nu}_{\pnu}hh+h^{\mu\be}h^{\be\al}_{\pal}h+
                    u^{\ro}_{\pro}\ti{u}^{\nu}h^{\nu\mu} \\
                  & -2\,u^{\mu}_{\pnu}\ti{u}^{\ro}h^{\ro\nu}+\frac{1}{2}\,u^{\mu}_{\pro}\ti{u}^{\ro}h-\frac{1}{2}\,u^{\ro}\ti{u}^{\ro}_{\pmu}h+\frac{1}{2}\,
                    u^{\ro}\ti{u}^{\ro}h^{}_{\pmu}\Bigr) \\
  \end{split} \\
  \begin{split}
    f^{-1}(v_9) = & -h^{\mu\nu}_{\pmu}h^{}_{\pnu}h+u^{\ro}_{\pro}\ti{u}^{\mu}_{\pmu}h \\
                = & -\frac{i}{2}\,d_Q\bigl(\ti{u}^{\mu}_{\pmu}hh\bigr)-\frac{1}{2}\,\partial_{\mu}\bigl(h^{\mu\nu}_{\pnu}hh\bigr) \\
  \end{split} \\
  \begin{split}
    f^{-1}(v_{10}) = & \ u^{\ro}_{\pmu}\ti{u}^{\mu}_{\pro}h-u^{\ro}\ti{u}^{\mu}_{\pmu}h^{}_{\pro}-u^{\ro}_{\pro}\ti{u}^{\mu}_{\pmu}h+
                       u^{\ro}\ti{u}^{\mu}_{\pro}h^{}_{\pmu} \\
                   = & \ \partial_{\mu}\bigl(u^{\ro}\ti{u}^{\mu}_{\pro}h-u^{\mu}\ti{u}^{\ro}_{\pro}h\bigr) \\
  \end{split} \\
  \begin{split}
    f^{-1}(v_{11}) = & -\frac{3}{2}\,h^{\al\be}_{\pal}h^{\be\mu}_{\pmu}h+h^{\mu\nu}_{\pal}h^{\nu\al}_{\pbe}h^{\mu\be}+h^{\mu\nu}_{\pal}h^{\nu\al}h^{\mu\be}_{\pbe}
                       +h^{\mu\nu}h^{\nu\al}_{\pal}h^{\mu\be}_{\pbe}+h^{\mu\nu}h^{\nu\al}_{\pbe}h^{\mu\be}_{\pal} \\
                     & -2\,u^{\ro}_{\pnu}\ti{u}^{\mu}h^{\mu\nu}_{\pro}-u^{\ro}\ti{u}^{\mu}_{\pnu}h^{\mu\nu}_{\pro}-u^{\ro}_{\pro}\ti{u}^{\mu}_{\pnu}h^{\mu\nu}
                       +u^{\ro}_{\pmu}\ti{u}^{\mu}_{\pnu}h^{\ro\nu}+u^{\mu}_{\pnu}\ti{u}^{\mu}_{\pro}h^{\ro\nu} \\
                   = & \ i\,d_Q\Bigl(2\,\ti{u}^{\mu}_{\psig}h^{\mu\nu}h^{\nu\si}+\frac{3}{2}\,\ti{u}^{\mu}h^{\mu\nu}_{\pnu}h-
                       \frac{1}{2}\,u^{\mu}\ti{u}^{\nu}_{\pmu}\ti{u}^{\nu}\Bigr)+\partial_{\al}\Bigl(h^{\mu\nu}h^{\nu\al}_{\pbe}h^{\mu\be} \\
                     & +h^{\mu\nu}h^{\nu\al}h^{\mu\be}_{\pbe}+\frac{3}{2}\,u^{\ro}_{\pro}\ti{u}^{\mu}h^{\mu\al}+\frac{1}{2}\,u^{\ro}\ti{u}^{\mu}_{\pro}h^{\mu\al}-
                       \frac{1}{2}\,u^{\ro}\ti{u}^{\mu}h^{\mu\al}_{\pro}-\frac{3}{2}\,u^{\al}_{\pnu}\ti{u}^{\mu}h^{\mu\nu} \\
                     & -\frac{1}{2}\,u^{\al}\ti{u}^{\mu}_{\pnu}h^{\mu\nu}-\frac{1}{2}\,u^{\ro}_{\pal}\ti{u}^{\mu}h^{\ro\mu}-
                       \frac{1}{2}\,u^{\ro}\ti{u}^{\mu}_{\pal}h^{\ro\mu}+\frac{1}{2}\,u^{\ro}\ti{u}^{\mu}h^{\ro\mu}_{\pal}\Bigr) \\
  \end{split} \\
  \begin{split}
    f^{-1}(v_{12}) = & \ u^{\ro}_{\pmu}\ti{u}^{\mu}h^{\ro\nu}_{\pnu}+u^{\ro}\ti{u}^{\mu}_{\pmu}h^{\ro\nu}_{\pnu}+u^{\mu}\ti{u}^{\mu}_{\pnu}h^{\ro\nu}_{\pro} \\
                   = & \ i\,d_Q\bigl(u^{\mu}\ti{u}^{\mu}_{\pnu}\ti{u}^{\nu}\bigr)+\partial_{\mu}\bigl(u^{\ro}\ti{u}^{\mu}h^{\ro\nu}_{\pnu}\bigr) \\
  \end{split} \\
  \begin{split}
    f^{-1}(v_{13}) = & \ u^{\ro}_{\pmu}\ti{u}^{\mu}h^{\ro\nu}_{\pnu}-u^{\mu}_{\pnu}\ti{u}^{\mu}h^{\ro\nu}_{\pro} \\
                   = & -i\,d_Q\bigl(u^{\mu}_{\pnu}\ti{u}^{\mu}\ti{u}^{\nu}\bigr) \\
  \end{split} \\
  \begin{split}
    f^{-1}(v_{14}) = & \ \frac{1}{2}\,h^{\al\be}_{\pal}h^{\be\mu}_{\pmu}h-h^{\mu\nu}_{\pal}h^{\nu\al}_{\pbe}h^{\mu\be}+h^{\mu\nu}_{\pal}h^{\nu\al}h^{\mu\be}_{\pbe}
                       -h^{\mu\nu}h^{\nu\al}_{\pbe}h^{\mu\be}_{\pal}+u^{\mu}_{\pnu}\ti{u}^{\mu}h^{\ro\nu}_{\pro} \\
                   = & \ i\,d_Q\Bigl(\ti{u}^{\mu}h^{\mu\nu}h^{\nu\si}_{\psig}-\frac{1}{2}\,\ti{u}^{\mu}h^{\mu\nu}_{\pnu}h+\frac{1}{2}\,u^{\mu}_{\pnu}\ti{u}^{\mu}
                       \ti{u}^{\nu}\Bigr) \\
                     & -\partial_{\al}\bigl(h^{\mu\nu}h^{\nu\al}_{\pbe}h^{\mu\be}-h^{\mu\nu}h^{\nu\al}h^{\mu\be}_{\pbe}\bigr) \\
  \end{split} \\
  \begin{split}
    f^{-1}(v_{15}) = & -u^{\ro}_{\pmu}\ti{u}^{\mu}_{\pnu}h^{\ro\nu}+u^{\ro}\ti{u}^{\mu}_{\pmu}h^{\ro\nu}_{\pnu}+u^{\ro}_{\pnu}\ti{u}^{\mu}_{\pmu}h^{\ro\nu}-
                       u^{\ro}\ti{u}^{\mu}_{\pnu}h^{\ro\nu}_{\pmu} \\
                   = & \ \partial_{\mu}\bigl(u^{\ro}\ti{u}^{\nu}_{\pnu}h^{\ro\mu}-u^{\ro}\ti{u}^{\mu}_{\pnu}h^{\ro\nu}\bigr) \\
  \end{split} \\
  \begin{split}
    f^{-1}(v_{16}) = & -\frac{1}{2}\,h^{\mu\nu}_{\pmu}h^{}_{\pnu}h+h^{\mu\nu}_{\pmu}h^{\al\be}_{\pnu}h^{\al\be}+u^{\ro}_{\pnu}\ti{u}^{\mu}_{\pmu}h^{\ro\nu} \\
                   = & \ i\,d_Q\Bigl(\frac{1}{2}\,\ti{u}^{\mu}_{\pmu}h^{\al\be}h^{\al\be}-\frac{1}{4}\,\ti{u}^{\mu}_{\pmu}hh\Bigr)+\partial_{\mu}\Bigl(
                       \frac{1}{2}\,h^{\mu\nu}_{\pnu}h^{\al\be}h^{\al\be}-\frac{1}{4}\,h^{\mu\nu}_{\pnu}hh\Bigr) \\
  \end{split} \\
  \begin{split}
    f^{-1}(v_{17}) = & \ u^{\ro}\ti{u}^{\mu}_{\pmu}h^{\ro\nu}_{\pnu}-u^{\ro}_{\pnu}\ti{u}^{\mu}h^{\ro\nu}_{\pmu}-u^{\ro}\ti{u}^{\mu}_{\pnu}h^{\ro\nu}_{\pmu}
                       +u^{\mu}_{\pnu}\ti{u}^{\mu}h^{\ro\nu}_{\pro} \\
                   = & \ i\,d_Q\Bigl(\frac{1}{4}\,\ti{u}^{\mu}_{\pmu}hh+\frac{1}{2}\,\ti{u}^{\mu}h^{}_{\pmu}h-\frac{1}{2}\,\ti{u}^{\mu}_{\pmu}h^{\al\be}
                       h^{\al\be}-\ti{u}^{\mu}h^{\al\be}_{\pmu}h^{\al\be}+u^{\mu}_{\pnu}\ti{u}^{\mu}\ti{u}^{\nu}\Bigr) \\
                     & +\partial_{\mu}\Bigl(\frac{1}{4}\,h^{\mu\nu}_{\pnu}hh-\frac{1}{2}\,h^{\mu\nu}_{\pnu}h^{\al\be}h^{\al\be}-u^{\ro}\ti{u}^{\mu}_{\pnu}
                       h^{\ro\nu}+u^{\ro}_{\pnu}\ti{u}^{\nu}h^{\ro\mu}+u^{\ro}\ti{u}^{\nu}_{\pnu}h^{\ro\mu}\Bigr) \\
  \end{split}
\end{align}
This shows explicitly that indeed all the basis vectors $v_i$ are a sum of divergences plus coboundaries. It should be noted that there is no possibility to write $v_{{\scriptscriptstyle EH}}$ in the form (\ref{eq:Corand-Divergenz-Summe}). Then the theorem is proven because all basis vectors except $v_{{\scriptscriptstyle EH}}$ have a form which lead to trivial $S$-matrix elements between physical states. Together with the discussion preceeding the theorem we claim that the only physically relevant theory is the coupling of Einstein-Hilbert.

Now there arise two questions: 

\emph{Question 1:} Will the statement of this theorem remains true in higher orders? To answer this question we have to show that in each order $n$ we can achieve the form
\begin{equation}
  T_n=T_n^{EH}+d_Q(X_n)+\text{divergences}
\end{equation}
where $T_n^{EH}$ will be constructed from $T_{j}^{EH},\,j=1,\ldots,n-1$ only. We are quite sure that this is indeed the case so that the divergence or coboundary contributions will have no physical effect.  

\emph{Question 2:} What about the gauge invariance of the Einstein-Hilbert coupling in higher orders? In~\cite{sch:giqgca} Schorn obtained the result that the Einstein-Hilbert coupling in combination with the Kugo-Ojima-coupling for the ghosts is gauge invariant to second order, see sect. 5.7 and 5.8 of \cite{sch:qgt}. There it was necessary to introduce normalization terms which coincide with the four graviton coupling obtained from the expansion of the Einstein-Hilbert Lagrangian. Higher than second order have not been investigated up to now.

To summarize, we have given a detailed analysis of the gauge properties of a quantized tensor field. Very strong restrictions on the admissible form of the interaction are obtained through the requirement of perturbative gauge invariance even to first order of perturbation theory. Among all solutions to our set of equations only the Einstein-Hilbert coupling remains as a physically relevant theory. This fact is very remarkable since in our approach only the gauge properties of a quantum field describing a spin-$2$ particle were considered and no use was made of any geometrical input from classical general relativity. In view of this and with the preceding work about Yang-Mills theories in mind we have seen that the principle of operator gauge invariance is really universal.
%
%
\section{Infrared Behaviour of Massive Scalar Matter Coupled to Gravity}
Now we come to the second part of this work in which we investigate the long range behaviour of massive scalar matter coupled to the quantized gravitational field. The study of the infrared problem in quantum gravity goes mainly back to the work of Weinberg \cite{wei:iphgrav} who has investigated the infrared behaviour of virtual and real soft graviton emission processes. The transition amplitude for a process in which an emission of a single graviton occurs is logarithmically divergent. In the sum over an infinite number of soft gravitons emited he claims that infrared divergencies in the transition amplitude cancel to every order of perturbation theory. In his treatment he considers virtual and real bremsstrahlung seperately. For the \emph{virtual} ones he obtained in the sum over an  infinte number of bremsstrahlungs processes a transition amplitude which is proportional to some power of the lower cutoff parameter $\lambda$. So this rate will vanish in the limit $\lambda\rightarrow 0$. For an infinite number of \emph{real} processes he obtained a transition amplitude which is proportional to the same negative power in $\lambda$. So the cutoff disappears if virtual and real processes are combined. From our point of view we think that the problem cannot be treated in this rather simple way for the following reasons. The structure of the problem is the same as in quantum electrodynamics (QED). There it is well known that the infrared divergencies cancel in lowest order but in higher orders one has to deal with subdivergences in an arbitrary Feynman diagram. The question of wheather or not these divergencies cancel to every order of perturbation theory is a topic which had long been investigated, see \cite{yfs:idp,gy:itidp}, but is still under discussion \cite{stei:pqed}. As far as we know the cancellation of these divergencies is not at all clear. So we think that one has to investigate these processes carefully order by order.      

We will reinvestigate the infrared problem in quantum gravity in the framework of causal perturbation theory. Especially we consider the emission of a real soft graviton in two particle scattering. We use the method of adiabatic switching with a test function from the Schwartz space which cuts off the interaction at large distances in spacetime. This makes all expressions well defined during the calculation and it turns out to be a natural infrared cutoff. We therefore avoid the introduction of a graviton mass. The latter, although widely used in the literature, is less satisfactory on physical grounds because it modifies the interaction at short distances, too. It is not at all clear that the different infrared regularizations give the same result for observable quantities. To study this question we want to calculate the differential cross section for bremsstrahlung from first principles.

In order to set our notation let us start with a very short review of the quantization of the massive scalar field $\vp(x)$. The field $\vp(x)$ obeys the Klein-Gordan equation 
\begin{equation}
  (\Box +m^2)\vp(x)=0
\end{equation}
 and it has the following representation in terms of annihilation- and creation-operators:
\begin{equation}
  \vp(x)=(2\pi)^{-3/2}\int \frac{d^3\vec{p}}{\sqrt{2\omega(\vec{p})}}\bigl(a(\vec{p})\exp(-ipx)+\ti{a}(\vec{p})^{\tot}\exp(+ipx)\bigr)
\label{phi-field}
\end{equation}
 The energy $\om(\vec{p})$ is as usual given by $\omega(\vec{p})=\sqrt{\vec{p_{}}^2+m^2}$. The operator $a(\vec{p})$ annihilates a particle with momentum $\vec{p}$ whereas $\ti{a}(\vec{p})^{\tot}$ creates an antiparticle with momentum $\vec{p}$. These operators obey the following commutation relations
\begin{equation}
  \begin{split}
    \bigl[a(\vec{p}),{a(\vec{p_{}}^{\prime})}^{\tot}\bigr] & =\delta^{(3)}(\vec{p}-\vec{p_{}}^{\prime}) \\
    \bigl[\ti{a}(\vec{p}),{\ti{a}(\vec{p_{}}^{\prime})}^{\tot}\bigr] & = \delta^{(3)}(\vec{p}-\vec{p_{}}^{\prime}) \\ 
  \end{split}
\label{phi-com}
\end{equation}
and all the other commutators vanish. The energy-momentum tensor of $\vp(x)$  is given by
\begin{equation}
  T^{\mu\nu}_m(x)=\ :\Bigl(\vp(x)^{\tot}_{\pmu}\vp(x)_{\pnu}+\vp(x)^{\tot}_{\pnu}\vp(x)_{\pmu}-\eta_{\mu\nu}\bigl(\vp(x)^{\tot}_{\pxi}\vp(x)_{\pxi}-m^2\vp(x)^{\tot}
                  \vp(x)\bigr)\Bigr):.
\end{equation}
The interaction with the gravitational field will be described through the following first order coupling \cite{gr:scmqg}
\begin{equation}
  T_1(x)=\frac{i}{2}\kappa :h^{\al\be}(x)b_{\al\be\mu\nu}T^{\mu\nu}_m(x):
\end{equation}
where $\kappa=\sqrt{32\pi G}$ with Newton's constant $G$. This leads to the following explicit form of the first order interaction
\begin{equation}
  T_1(x)=\frac{i}{2}\kappa\Bigl[2:h^{\al\be}(x)\vp(x)^{\tot}_{\pal}\vp(x)_{\pbe}:-m^2:h(x)\vp(x)^{\tot}\vp(x):\Bigr].
  \label{T1}
\end{equation}
This first order coupling will be used in the following subsections for the calculation of cross-sections.
%
%
\subsection{Bremsstrahlung}
For the calculation of the bremsstrahlung process we need to know the explicit form of the $S$-matrix up to third order in the coupling constant $\kappa$. This means that we have to construct the time-ordered product $T_3(x_1,x_2,x_3)$. According to the inductive construction of Epstein and Glaser we proceed by first calculating $T_2(x_1,x_2)$. We start with the first order term(\ref{T1}) and apply the inductive construction as described in section 2. First of all we have to build the two distributions
\begin{equation}
  A_2^{\prime}(x_1,x_2)=-T_1(x_1)T_1(x_2)
\end{equation}
and 
\begin{equation}
  R_2^{\prime}(x_1,x_2)=-T_1(x_2)T_1(x_1).
\end{equation}
Since $A_2^{\prime}$ and $R_2^{\prime}$ are products of normally ordered field operators we apply Wick's theorem to obtain a normally ordered expression. We are interested in scattering processes so we have to concentrate on terms with zero or one contraction only. The contraction of the free field operators are defined by
\begin{equation}
  \begin{split}
    \contract{h^{\al\be}(x_1)h^{\mu\nu}}(x_2) := & -ib^{\al\be\mu\nu}D^{(+)}_0(x_1-x_2)\\
    \contract{\vp(x_1)\vp}(x_2) := & -iD^{(+)}_m(x_1-x_2)\\
  \end{split}
\end{equation}
where $D^{(+)}_m(x)$ is the positive frequency part of the massive Pauli-Jordan distribution. For the splitting operation we need the difference 
\begin{equation}
  D_2(x_1,x_2)=R_2^{\prime}(x_1,x_2)-A_2^{\prime}(x_1,x_2).
\end{equation}

This distribution has causal support and it is given by
\begin{equation}
  \begin{split}
    D_2(x_1,x_2) = & \ i\frac{\kappa^2}{4}\Bigl[4:h^{\al\be}(x_2)h^{\ro\si}(x_1)\vp(x_2)_{\pbe}\vp(x_1)^{\tot}_{\pro}:\partial^{x_2}_{\al}\partial^{x_1}_{\si}D_m(x_1-x_2) \\
                   & +4:h^{\al\be}(x_2)h^{\ro\si}(x_1)\vp(x_2)^{\tot}_{\pal}\vp(x_1)_{\psig}:\partial^{x_2}_{\be}\partial^{x_1}_{\ro}D_m(x_1-x_2) \\
                   & -2m^2:h^{\al\be}(x_2)h(x_1)\vp(x_2)_{\pbe}\vp(x_1)^{\tot}:\partial^{x_2}_{\al}D_m(x_1-x_2) \\
                   & -2m^2:h^{\al\be}(x_2)h(x_1)\vp(x_2)^{\tot}_{\pal}\vp(x_1):\partial^{x_2}_{\be}D_m(x_1-x_2) \\
                   & -2m^2:h(x_2)h^{\ro\si}(x_1)\vp(x_2)\vp(x_1)^{\tot}_{\pro}:\partial^{x_1}_{\si}D_m(x_1-x_2) \\
                   & -2m^2:h(x_2)h^{\ro\si}(x_1)\vp(x_2)^{\tot}\vp(x_1)_{\psig}:\partial^{x_1}_{\ro}D_m(x_1-x_2) \\
                   & +m^4:h(x_2)h(x_1)\vp(x_2)\vp(x_1)^{\tot}:D_m(x_1-x_2) \\
                   & +m^4:h(x_2)h(x_1)\vp(x_2)^{\tot}\vp(x_1):D_m(x_1-x_2) \\
                   & + 4:\vp(x_2)^{\tot}_{\pro}\vp(x_2)_{\psig}\vp(x_1)^{\tot}_{\pal}\vp(x_1)_{\pbe}:b^{\al\be\ro\si}D_0(x_1-x_2) \\
                   & +2m^2:\vp(x_2)^{\tot}_{\pro}\vp(x_2)_{\pro}\vp(x_1)^{\tot}\vp(x_1):D_0(x_1-x_2) \\
                   & +2m^2:\vp(x_2)^{\tot}\vp(x_2)\vp(x_1)^{\tot}_{\pal}\vp(x_1)_{\pal}:D_0(x_1-x_2) \\
                   & -4m^4:\vp(x_2)^{\tot}\vp(x_2)\vp(x_1)^{\tot}\vp(x_1):D_0(x_1-x_2)\Bigr]. \\
  \end{split}
\end{equation}
To obtain the time-ordered product $T_2$ we have to split the numerical distribution in $D_2$ according to it's singular order. The singular order of the Pauli-Jordan distribution is $\omega=-2$, so the splitting is trivial and we can use the formula
\begin{equation}
  D_m(x)=D^{ret}_m(x)-D^{av}_m(x).
\end{equation}
The retarded distribution $R_2(x_1,x_2)$ is then given by $D_2(x_1,x_2)$ if we replace all Pauli-Jordan distributions by the retarded ones $D^{ret}$. Finally the time-ordered distribution $T_2(x_1,x_2)$ is given by
\begin{equation}
  T_2(x_1,x_2)=R_2(x_1,x_2)-R_2^{\prime}(x_1,x_2).
\end{equation}
For later reference we here give the explicit expression
\begin{equation}
  \begin{split}
     T_2(x_1,x_2) = & \ i\frac{\kappa^2}{4}\Bigl[4:h^{\al\be}(x_2)h^{\ro\si}(x_1)\vp(x_2)_{\pbe}\vp(x_1)^{\tot}_{\pro}:\partial^{x_2}_{\al}\partial^{x_1}_{\si}D^F_m(x_1-x_2) \\
                   & +4:h^{\al\be}(x_2)h^{\ro\si}(x_1)\vp(x_2)^{\tot}_{\pal}\vp(x_1)_{\psig}:\partial^{x_2}_{\be}\partial^{x_1}_{\ro}D^F_m(x_1-x_2) \\
                   & -2m^2:h^{\al\be}(x_2)h(x_1)\vp(x_2)_{\pbe}\vp(x_1)^{\tot}:\partial^{x_2}_{\al}D^F_m(x_1-x_2) \\
                   & -2m^2:h^{\al\be}(x_2)h(x_1)\vp(x_2)^{\tot}_{\pal}\vp(x_1):\partial^{x_2}_{\be}D^F_m(x_1-x_2) \\
                   & -2m^2:h(x_2)h^{\ro\si}(x_1)\vp(x_2)\vp(x_1)^{\tot}_{\pro}:\partial^{x_1}_{\si}D^F_m(x_1-x_2) \\
                   & -2m^2:h(x_2)h^{\ro\si}(x_1)\vp(x_2)^{\tot}\vp(x_1)_{\psig}:\partial^{x_1}_{\ro}D^F_m(x_1-x_2) \\
                   & +m^4:h(x_2)h(x_1)\vp(x_2)\vp(x_1)^{\tot}:D^F_m(x_1-x_2) \\
                   & +m^4:h(x_2)h(x_1)\vp(x_2)^{\tot}\vp(x_1):D^F_m(x_1-x_2) \\
                   & + 4:\vp(x_2)^{\tot}_{\pro}\vp(x_2)_{\psig}\vp(x_1)^{\tot}_{\pal}\vp(x_1)_{\pbe}:b^{\al\be\ro\si}D^F_0(x_1-x_2) \\
                   & +2m^2:\vp(x_2)^{\tot}_{\pro}\vp(x_2)_{\pro}\vp(x_1)^{\tot}\vp(x_1):D^F_0(x_1-x_2) \\
                   & +2m^2:\vp(x_2)^{\tot}\vp(x_2)\vp(x_1)^{\tot}_{\pal}\vp(x_1)_{\pal}:D^F_0(x_1-x_2) \\
                   & -4m^4:\vp(x_2)^{\tot}\vp(x_2)\vp(x_1)^{\tot}\vp(x_1):D^F_0(x_1-x_2) \\
                   & -4:h^{\al\be}(x_1)\vp(x_1)^{\tot}_{\pal}\vp(x_1)_{\pbe}h^{\ro\si}(x_2)\vp(x_2)^{\tot}_{\pro}\vp(x_2)_{\psig}: \\
                   & +2m^2:h^{\al\be}(x_1)\vp(x_1)^{\tot}_{\pal}\vp(x_1)_{\pbe}h(x_2)\vp(x_2)^{\tot}\vp(x_2): \\
                   & +2m^2:h(x_1)\vp(x_1)^{\tot}\vp(x_1)h^{\ro\si}(x_2)\vp(x_2)^{\tot}_{\pro}\vp(x_2)_{\psig}: \\
                   & -m^4:h(x_1)\vp(x_1)^{\tot}\vp(x_1)h(x_2)\vp(x_2)^{\tot}\vp(x_2):\Bigr]. \\
  \end{split}
\label{T2}
\end{equation}

To obtain the time-ordered product to third order we proceed in much the same way as in the calculation of $T_2$. We calculate the distributions $R_3^{\prime}(x_1,x_2,x_3)$ and $A_3^{\prime}(x_1,x_2,x_3)$ which are given by
\begin{equation}
  R_3^{\prime}(x_1,x_2,x_3)=T_2(x_1,x_3)\wti{T}_1(x_2)+T_2(x_2,x_3)\wti{T}_1(x_1)+T_1(x_3)\wti{T}_2(x_1,x_2)
\end{equation}
and
\begin{equation}
  A_3^{\prime}(x_1,x_2,x_3)=\wti{T}_1(x_2)T_2(x_1,x_3)+\wti{T}_1(x_1)T_2(x_2,x_3)+\wti{T}_2(x_1,x_2)T_1(x_3)
\end{equation}
where $\wti{T}_1$ and $\wti{T}_2$ are defined by
\begin{equation}
  \wti{T}_1(x_i):=-T_1(x_i),\quad i=1,2,3
\end{equation}
and
\begin{equation}
 \wti{T}_2(x_i,x_j):=-T_2(x_i,x_j)+T_1(x_i)T_1(x_j)+T_1(x_j)T_1(x_i),\quad i,j=1,2,3.
\end{equation}
It should be noticed that we only collect those terms which have two contractions involving all the spacetime points $x_1,x_2$ and $x_3$, i.e. there is no product of contraction functions with the same argument. After we have build the causal difference $D_3=R_3^{\prime}-A_3^{\prime}$ we split the numerical part into $R_3$ with support in the backward light cone and $A_3$ with support in the forward light cone. The time-ordered distribution $T_3$ is then given by
\begin{equation}
  T_3(x_1,x_2,x_3)=R_3(x_1,x_2,x_3)-R_3^{\prime}(x_1,x_2,x_3).
\end{equation}
Because the calculation is staightforward and not very enlightning we only give the result
\begin{equation*}
  \begin{split}
    T_3(x_1,x_2,x_3) = & \ \frac{i\kappa^3}{8}\sum_{\pi\in S_3}\Biggl[8\Bigl[:h^{\ro\si}(x_{\pi(1)})\vp(x_{\pi(1)})^{\tot}_{\psig}\vp(x_{\pi(2)})^{\tot}_{\pga}
                         \vp(x_{\pi(2)})_{\pep}\vp(x_{\pi(3)})_{\pal}:\\
                       & \ +:h^{\ro\si}(x_{\pi(1)})\vp(x_{\pi(1)})_{\psig}\vp(x_{\pi(2)})^{\tot}_{\pga}\vp(x_{\pi(2)})_{\pep}\vp(x_{\pi(3)})^{\tot}_{\pal}:\Bigr]\\ 
                       & \ \!\times b^{\al\be\ga\vep}D^F_0(x_{\pi(2)}-x_{\pi(3)})\partial^{x_{\pi(3)}}_{\be}\partial^{x_{\pi(1)}}_{\ro}D^F_m(x_{\pi(1)}-x_{\pi(3)})\\
                       & \ \!+4m^2\Bigl[:h^{\ro\si}(x_{\pi(1)})\vp(x_{\pi(1)})^{\tot}_{\pro}\vp(x_{\pi(2)})^{\tot}\vp(x_{\pi(2)})\vp(x_{\pi(3)})_{\pal}:\\
                       & \ +:h^{\ro\si}(x_{\pi(1)})\vp(x_{\pi(1)})_{\pro}\vp(x_{\pi(2)})^{\tot}\vp(x_{\pi(2)})\vp(x_{\pi(3)})^{\tot}_{\pal}:\Bigr]\\
                       & \ \!\times D^F_0(x_{\pi(2)}-x_{\pi(3)})\partial^{x_{\pi(3)}}_{\al}\partial^{x_{\pi(1)}}_{\si}D^F_m(x_{\pi(1)}-x_{\pi(3)})\\
                       & \ \!+4m^2\Bigl[:h^{\ro\si}(x_{\pi(1)})\vp(x_{\pi(1)})^{\tot}_{\pro}\vp(x_{\pi(2)})^{\tot}_{\pep}\vp(x_{\pi(2)})_{\pep}\vp(x_{\pi(3)}):\\
                       & \ +:h^{\ro\si}(x_{\pi(1)})\vp(x_{\pi(1)})_{\pro}\vp(x_{\pi(2)})^{\tot}_{\pep}\vp(x_{\pi(2)})_{\pep}\vp(x_{\pi(3)})^{\tot}:\Bigr]\\
                       & \ \!\times D^F_0(x_{\pi(3)}-x_{\pi(2)})\partial^{x_{\pi(1)}}_{\si}D^F_m(x_{\pi(1)}-x_{\pi(3)})\\
                       & \ \!+4m^2\Bigl[:h(x_{\pi(1)})\vp(x_{\pi(1)})^{\tot}\vp(x_{\pi(2)})^{\tot}_{\pga}\vp(x_{\pi(2)})_{\pep}\vp(x_{\pi(3)})_{\pbe}:\\
                       & \ +:h(x_{\pi(1)})\vp(x_{\pi(1)})\vp(x_{\pi(2)})^{\tot}_{\pga}\vp(x_{\pi(2)})_{\pep}\vp(x_{\pi(3)})^{\tot}_{\pbe}:\Bigr]\\
                       & \ \!\times b^{\al\be\ga\vep}D^F_0(x_{\pi(3)}-x_{\pi(2)})\partial^{x_{\pi(3)}}_{\al}D^F_m(x_{\pi(1)}-x_{\pi(3)})\\
                       & \ \!+2m^4\Bigl[:h(x_{\pi(1)})\vp(x_{\pi(1)})^{\tot}\vp(x_{\pi(2)})^{\tot}_{\pga}\vp(x_{\pi(2)})_{\pga}\vp(x_{\pi(3)}):\\
                       & \ +:h(x_{\pi(1)})\vp(x_{\pi(1)})\vp(x_{\pi(2)})^{\tot}_{\pga}\vp(x_{\pi(2)})_{\pga}\vp(x_{\pi(3)})^{\tot}:\Bigr]\\
                       & \ \!\times D^F_0(x_{\pi(3)}-x_{\pi(2)})D^F_m(x_{\pi(1)}-x_{\pi(3)})\\
                       & \ \!+2m^4\Bigl[:h(x_{\pi(1)})\vp(x_{\pi(1)})^{\tot}\vp(x_{\pi(2)})^{\tot}\vp(x_{\pi(2)})\vp(x_{\pi(3)})_{\pbe}:\\
                       & \ +:h(x_{\pi(1)})\vp(x_{\pi(1)})\vp(x_{\pi(2)})^{\tot}\vp(x_{\pi(2)})\vp(x_{\pi(3)})^{\tot}_{\pbe}:\Bigr]\\
                       & \ \!\times D^F_0(x_{\pi(3)}-x_{\pi(2)})\partial^{x_{\pi(3)}}_{\be}D^F_m(x_{\pi(1)}-x_{\pi(3)})\\
                       & \ \!+8m^4\Bigl[:h^{\ro\si}(x_{\pi(1)})\vp(x_{\pi(1)})^{\tot}_{\pro}\vp(x_{\pi(2)})^{\tot}\vp(x_{\pi(2)})\vp(x_{\pi(3)}):\\
                       & \ +:h^{\ro\si}(x_{\pi(1)})\vp(x_{\pi(1)})_{\pro}\vp(x_{\pi(2)})^{\tot}\vp(x_{\pi(2)})\vp(x_{\pi(3)})^{\tot}:\Bigr]\\
                       & \ \!\times D^F_0(x_{\pi(3)}-x_{\pi(2)})\partial^{x_{\pi(1)}}_{\si}D^F_m(x_{\pi(1)}-x_{\pi(3)})\\
  \end{split}
\end{equation*}
\begin{equation}
  \begin{split}
                 \quad & \ \!+4m^6\Bigl[:h(x_{\pi(1)})\vp(x_{\pi(1)})^{\tot}\vp(x_{\pi(2)})^{\tot}\vp(x_{\pi(2)})\vp(x_{\pi(3)}):\\
                       & \ +:h(x_{\pi(1)})\vp(x_{\pi(1)})\vp(x_{\pi(2)})^{\tot}\vp(x_{\pi(2)})\vp(x_{\pi(3)})^{\tot}:\Bigr]\\
                       & \ \!\times D^F_0(x_{\pi(3)}-x_{\pi(2)})D^F_m(x_{\pi(1)}-x_{\pi(3)})\Biggr].\\  
  \label{T3} 
  \end{split}
\end{equation}

Due to the sum over all permutations of the indices this expression for $T_3$ is obviously symmetric in it's arguments as it is required by the definition of the time-ordered product \cite{iz:qft}. The third order $S$-matrix for the bremsstrahlungs processes is then given by
\begin{equation}
  S_3(g)=\int d^4x_1\ldots d^4x_3T_3(x_1,x_2,x_3)g(x_1)g(x_2)g(x_3),\quad g\in\mathcal{S}(\mathbb{R}^4).
\label{S-matrix}
\end{equation}

Now we consider the scattering process of two massive scalar particles in which a bremsstrahlungs graviton is emited in the final state. That is, we want to calculate the expectation value of $S_3(g)$ between the following initial and final states
\begin{align}
   \Phi_i & = \int d^3\vec{p}_1d^3\vec{q}_1\hat{\psi}_1(\vec{p}_1)\hat{\psi}_2(\vec{q}_1)a(\vec{p}_1)^{\tot}a(\vec{q}_1)^{\tot}\Omega \\
   \Phi_f & = \int d^3\vec{k} d^3\vec{p}_2d^3\vec{q}_2\hat{\Psi}(\vec{p}_2,\vec{q}_2)\hat{\phi}^{\mu\nu}(\vec{k})a^{\mu\nu}(\vec{k})^{\tot}a(\vec{p}_2)^{\tot}
              a(\vec{q}_2)^{\tot}\Omega.
\label{States}
\end{align}
These are wave packets in momentum space where $\hat{\psi}_i,\ i=1,2$ are one particle wave functions from $L^2(\mathbb{R}^3)$, $\hat{\Psi}$ is a two particle wave function from $L^2(\mathbb{R}^3\otimes\mathbb{R}^3)$ and $\hat{\phi}^{\mu\nu}$ is a tensor-valued square integrable wave function on $\mathbb{R}^3$. The vector $\Omega$ is the vacuum vector in the asymptotic Fock-Hilbert space. We want to calculate the following $S$-matrix element: 
\begin{equation}
  S_{fi}=(\Phi_f,S_3(g)\Phi_i).
\label{Sfi1}
\end{equation}
To show the details of the calculation we restrict ourself to the following term from the above computed $T_3$ (\ref{T3}) 
\begin{equation}
  \begin{split}
    T^{(1)}_3(x_1,x_2,x_3) := & \ i\kappa^3:h^{\ro\si}(x_1)\vp(x_3)_{\pbe}\vp(x_1)^{\tot}_{\pro}\vp(x_2)^{\tot}_{\pga}\vp(x_2)_{\pep}:\\
                              & \ b^{\al\be\ga\vep}D^F_0(x_2-x_3)\partial^{x_3}_{\al}\partial^{x_1}_{\si}D^F_m(x_1-x_3)\\
  \end{split}
\label{T3Term1}
\end{equation}
where the $(1)$ refers to this particular term of $T_3$. The $S$-matrix element corresponding to this contribution is then given by
\begin{equation}
  \begin{split}
    S_{fi}^{(1)} = & \int d^3\vec{p}_1d^3\vec{q_1}d^3\vec{k}d^3\vec{p}_2d^3\vec{q}_2d^4x_1\ldots d^4x_3\hat{\psi}_1(\vec{p}_1)\hat{\psi}_2(\vec{q}_1)
                     \hat{\Psi}(\vec{p}_2,\vec{q}_2)\hat{\phi}^{\mu\nu}(\vec{k}) \\
                   & \times g(x_1)\ldots g(x_3)\bigl(a^{\mu\nu}(\vec{k})^{\tot}a(\vec{p}_2)^{\tot}a(\vec{q}_2)^{\tot}\Omega , T^{(1)}_3(x_1,x_2,x_3)
                     a(\vec{p}_1)^{\tot}a(\vec{q}_1)^{\tot}\Omega\bigr).\\
  \end{split}
\label{Sfi2}
\end{equation}
Then we have to compute the vacuum expectation value
\begin{equation}
  \bigl(\Omega ,a(\vec{q_2})a(\vec{p_2})a^{\mu\nu}(\vec{k}):h^{\ro\si}(x_1)\vp(x_3)_{\pbe}\vp(x_1)^{\tot}_{\pro}\vp(x_2)^{\tot}_{\pga}\vp(x_2)_{\pep}:a(\vec{p_1})^{\tot}
  a(\vec{q_1})^{\tot}\Omega\bigr).
\label{vep}
\end{equation}
This can easily be evaluated by inserting the Fourier representations of the free field operators (\ref{h-field}) and (\ref{phi-field}) and the use of the commutation relations in momentum space (\ref{h-com}),(\ref{phi-com}). The result is given by
\begin{equation}
  \begin{split}
    \bigl(\Omega ,\ldots\Omega\bigr) = & \ (2\pi)^{-15/2}b^{\mu\nu\ro\si}\bigl(32\omega(\vec{k})\omega(\vec{p}_1)\omega(\vec{p}_2)\omega(\vec{q}_1)\omega(\vec{q}_2)
                                         \bigr)^{-1/2}\\
                                       & \Bigl[\exp\bigl[+i(kx_1-p_1x_3+q_2x_1+p_2x_2-q_1x_2)\bigr]\,p_1^{\be}\,q_2^{\ro}\,p_2^{\ga}\,q_1^{\vep} \\
                                       & +\exp\bigl[+i(kx_1-q_1x_3+q_2x_1+p_2x_2-p_1x_2)\bigr]\,q_1^{\be}\,q_2^{\ro}\,p_2^{\ga}\,p_1^{\vep}\\
                                       & +\exp\bigl[+i(kx_1-p_1x_3+p_2x_1+q_2x_2-q_1x_2)\bigr]\,p_1^{\be}\,p_2^{\ro}\,q_2^{\ga}\,q_1^{\vep}\\
                                       & +\exp\bigl[+i(kx_1-q_1x_3+p_2x_1+q_2x_2-p_1x_2)\bigr]\,q_1^{\be}\,p_2^{\ro}\,q_2^{\ga}\,p_1^{\vep}\Bigr].\\
  \end{split}
\end{equation}
The $S$-matrix element then reads
\begin{equation}
  \begin{split}
    S_{fi}^{(1)} = & \ i(2\pi)^{-15/2}\kappa^3\int d^3\vec{p}_1d^3\vec{q_1}d^3\vec{k}d^3\vec{p}_2d^3\vec{q}_2d^4x_1\ldots d^4x_3\hat{\psi}_1(\vec{p}_1)\hat{\psi}_2(\vec{q}_1)
                     \hat{\Psi}(\vec{p}_2,\vec{q}_2) \\
                   & \ \hat{\phi}^{\mu\nu}(\vec{k})g(x_1)\ldots g(x_3)b^{\al\be\ga\vep}D^F_0(x_2-x_3)\partial^{x_3}_{\al}\partial^{x_1}_{\si}D^F_m(x_1-x_3)b^{\mu\nu\ro\si}\\
                   & \ \bigl(32\omega(\vec{k})\omega(\vec{p}_1)\omega(\vec{p}_2)\omega(\vec{q}_1)\omega(\vec{q_2})\bigr)^{-1/2}\\
                   & \Bigl[\exp\bigl[+i(kx_1-p_1x_3+q_2x_1+p_2x_2-q_1x_2)\bigr]\,p_1^{\be}\,q_2^{\ro}\,p_2^{\ga}\,q_1^{\vep} \\
                   & +\exp\bigl[+i(kx_1-q_1x_3+q_2x_1+p_2x_2-p_1x_2)\bigr]\,q_1^{\be}\,q_2^{\ro}\,p_2^{\ga}\,p_1^{\vep}\\
                   & +\exp\bigl[+i(kx_1-p_1x_3+p_2x_1+q_2x_2-q_1x_2)\bigr]\,p_1^{\be}\,p_2^{\ro}\,q_2^{\ga}\,q_1^{\vep}\\
                   & +\exp\bigl[+i(kx_1-q_1x_3+p_2x_1+q_2x_2-p_1x_2)\bigr]\,q_1^{\be}\,p_2^{\ro}\,q_2^{\ga}\,p_1^{\vep}\Bigr].\\
  \end{split}
\label{Sfi3}
\end{equation} 
We introduce the abbreviations $I_1,\ldots, I_4$ for the four parts of $S_{fi}^{(1)}$. Then we consider the spatial integrations in $I_1$. They can be carried out by inserting the Fourier transforms of the test functions as well as the Feynman propagators:
\begin{equation}
  \begin{split}
    I_1 = & \int d^4x_1\ldots d^4x_3g(x_1)\ldots g(x_3)D^F_0(x_2-x_3)\partial^{x_3}_{\al}\partial^{x_1}_{\si}D^F_m(x_1-x_3)\\
          & \ \exp\bigl[+i(kx_1-p_1x_3+q_2x_1+p_2x_2-q_1x_2)\bigr]\\
        = & \ (2\pi)^{-10}\int d^4x_1\ldots d^4x_3d^4l_1\ldots d^4l_5\hat{g}(l_1)\ldots\hat{g}(l_3)\exp\bigl[-i(l_1x_1+l_2x_2+l_3x_3)\bigr] \\
          & \ \hat{D}^F_0(l_4)\exp\bigl[-il_4(x_2-x_3)\bigr]\partial^{x_3}_{\al}\partial^{x_1}_{\si}\hat{D}^F_m(l_5)\exp\bigl[-il_5(x_1-x_3)\bigr] \\
          & \ \exp\bigl[+i(kx_1-p_1x_3+q_2x_1+p_2x_2-q_1x_2)\bigr]\\
        = & \ (2\pi)^{-10}\int d^4x_1\ldots d^4x_3d^4l_1\ldots d^4l_5\hat{g}(l_1)\ldots\hat{g}(l_3)\exp\bigl[-i(l_1x_1+l_2x_2+l_3x_3)\bigr] \\
          & \ \exp\bigl[-il_4(x_2-x_3)\bigr]l_5^{\al}l_5^{\si}\hat{D}^F_m(l_5)\exp\bigl[-il_5(x_1-x_3)\bigr] \\
          & \ \exp\bigl[+i(kx_1-p_1x_3+q_2x_1+p_2x_2-q_1x_2)\bigr]\\
        = & \ (2\pi)^2\int d^4l_1\ldots d^4l_5\hat{g}(l_1)\ldots\hat{g}(l_3)\hat{D}^F_0(l_4)l_5^{\al}l_5^{\si}\hat{D}^F_m(l_5)\delta^{(4)}(l_1+l_5-k-q_2) \\
          & \ \delta^{(4)}(l_2+l_4-p_2+q_1)\delta^{(4)}(l_3-l_4-l_5+p_1).\\
  \end{split}
\label{I1}
\end{equation}
We can do the integrations w.r.t. $l_4$ and $l_5$ immediately and get
\begin{equation}
  \begin{split}
    I_1 = & \ (2\pi)^2\int d^4l_1\ldots d^4l_3\hat{g}(l_1)\ldots\hat{g}(l_3)\hat{D}^F_0(l_1+l_3+p_1-k-q_2)\\
          & \ [l_1^{\al}-p_1^{\al}+k^{\al}+q_2^{\al}][-l_1^{\si}-p_1^{\si}+k^{\si}+q_2^{\si}]\hat{D}^F_m(-l_1+k+q_2)\\
          & \ \delta^{(4)}(l_1+l_2+l_3+p_1+q_1-q_2-p_2-k).\\
  \end{split}
\end{equation}
Then the first term of $S^{(1)}_{fi}$ becomes
\begin{equation}
  \begin{split}
    S^{(1,I_1)}_{fi} = & \ (2\pi)^{-11/2}i\kappa^3\int d^3\vec{p}_1d^3\vec{q_1}d^3\vec{k}d^3\vec{p}_2d^3\vec{q}_2\hat{\psi}_1(\vec{p}_1)\hat{\psi}_2(\vec{q}_1)
                         \hat{\Psi}(\vec{p}_2,\vec{q}_2)\hat{\phi}^{\mu\nu}(\vec{k})\\
                       & \ \int d^4l_1\ldots d^4l_3\hat{g}(l_1)\ldots\hat{g}(l_3)b^{\al\be\ga\vep}\hat{D}^F_0(l_1+l_3+p_1-k-q_2)\\
                       & \ [-l_1^{\al}-p_1^{\al}+k^{\al}+q_2^{\al}][-l_1^{\si}-p_1^{\si}+k^{\si}+q_2^{\si}]\hat{D}^F_m(-l_1+k+q_2)\\
                       & \ b^{\mu\nu\ro\si}p_1^{\be}q_2^{\ro}p_2^{\ga}q_1^{\vep}\bigl(32\omega(\vec{k})\omega(\vec{p}_1)\omega(\vec{p}_2)\omega(\vec{q}_1)
                         \omega(\vec{q_2})\bigr)^{-1/2}\\
                       & \ \delta^{(4)}(l_1+l_2+l_3+p_1+q_1-q_2-p_2-k)\\
                     = & -(2\pi)^{-19/2}i\kappa^3\int d^3\vec{p}_1d^3\vec{q_1}d^3\vec{k}d^3\vec{p}_2d^3\vec{q}_2\hat{\psi}_1(\vec{p}_1)\hat{\psi}_2(\vec{q}_1)
                         \hat{\Psi}(\vec{p}_2,\vec{q}_2)\hat{\phi}^{\mu\nu}(\vec{k})\\
                       & \ \int d^4l_1\ldots d^4l_3\hat{g}(l_1)\ldots\hat{g}(l_3)b^{\al\be\ga\vep}\frac{-l_1^{\al}-p_1^{\al}+k^{\al}+q_2^{\al}}{(l_1+l_3+p_1-k-q_2)^2+i0}
                         b^{\mu\nu\ro\si}\\
                       & \ \frac{-l_1^{\si}-p_1^{\si}+k^{\si}+q_2^{\si}}{m^2-(-l_1+k+q_2)^2-i0}p_1^{\be}q_2^{\ro}p_2^{\ga}q_1^{\vep}\bigl(32\omega(\vec{k})
                         \omega(\vec{p}_1)\omega(\vec{p}_2)\omega(\vec{q}_1)\omega(\vec{q_2})\bigr)^{-1/2}\\
                       & \ \delta^{(4)}(l_1+l_2+l_3+p_1+q_1-q_2-p_2-k)\\ 
  \end{split}
\label{Sfi4}
\end{equation}
where we have inserted the propagators explicitely. In this expression we can carry out the adiabatic limit in the variable $l_2$ since it doesn't appear in the argument of the propagators. We do this in the following way. First of all we choose a fixed test function $g_0\in \mathcal{S}(\mathbb{R}^4)$ with the property $g_0(0)=1$. Then a new test function is defined by
\begin{equation}
  g_{\vep}(x):=g_0(\vep x).
\end{equation}
The adiabatic limit $g_{\vep}\rightarrow 1$ is then equivalent to the limit $\vep\rightarrow 0$. The Fourier transform of $g_{\vep}$ is given by
\begin{equation}
  \hat{g}_{\vep}(p)=\frac{1}{\vep^4}\hat{g}_0\Bigl(\frac{p}{\vep}\Bigr).
\end{equation}
Then we have
\begin{equation}
  \lim_{\vep\rightarrow 0} \frac{1}{\vep^4}\hat{g}_0\Bigl(\frac{p}{\vep}\Bigr)=(2\pi)^2\delta^{(4)}(p).
\label{adia}
\end{equation}
Using this in (\ref{Sfi4}) the integration w.r.t. $l_2$ gives just the factor $(2\pi)^2$. We are interested in the $S$-matrix element in the limit where only gravitons at low momenta (soft gravitons) are emited. So we omit all small quantities in the numerators, and  we take only the leading terms in the denominators, neglecting terms $o(\ep)$. This is known as the eikonal approximation in the literature, see e.g. \cite{css:qcd}. With this approximation we obtain for the argument of the massless propagator
\begin{equation}
  \begin{split}
    (p_1-q_2-k+l_1+l_3)^2 = & \ (p_1-q_2)^2+(-k+l_1+l_3)^2-2(p_1-q_2)(k-l_1-l_3)\\
                          = & \ (p_1-q_2)^2+o(\vep).\\
  \end{split}
\end{equation}
where $\vep$ is the scaling parameter from the adiabatic limit. The argument of the massive propagator becomes
\begin{equation}
  \begin{split}
    m^2-(-l_1+k+q_2)^2 = & \ m^2-q_2^2-(-l_1+k)^2+2q_2(l_1-k)\\
                       = & \ 2q_2(l_1-k)+o(\vep^2)\\
  \end{split}
\end{equation}
since the momentum $q_2$ is on the mass shell. Then the $S$-matrix element becomes
\begin{equation}
  \begin{split}
     S^{(1,I_1)}_{fi} = & -(2\pi)^{-15/2}i\kappa^3\int d^3\vec{p}_1d^3\vec{q_1}d^3\vec{k}d^3\vec{p}_2d^3\vec{q}_2\hat{\psi}_1(\vec{p}_1)\hat{\psi}_2(\vec{q}_1)
                          \hat{\Psi}(\vec{p}_2,\vec{q}_2)\hat{\phi}^{\mu\nu}(\vec{k})\\
                        & \ \int d^4l_1d^4l_3\hat{g}(l_1)\hat{g}(l_3)b^{\al\be\ga\vep}\frac{(q_2^{\al}-p_1^{\al})}{(p_1-q_2)^2}\frac{(q_2^{\si}-p_1^{\si})}{2q_2(l_1-k)}
                          b^{\mu\nu\ro\si}\\
                        & \ p_1^{\be}q_2^{\ro}p_2^{\ga}q_1^{\vep}\bigl(32\omega(\vec{k})\omega(\vec{p}_1)\omega(\vec{p}_2)\omega(\vec{q}_1)\omega(\vec{q}_2)\bigr)^{-1/2}\\
                        & \ \delta^{(4)}(l_1+l_3+p_1+q_1-q_2-p_2-k).\\
  \end{split}
\label{Sfi5}
\end{equation}
Now the adiabatic limit in $l_3$ can be done as above and we arrive at the result for the $S$-matrix element
\begin{equation}
  \begin{split}
    S^{(1,I_1)}_{fi} = & -(2\pi)^{-11/2}i\kappa^3\int d^3\vec{p}_1d^3\vec{q_1}d^3\vec{k}d^3\vec{p}_2d^3\vec{q}_2\hat{\psi}_1(\vec{p}_1)\hat{\psi}_2(\vec{q}_1)
                          \hat{\Psi}(\vec{p}_2,\vec{q}_2)\hat{\phi}^{\mu\nu}(\vec{k})\\
                       & \ \int d^4l_1\hat{g}(l_1)b^{\al\be\ga\vep}\frac{(q_2^{\al}-p_1^{\al})}{(p_1-q_2)^2}\frac{(q_2^{\si}-p_1^{\si})}{2q_2(l_1-k)}b^{\mu\nu\ro\si} \\
                       & \ p_1^{\be}q_2^{\ro}p_2^{\ga}q_1^{\vep}\bigl(32\omega(\vec{k})\omega(\vec{p}_1)\omega(\vec{p}_2)\omega(\vec{q}_1)\omega(\vec{q}_2)\bigr)^{-1/2}\\
                       & \ \delta^{(4)}(l_1+p_1+q_1-q_2-p_2-k).\\
  \end{split}
\label{Sfi6}
\end{equation}
There remains to calculate the other three parts $S^{(1,I_2)}_{fi},\ldots, S^{(1,I_4)}_{fi}$. The calculations are similiar to the one above so we only present the results
\begin{equation}
  \begin{split}
    S^{(1,I_2)}_{fi} = & -(2\pi)^{-11/2}i\kappa^3\int d^3\vec{p}_1d^3\vec{q_1}d^3\vec{k}d^3\vec{p}_2d^3\vec{q}_2\hat{\psi}_1(\vec{p}_1)\hat{\psi}_2(\vec{q}_1)
                          \hat{\Psi}(\vec{p}_2,\vec{q}_2)\hat{\phi}^{\mu\nu}(\vec{k})\\
                       & \ \int d^4l_1\hat{g}(l_1)b^{\al\be\ga\vep}\frac{(q_2^{\al}-q_1^{\al})}{(q_1-q_2)^2}\frac{(q_2^{\si}-q_1^{\si})}{2q_2(l_1-k)}b^{\mu\nu\ro\si} \\
                       & \ q_1^{\be}q_2^{\ro}p_2^{\ga}p_1^{\vep}\bigl(32\omega(\vec{k})\omega(\vec{p}_1)\omega(\vec{p}_2)\omega(\vec{q}_1)\omega(\vec{q}_2)\bigr)^{-1/2}\\
                       & \ \delta^{(4)}(l_1+q_1+p_1-q_2-p_2-k)\\
    S^{(1,I_3)}_{fi} = & -(2\pi)^{-11/2}i\kappa^3\int d^3\vec{p}_1d^3\vec{q_1}d^3\vec{k}d^3\vec{p}_2d^3\vec{q}_2\hat{\psi}_1(\vec{p}_1)\hat{\psi}_2(\vec{q}_1)
                          \hat{\Psi}(\vec{p}_2,\vec{q}_2)\hat{\phi}^{\mu\nu}(\vec{k})\\
                       & \ \int d^4l_1\hat{g}(l_1)b^{\al\be\ga\vep}\frac{(p_2^{\al}-p_1^{\al})}{(p_1-p_2)^2}\frac{(p_2^{\si}-p_1^{\si})}{2p_2(l_1-k)}b^{\mu\nu\ro\si} \\
                       & \ p_1^{\be}p_2^{\ro}q_2^{\ga}q_1^{\vep}\bigl(32\omega(\vec{k})\omega(\vec{p}_1)\omega(\vec{p}_2)\omega(\vec{q}_1)\omega(\vec{q}_2)\bigr)^{-1/2}\\
                       & \ \delta^{(4)}(l_1+q_1+p_1-q_2-p_2-k)\\
    S^{(1,I_4)}_{fi} = & -(2\pi)^{-11/2}i\kappa^3\int d^3\vec{p}_1d^3\vec{q_1}d^3\vec{k}d^3\vec{p}_2d^3\vec{q}_2\hat{\psi}_1(\vec{p}_1)\hat{\psi}_2(\vec{q}_1)
                          \hat{\Psi}(\vec{p}_2,\vec{q}_2)\hat{\phi}^{\mu\nu}(\vec{k})\\
                       & \ \int d^4l_1\hat{g}(l_1)b^{\al\be\ga\vep}\frac{(p_2^{\al}-q_1^{\al})}{(q_1-p_2)^2}\frac{(p_2^{\si}-q_1^{\si})}{2p_2(l_1-k)}b^{\mu\nu\ro\si} \\
                       & \ q_1^{\be}p_2^{\ro}q_2^{\ga}p_1^{\vep}\bigl(32\omega(\vec{k})\omega(\vec{p}_1)\omega(\vec{p}_2)\omega(\vec{q}_1)\omega(\vec{q}_2)\bigr)^{-1/2}\\
                       & \ \delta^{(4)}(l_1+q_1+p_1-q_2-p_2-k).\\ 
  \end{split}
\end{equation}
Now we are ready to consider the differential cross section. 

We have to calculate the absolute square of the $S$-matrix element $S^{(1)}_{fi}$. We do this again separately for the four parts $S^{(1,I_1)}_{fi},\ldots, S^{(1,I_4)}_{fi}$. For the first part we start with (\ref{Sfi6}) and write
\begin{equation}
  \begin{split}
    \big\lvert S^{(1,I_1)}_{fi}\big\rvert^2 = & \ (2\pi)^{-11}\kappa^6\int d^3\vec{p}_1d^3\vec{q}_1d^3\vec{k}d^3\vec{p}_2d^3\vec{q}_2d^3\vec{p_1}^{\prime}d^3\vec{q_1}^{\prime}
                                                d^3\vec{k}^{\prime}d^3\vec{p_2}^{\prime}d^3\vec{q_2}^{\prime}\\
                                              & \ \hat{\psi}_1(\vec{p}_1)\hat{\psi}_2(\vec{q_1})\hat{\Psi}(\vec{p}_2,\vec{q}_2)\hat{\phi}^{\mu\nu}(\vec{k})
                                                \hat{\psi}_1(\vec{p_1}^{\prime})^{\ast}\hat{\psi}_2(\vec{q_1}^{\prime})^{\ast}\hat{\Psi}(\vec{p_2}^{\prime},
                                                \vec{q_2}^{\prime})^{\ast}\hat{\phi}^{\mu^{\prime}\nu^{\prime}}(\vec{k}^{\prime})^{\ast}\\
                                              & \ \int d^4l_1d^4l_2\hat{g}(l_1)\hat{g}(l_2)^{\ast}b^{\al\be\ga\vep}\frac{(q_2^{\al}-p_1^{\al})}{(q_2-p_1)^2}
                                                \frac{(q_2^{\si}-p_1^{\si})}{2q_2(l_1-k)}b^{\mu\nu\ro\si}\\
                                              & \ p_1^{\be}q_2^{\ro}p_2^{\ga}q_1^{\vep}b^{\al^{\prime}\be^{\prime}\ga^{\prime}\vep^{\prime}}\frac{({q_2^{\prime}}^{\al^{\prime}}-
                                                {p_1^{\prime}}^{\al^{\prime}})}{(q_2^{\prime}-p_1^{\prime})^2}\frac{({q_2^{\prime}}^{\si^{\prime}}
                                                -{p_1^{\prime}}^{\si^{\prime}})}{2q_2^{\prime}(l_2-k^{\prime})}b^{\mu^{\prime}\nu^{\prime}\ro^{\prime}\si^{\prime}}\\
                                              & \ {p_1^{\prime}}^{\be^{\prime}}{q_2^{\prime}}^{\ro^{\prime}}{p_2^{\prime}}^{\ga^{\prime}}{q_1^{\prime}}^{\vep^{\prime}}
                                                \bigl(32\omega(\vec{k})\omega(\vec{p}_1)\omega(\vec{p}_2)\omega(\vec{q}_1)\omega(\vec{q}_2)\bigr)^{-1/2}\\
                                              & \ \bigl(32\omega(\vec{k^{\prime}})\omega(\vec{p_1}^{\prime})\omega(\vec{p_2}^{\prime})\omega(\vec{q_1}^{\prime})
                                                \omega(\vec{q_2}^{\prime})\bigr)^{-1/2}\\
                                              & \ \delta^{(4)}(l_1+p_1+q_1-q_2-p_2-k)\delta^{(4)}(l_2+p_1^{\prime}+q_1^{\prime}-q_2^{\prime}-p_2^{\prime}-k^{\prime})\\
  \end{split}
\label{sigma1}
\end{equation}
where a $^{\ast}$ indicates complex conjugation. We use the completeness relations of the final states, i.e.
\begin{align}
  \sum_f \hat{\Psi}(\vec{p_2},\vec{q_2})\hat{\Psi}(\vec{p_2}^{\prime},\vec{q_2}^{\prime})^{\ast} & =\delta^{(3)}(\vec{p_2}-\vec{p_2}^{\prime})
                                                                                                  \delta^{(3)}(\vec{q_2}-\vec{q_2}^{\prime})\\
  \sum_f \hat{\phi}^{\mu\nu}(\vec{k})\hat{\phi}^{\mu^{\prime}\nu^{\prime}}(\vec{k}^{\prime})^{\ast} & =\eta^{\mu\nu}\eta^{\mu^{\prime}\nu^{\prime}}
                                                                                                     \delta^{(3)}(\vec{k}-\vec{k}^{\prime})
\end{align}
where the sum is taken over all final states. Then we have
\begin{equation}
  \begin{split}
    \sum_f \big\lvert S^{(1,I_1)}_{fi}\big\rvert^2 = & \ (2\pi)^{-11}\kappa^6\int d^3\vec{p}_1d^3\vec{q}_1d^3\vec{k}d^3\vec{p}_2d^3\vec{q}_2d^3\vec{p_1}^{\prime}d^3
                                                    \vec{q_1}^{\prime}\hat{\psi}_1(\vec{p}_1)\hat{\psi}_2(\vec{q_1})\\
                                                  & \ \hat{\psi}_1(\vec{p_1}^{\prime})^{\ast}\hat{\psi}_2(\vec{q_1}^{\prime})^{\ast}\int d^4l_1d^4l_2\hat{g}(l_1)
                                                    \hat{g}(l_2)^{\ast}b^{\al\be\ga\vep}\frac{(q_2^{\al}-p_1^{\al})}{(q_2-p_1)^2}\frac{(q_2^{\ro}-p_1^{\ro})}{2q_2(l_1-k)}\\
                                                  & \ p_1^{\be}q_2^{\ro}p_2^{\ga}q_1^{\vep}b^{\al^{\prime}\be^{\prime}\ga^{\prime}\vep^{\prime}}\frac{(q_2^{\al^{\prime}}-
                                                    {p_1^{\prime}}^{\al^{\prime}})}{(q_2-p_1^{\prime})^2}\frac{(q_2^{\ro^{\prime}}-{p_1^{\prime}}^{\ro^{\prime}})}
                                                    {2q_2(l_2-k)}{p_1^{\prime}}^{\be^{\prime}}q_2^{\ro^{\prime}}p_2^{\ga^{\prime}}{q_1^{\prime}}^{\vep^{\prime}}\\
                                                  & \ \frac{\delta^{(4)}(l_1+p_1+q_1-q_2-p_2-k)\delta^{(4)}(l_2+p_1^{\prime}+q_1^{\prime}-q_2-p_2-k)}
                                                    {\sqrt{\bigl(32\omega(\vec{k})\omega(\vec{p}_1)\omega(\vec{p}_2)\omega(\vec{q}_1)\omega(\vec{q}_2)\bigr)\bigl(32
                                                    \omega(\vec{k})\omega(\vec{p_1}^{\prime})\omega(\vec{p_2})\omega(\vec{q_1}^{\prime})\omega(\vec{q_2})\bigr)}}.\\
  \end{split}
\label{sigme2}
\end{equation}
Now we assume that the wavefunctions $\hat{\psi}_1$ and $\hat{\psi}_2$ are sharply concentrated around the initial momenta $p_i$ and $q_i$, so that we can replace the momenta in the propagators by these initial momenta. Then the expression can be simplified to
\begin{equation}
  \begin{split}
    \sum_f \big\lvert S^{(1,I_1)}_{fi}\big\rvert^2 = & \ (2\pi)^{-11}\kappa^6\int d^4l_1d^4l_2\hat{g}(l_1)\hat{g}(l_2)^{\ast}\int d^3\vec{k}d^3\vec{p_2}d^3\vec{q_2}
                                                    b^{\al\be\ga\vep}\\
                                                  & \ \frac{(q_2^{\al}-p_i^{\al})}{(q_2-p_i)^2}\frac{(q_2^{\ro}-p_i^{\ro})}{2q_2(l_1-k)}p_i^{\be}q_2^{\ro}p_2^{\ga}q_i^{\vep}
                                                    b^{\al^{\prime}\be^{\prime}\ga^{\prime}\vep^{\prime}}\frac{(q_2^{\al^{\prime}}-p_i^{\al^{\prime}})}{(q_2-p_i)^2}
                                                    \frac{(q_2^{\ro^{\prime}}-p_i^{\ro^{\prime}})}{2q_2(l_2-k)}\\
                                                  & \ p_i^{\be^{\prime}}q_2^{\ro^{\prime}}p_2^{\ga^{\prime}}q_i^{\vep^{\prime}}\bigl[32\omega(\vec{k})\omega(\vec{p_i})
                                                    \omega(\vec{p_2})\omega(\vec{q_i})\omega(\vec{q_2})\bigr]^{-1}\\
                                                  & \ \int d^3\vec{p}_1d^3\vec{q}_1d^3\vec{p_1}^{\prime}d^3\vec{q_1}^{\prime}\hat{\psi}_1(\vec{p}_1)\hat{\psi}_2(\vec{q_1})
                                                    \hat{\psi}_1(\vec{p_1}^{\prime})^{\ast}\hat{\psi}_2(\vec{q_1}^{\prime})^{\ast}\\
                                                  & \ \delta^{(4)}(l_1+p_1+q_1-q_2-p_2-k)\delta^{(4)}(l_2+p_1^{\prime}+q_1^{\prime}-q_2-p_2-k).\\
  \end{split}
\label{sigma3}
\end{equation}
We observe that the last integral herein depends on the initial state only. We denote it by $F(P,l_1,l_2)$, where we have introduced $P=p_2+q_2+k$. We have
\begin{equation}
  \begin{split}
    F(P,l_1,l_2) = & \ \int d^3\vec{p}_1d^3\vec{q}_1d^3\vec{p_1}^{\prime}d^3\vec{q_1}^{\prime}\hat{\psi}_1(\vec{p}_1)\hat{\psi}_2(\vec{q_1})
                     \hat{\psi}_1(\vec{p_1}^{\prime})^{\ast}\hat{\psi}_2(\vec{q_1}^{\prime})^{\ast}\\
                   & \ \delta^{(4)}(l_1+p_1+q_1-q_2-p_2-k)\delta^{(4)}(l_2+p_1^{\prime}+q_1^{\prime}-q_2-p_2-k)\\
                 = & \ (2\pi)^{-8}\int d^4y_1d^4y_2d^3\vec{p}_1\ldots d^3\vec{q_1}^{\prime}\hat{\psi}_1(\vec{p}_1)\hat{\psi}_2(\vec{q_1})
                     \hat{\psi}_1(\vec{p_1}^{\prime})^{\ast}\hat{\psi}_2(\vec{q_1}^{\prime})^{\ast}\\
                   & \ \exp\bigl[-i(l_1+p_1+q_1-P)y_1\bigr]\exp\bigl[+i(l_2+p_1^{\prime}+q_1^{\prime}-P)y_2\bigr]\\
                 = & \ \int d^4y_1d^4y_2\psi_1(y_1)\psi_2(y_1)\psi_1(y_2)^{\ast}\psi_2(y_2)^{\ast}\exp[-il_1y_1]\exp[+il_2y_2]\\
                   & \ \exp\bigl[-i(y_1-y_2)P\bigr].\\
  \end{split}
\end{equation}
The function $F$ is normalized according to
\begin{equation}
  \int F(P,l_1,l_2)d^4P=(2\pi)^2\int d^4y|\psi_1(y)|^2|\psi_2(y)|^2\exp\bigl[-i(l_1-l_2)y\bigr]
\end{equation}
and it is concentrated around $P=p_2+q_2+k\approx p_i+q_i$. In the limit of infintely sharp wave packets we may represent it by
\begin{equation}
  F(P,l_1,l_2)=(2\pi)^2\delta^{(4)}(P-p_i-q_i)\int d^4y|\psi_1(y)|^2|\psi_2(y)|^2\exp\bigl[-i(l_1-l_2)y\bigr].
\end{equation}
This will be inserted in (\ref{sigma3}) and we obtain
\begin{equation} 
  \begin{split}
    \sum_f \big\lvert S^{(1,I_1)}_{fi}\big\rvert^2 = & \ (2\pi)^{-9}\kappa^6\int d^4l_1d^4l_2\hat{g}(l_1)\hat{g}(l_2)^{\ast}\int d^3\vec{k}d^3\vec{p_2}d^3\vec{q_2}
                                                    b^{\al\be\ga\vep}\\
                                                  & \ \frac{(q_2^{\al}-p_i^{\al})}{(q_2-p_i)^2}\frac{(q_2^{\ro}-p_i^{\ro})}{2q_2(l_1-k)}p_i^{\be}q_2^{\ro}p_2^{\ga}q_i^{\vep}
                                                    b^{\al^{\prime}\be^{\prime}\ga^{\prime}\vep^{\prime}}\frac{(q_2^{\al^{\prime}}-p_i^{\al^{\prime}})}{(q_2-p_i)^2}
                                                    \frac{(q_2^{\ro^{\prime}}-p_i^{\ro^{\prime}})}{2q_2(l_2-k)}\\
                                                  & \ p_i^{\be^{\prime}}q_2^{\ro^{\prime}}p_2^{\ga^{\prime}}q_i^{\vep^{\prime}}\bigl[32\omega(\vec{k})\omega(\vec{p_i})
                                                    \omega(\vec{p_2})\omega(\vec{q_i})\omega(\vec{q_2})\bigr]^{-1}\\
                                                  & \ \delta^{(4)}(p_2+q_2+k-p_i-q_i)\int d^4y|\psi_1(y)|^2|\psi_2(y)|^2\exp\bigl[-i(l_1-l_2)y\bigr].\\
  \end{split}
\label{sigma4}
\end{equation}
We can set $l_1=l_2$ in the last integral because we neglect all contributions $o(\vep)$ in the numerator, see the remarks after (\ref{Sfi4}). We denote the absolute square of $S_{fi}$ by $p_{fi}:=|S_{fi}|^2$. Then the definition of the cross section is as follows: We consider a beam of incoming particles of radius $R$ incident on the target which is assumed to be at rest. The transition amplitude $p_{fi}$ must then be averaged in space over the cylinder of the beam of incoming particles. Then the cross section is given by
\begin{equation}
  \sigma=\lim_{R\rightarrow\infty}\pi R^2\sum_fp_{fi}(R)
\end{equation}  
where we sum over a complete set of final states. Then we restrict ourselves again to (\ref{T3Term1}) of $T_3$ and furthermore to the first integral $I_1$ (\ref{I1}) of $S_{fi}$. If we denote the integration variables $p_2$ and $q_2$ by $p_f$ and $q_f$, respectively, then we obtain for the cross section
\begin{equation}
  \begin{split}
    \sigma = & \ (2\pi)^{-9}\kappa^6\bigl[4\omega(\vec{p_i})\omega(\vec{q_i})\bigr]^{-1}\frac{\omega(\vec{p_i})\omega(\vec{q_i})}{\sqrt{\bigl((p_iq_i)^2-m^4\bigr)}}\int 
               d^4l_1d^4l_2\hat{g}(l_1)\hat{g}(l_2)^{\ast}\\
             & \ \int\frac{d^3\vec{k}}{2\omega(\vec{k})}\frac{d^3\vec{q_f}}{2\omega(\vec{q_f})}\frac{d^3\vec{p_f}}{2\omega(\vec{p_f})}b^{\al\be\ga\vep}\frac{(q_f^{\al}-
               p_i^{\al})}{(q_f-p_i)^2}\frac{(q_f^{\ro}-p_i^{\ro})}{2q_f(l_1-k)}p_i^{\be}q_f^{\ro}p_f^{\ga}q_i^{\vep}b^{\al^{\prime}\be^{\prime}\ga^{\prime}\vep^{\prime}}\\
             & \ \frac{(q_f^{\al^{\prime}}-p_i^{\al^{\prime}})}{(q_f-p_i)^2}\frac{(q_f^{\ro^{\prime}}-p_i^{\ro^{\prime}})}{2q_f(l_2-k)}p_i^{\be^{\prime}}q_f^{\ro^{\prime}}
               p_f^{\ga^{\prime}}q_i^{\vep^{\prime}}\delta^{(3)}(\vec{p_f}+\vec{q_f}+\vec{k}-\vec{p_i}-\vec{q_i})\\
             & \ \delta(\omega(\vec{p_f})+\omega(\vec{q_f})+\omega(\vec{k})-\omega(\vec{p_i})-\omega(\vec{q_i})).\\
  \end{split}
\end{equation}
To carry out the remainig integrations we have to know the integrand explicitely. So at this point of the calculation the tensor structure of the integrand becomes important and we have to write down the various contractions between the four-vectors $p_i,q_i,p_f,q_f$. With the tensor $b^{\mu\nu\ro\si}$ (\ref{eq:b-tensor}) we get for the cross section 
\begin{equation}
  \begin{split}
    \sigma = & \ \frac{1}{4}\frac{(2\pi)^{-9}\kappa^6}{\sqrt{\bigl((p_iq_i)^2-m^4\bigr)}}\int d^4l_1d^4l_2\hat{g}(l_1)\hat{g}(l_2)^{\ast}\int\frac{d^3\vec{k}}
               {2\om(\vec{k})}\frac{d^3\vec{q_f}}{2\om(\vec{q_f})}\frac{d^3\vec{p_f}}{2\om(\vec{p_f})}\\
             & \ \Biggl[\Bigl[\bigl(\om(\vec{q_f})\om(\vec{p_f})-\vec{q_f}\vec{p_f}\bigr)(p_iq_i)+\bigl(\om(\vec{q_f})\om(\vec{q_i})-\vec{q_f}\vec{q_i}\bigr)\bigl(\om(\vec{p_i})
               \om(\vec{p_f})-\vec{p_i}\vec{p_f}\bigr)\\
             & -\bigl(\om(\vec{q_f})\om(\vec{p_i})-\vec{q_f}\vec{p_i}\bigr)\bigl(\om(\vec{p_f})\om(\vec{q_i})-\vec{p_f}\vec{q_i}\bigr)-\bigl(\om(\vec{p_i})\om(\vec{p_f})-
               \vec{p_i}\vec{p_f}\bigr)(p_iq_i)\\
             & -(p_iq_i)\bigl(\om(\vec{p_i})\om(\vec{p_f})-\vec{p_i}\vec{p_f}\bigr)+p_i^2\bigl(\om(\vec{p_f})\om(\vec{q_i})-\vec{p_f}\vec{q_i}\bigr)\Bigr]\bigl(\om(\vec{q_f})^2
               -\vec{q_f}^2\bigr)\\
             & -\Bigl[\bigl(\om(\vec{q_f})\om(\vec{p_f})-\vec{q_f}\vec{p_f}\bigr)(p_iq_i)-\bigl(\om(\vec{q_f})\om(\vec{q_i})-\vec{q_f}\vec{q_i}\bigr)\bigl(\om(\vec{p_i})
               \om(\vec{p_f})-\vec{p_i}\vec{p_f}\bigr)\\
             & +\bigl(\om(\vec{q_f})\om(\vec{p_i})-\vec{q_f}\vec{p_i}\bigr)\bigl(\om(\vec{p_f})\om(\vec{q_i})-\vec{p_f}\vec{q_i}\bigr)+\bigl(\om(\vec{p_i})\om(\vec{p_f})-
               \vec{p_i}\vec{p_f}\bigr)(p_iq_i)\\
             & +(p_iq_i)\bigl(\om(\vec{p_i})\om(\vec{p_f})-\vec{p_i}\vec{p_f}\bigr)-p_i^2\bigl(\om(\vec{p_f})\om(\vec{q_i})-\vec{p_f}\vec{q_i}\bigr)\Bigr]\bigl(\om(\vec{q_f})
               \om(\vec{p_i})-\vec{q_f}\vec{p_i}\bigr)\Biggr]^2\\
             & \ \frac{\delta^{(3)}(\vec{p_f}+\vec{q_f}+\vec{k}-\vec{p_i}-\vec{q_i})\delta(\omega(\vec{p_f})+\omega(\vec{q_f})+\omega(\vec{k})-\omega(\vec{p_i})
               -\omega(\vec{q_i}))}{(q_f-p_i)^22q_f(l_1-k)(q_f-p_i)^22q_f(l_2-k)}.\\
  \end{split}
\end{equation}
It is convenient to choose the center-of-mass system, defined by $\vec{p_i}+\vec{q_i}=0$. We can easily do the integration w.r.t. $\vec{p_f}$:
\begin{equation}
  \begin{split}
    \sigma = & \ \frac{1}{4}\frac{(2\pi)^{-9}\kappa^6}{\sqrt{\bigl((p_iq_i)^2-m^4\bigr)}}\int d^4l_1d^4l_2\hat{g}(l_1)\hat{g}(l_2)^{\ast}\int\frac{d^3\vec{k}}
               {2\om(\vec{k})}\frac{d^3\vec{q_f}}{2\om(\vec{q_f})}\frac{1}{2\om(\vec{q_f}+\vec{k})}\\
             & \ \Biggl[\Bigl[\bigl(\om(\vec{q_f})\om(\vec{q_f}+\vec{k})+\vec{q_f}(\vec{q_f}+\vec{k})\bigr)(p_iq_i)+\bigl(\om(\vec{q_f})\om(\vec{q_i})-\vec{q_f}\vec{q_i}\bigr)
               \bigl(\om(\vec{p_i})\om(\vec{q_f}+\vec{k})\\
             & +\vec{p_i}(\vec{q_f}+\vec{k})\bigr)-\bigl(\om(\vec{q_f})\om(\vec{p_i})-\vec{q_f}\vec{p_i}\bigr)\bigl(\om(\vec{q_f}+\vec{k})\om(\vec{q_i})+(\vec{q_f}
               +\vec{k})\vec{q_i}\bigr)\\
             & -\bigl(\om(\vec{p_i})\om(\vec{q_f}+\vec{k})+\vec{p_i}(\vec{q_f}+\vec{k})\bigr)(p_iq_i)-(p_iq_i)\bigl(\om(\vec{p_i})\om(\vec{q_f}+\vec{k})+\vec{p_i}
               (\vec{q_f}+\vec{k})\bigr)\\
             & +\vec{p_i}^2\bigl(\om(\vec{q_f}+\vec{k})\om(\vec{q_i})+(\vec{q_f}+\vec{k})\vec{q_i}\bigr)\Bigr]\bigl(\om(\vec{q_f})^2-\vec{q_f}^2\bigr)
               -\Bigl[\bigl(\om(\vec{q_f})\om(\vec{q_f}+\vec{k})\\
             & +\vec{q_f}(\vec{q_f}+\vec{k})\bigr)(p_iq_i)-\bigl(\om(\vec{q_f})\om(\vec{q_i})-\vec{q_f}\vec{q_i}\bigr)\bigl(\om(\vec{p_i})\om(\vec{q_f}+\vec{k})
               +\vec{p_i}(\vec{q_f}+\vec{k})\bigr)\\
             & +\bigl(\om(\vec{q_f})\om(\vec{p_i})-\vec{q_f}\vec{p_i}\bigr)\bigl(\om(\vec{q_f}+\vec{k})\om(\vec{q_i})+(\vec{q_f}+\vec{k})\vec{q_i}\bigr)+\bigl(\om(\vec{p_i})
               \om(\vec{q_f}+\vec{k})\\
             & +\vec{p_i}(\vec{q_f}+\vec{k})\bigr)(p_iq_i)+(p_iq_i)\bigl(\om(\vec{p_i})\om(\vec{q_f}+\vec{k})+\vec{p_i}(\vec{q_f}+\vec{k})\bigr)\\
             & -p_i^2\bigl(\om(\vec{q_f}+\vec{k})\om(\vec{q_i})+(\vec{q_f}+\vec{k})\vec{q_i}\bigr)\Bigr]\bigl(\om(\vec{q_f})\om(\vec{p_i})-\vec{q_f}\vec{p_i}\bigr)\Biggr]^2\\
             & \ \frac{\delta(\om(\vec{q_f})+\om(\vec{q_f}+\vec{k})+\om(\vec{k})-2\om(\vec{p_i}))}{(q_f-p_i)^22q_f(l_1-k)(q_f-p_i)^22q_f(l_2-k)}.\\
  \end{split}
\end{equation}
Now we are left with the integration w.r.t. $\vec{q_f}$. In order to be able to do this we have to rewrite the argument of the delta distribution showing the dependence on 
$|\vec{q_f}|$ explicitely. We want to use the formula
\begin{equation}
  \delta(f(x))=\sum_{i=1}^n\frac{1}{|f^{\prime}(x_i)|}\delta(x-x_i)
\end{equation}
where the $x_i$ are simple zeros of the function $f$. For that purpose we define
\begin{equation}
  \begin{split}
    f(\abs{q_f}) = & \ \sqrt{\vec{q_f}^2+m^2}+\sqrt{(\vec{q_f}+\vec{k})^2+m^2}+\om(\vec{k})-2\om(\vec{p_i})\\
                   = & \ \sqrt{\abs{q_f}^2+m^2}+\sqrt{\abs{q_f}^2+\abs{k}^2+2\abs{q_f}\abs{k}\cos\theta+m^2}+\abs{k}-2\om(\vec{p_i})\\
  \end{split}
\end{equation}
where $\theta=\sphericalangle (\vec{q_f},\vec{k})$ and we have used the notation $\abs{k}=|\vec{k}|,\,\abs{q_f}=|\vec{q_f}|$. In order to simplify the notation in the following formulas we write $\om=2\om(\vec{p_i})$. The zeros of the function $f$ are given by
\begin{equation}
  \begin{split}
    \abs{q_f}^{(1)}(\abs{k},\theta) = & \ A(\abs{k},\theta)+B(\abs{k},\theta)\\
    \abs{q_f}^{(2)}(\abs{k},\theta) = & \ A(\abs{k},\theta)-B(\abs{k},\theta)\\
  \end{split}
\label{zeros}
\end{equation}
where we have introduced
\begin{equation}
  \begin{split}
    A(\abs{k},\theta) = & \ \frac{\om(2\abs{k}-\om)\abs{k}\cos\theta}{2(\om-\abs{k})^2-2\abs{k}^2\cos^2\theta} \\
    B(\abs{k},\theta) = & \ \frac{(\om-\abs{k})\sqrt{\om^4-4\om^3\abs{k}+8\om\abs{k}m^2-2\abs{k}^2m^2+4\om^2(\abs{k}^2-m^2)+2\abs{k}^2m^2\cos 2\theta}}
                          {2(\om-\abs{k})^2-2\abs{k}^2\cos^2\theta}. \\
  \end{split}
\end{equation}
Clearly these solutions are themselves functions of the momentum $\abs{k}$ and the angle $\theta$. If we consider the limit of these functions (\ref{zeros}) as $\abs{k}$ goes to zero we obtain
\begin{equation}
  \lim_{\abs{k}\rightarrow 0}\abs{q_f}^{(j)}=\pm\abs{p_i},\ j=1,2    
\label{limitqf}
\end{equation}
where $\abs{p_i}=|\vec{p_i}|$. The derivative of $f$ with respect to $\abs{q_f}$ is given by
\begin{equation}
  \frac{df}{d\abs{q_f}}=\frac{\abs{q_f}}{\sqrt{m^2+\abs{q_f}^2}}+\frac{\abs{q_f}+\abs{k}\cos\theta}{\sqrt{\abs{k}^2+m^2+\abs{q_f}^2+2\abs{k}\abs{q_f}\cos\theta}}.
\label{fprime}
\end{equation}
By inserting the two zeros of $f$ into (\ref{fprime}) we observe that, by taking the limit $\abs{k}\rightarrow 0$, we get a finite result, namely
\begin{equation}
  \lim_{\abs{k}\rightarrow 0}\frac{df(\abs{q_f})}{d\abs{q_f}}\bigg{\rvert}_{\abs{q_f}=\abs{q_f}^{(j)}}=\pm\frac{2\abs{p_i}}{\om(\vec{p_i})},\ j=1,2.
\end{equation}
Now we can rewrite the cross section as
\begin{equation}
  \begin{split}
    \sigma = & \ \frac{1}{4}\frac{(2\pi)^{-9}\kappa^6}{\sqrt{\bigl((p_iq_i)^2-m^4\bigr)}}\int d^4l_1d^4l_2\hat{g}(l_1)\hat{g}(l_2)^{\ast}\int\frac{d^3\vec{k}}
               {2\om(\vec{k})}\frac{\abs{q_f}^2d\abs{q_f}d\Omega}{2\om(\vec{q_f})}\frac{1}{2\om(\vec{q_f}+\vec{k})}\\
             & \ \Biggl[\Bigl[\bigl(\om(\vec{q_f})\om(\vec{q_f}+\vec{k})+\vec{q_f}(\vec{q_f}+\vec{k})\bigr)(p_iq_i)+\bigl(\om(\vec{q_f})\om(\vec{q_i})-\vec{q_f}\vec{q_i}\bigr)
               \bigl(\om(\vec{p_i})\om(\vec{q_f}+\vec{k})\\
             & +\vec{p_i}(\vec{q_f}+\vec{k})\bigr)-\bigl(\om(\vec{q_f})\om(\vec{p_i})-\vec{q_f}\vec{p_i}\bigr)\bigl(\om(\vec{q_f}+\vec{k})\om(\vec{q_i})+(\vec{q_f}
               +\vec{k})\vec{q_i}\bigr)\\
             & -\bigl(\om(\vec{p_i})\om(\vec{q_f}+\vec{k})+\vec{p_i}(\vec{q_f}+\vec{k})\bigr)(p_iq_i)-(p_iq_i)\bigl(\om(\vec{p_i})\om(\vec{q_f}+\vec{k})+\vec{p_i}
               (\vec{q_f}+\vec{k})\bigr)\\
             & +\vec{p_i}^2\bigl(\om(\vec{q_f}+\vec{k})\om(\vec{q_i})+(\vec{q_f}+\vec{k})\vec{q_i}\bigr)\Bigr]\bigl(\om(\vec{q_f})^2-\vec{q_f}^2\bigr)
               -\Bigl[\bigl(\om(\vec{q_f})\om(\vec{q_f}+\vec{k})\\
             & +\vec{q_f}(\vec{q_f}+\vec{k})\bigr)(p_iq_i)-\bigl(\om(\vec{q_f})\om(\vec{q_i})-\vec{q_f}\vec{q_i}\bigr)\bigl(\om(\vec{p_i})\om(\vec{q_f}+\vec{k})
               +\vec{p_i}(\vec{q_f}+\vec{k})\bigr)\\
             & +\bigl(\om(\vec{q_f})\om(\vec{p_i})-\vec{q_f}\vec{p_i}\bigr)\bigl(\om(\vec{q_f}+\vec{k})\om(\vec{q_i})+(\vec{q_f}+\vec{k})\vec{q_i}\bigr)+\bigl(\om(\vec{p_i})
               \om(\vec{q_f}+\vec{k})\\
             & +\vec{p_i}(\vec{q_f}+\vec{k})\bigr)(p_iq_i)+(p_iq_i)\bigl(\om(\vec{p_i})\om(\vec{q_f}+\vec{k})+\vec{p_i}(\vec{q_f}+\vec{k})\bigr)\\
             & -p_i^2\bigl(\om(\vec{q_f}+\vec{k})\om(\vec{q_i})+(\vec{q_f}+\vec{k})\vec{q_i}\bigr)\Bigr]\bigl(\om(\vec{q_f})\om(\vec{p_i})-\vec{q_f}\vec{p_i}\bigr)\Biggr]^2\\ 
             & \ \frac{1}{(q_f-p_i)^22q_f(l_1-k)(q_f-p_i)^22q_f(l_2-k)}\sum_{j=1}^{2}\frac{1}{\lvert f^{\prime}(\abs{q_f})\rvert_{\abs{q_f}=\abs{q_f}^{(j)}}\rvert}
               \delta(\abs{q_f}-\abs{q_f}^{(j)}).\\
  \end{split}
\end{equation}
Before we do the integration w.r.t. $\abs{q_f}$ let us abbreviate by $F_1(\abs{q_f},\abs{k},\abs{k}\cos\theta)$ the expression which comes from the tensor part in the numerator, i.e.
\begin{equation}
  \begin{split}
    F_1(\abs{q_f},\abs{k},\abs{k}\cos\theta) = & \ \Biggl[\Bigl[\bigl(\om(\vec{q_f})\om(\vec{q_f}+\vec{k})+\vec{q_f}(\vec{q_f}+\vec{k})\bigr)(p_iq_i)+\bigl(\om(\vec{q_f})
                                                 \om(\vec{q_i})-\vec{q_f}\vec{q_i}\bigr)\\
                                               & \ \bigl(\om(\vec{p_i})\om(\vec{q_f}+\vec{k})+\vec{p_i}(\vec{q_f}+\vec{k})\bigr)-\bigl(\om(\vec{q_f})\om(\vec{p_i})
                                                 -\vec{q_f}\vec{p_i}\bigr)\bigl(\om(\vec{q_f}+\vec{k})\\
                                               & \ \om(\vec{q_i})+(\vec{q_f}+\vec{k})\vec{q_i}\bigr)-\bigl(\om(\vec{p_i})\om(\vec{q_f}+\vec{k})+\vec{p_i}(\vec{q_f}
                                                 +\vec{k})\bigr)(p_iq_i)-(p_iq_i)\\
                                               & \bigl(\om(\vec{p_i})\om(\vec{q_f}+\vec{k})+\vec{p_i}(\vec{q_f}+\vec{k})\bigr)+\vec{p_i}^2\bigl(\om(\vec{q_f}
                                                 +\vec{k})\om(\vec{q_i})+(\vec{q_f}+\vec{k})\vec{q_i}\bigr)\Bigr]\\
                                               & \ \bigl(\om(\vec{q_f})^2-\vec{q_f}^2\bigr)-\Bigl[\bigl(\om(\vec{q_f})\om(\vec{q_f}+\vec{k})+\vec{q_f}(\vec{q_f}+\vec{k})\bigr)
                                                 \ (p_iq_i)-\bigl(\om(\vec{q_f})\\
                                               & \ \om(\vec{q_i})-\vec{q_f}\vec{q_i}\bigr)\bigl(\om(\vec{p_i})\om(\vec{q_f}+\vec{k})+\vec{p_i}(\vec{q_f}+\vec{k})\bigr)
                                                 +\bigl(\om(\vec{q_f})\om(\vec{p_i})-\vec{q_f}\vec{p_i}\bigr)\\
                                               & \ \bigl(\om(\vec{q_f}+\vec{k})\om(\vec{q_i})+(\vec{q_f}+\vec{k})\vec{q_i}\bigr)+\bigl(\om(\vec{p_i})\om(\vec{q_f}+\vec{k})
                                                 +\vec{p_i}(\vec{q_f}+\vec{k})\bigr)\\
                                               & \ (p_iq_i)+(p_iq_i)\bigl(\om(\vec{p_i})\om(\vec{q_f}+\vec{k})+\vec{p_i}(\vec{q_f}+\vec{k})\bigr)-p_i^2\bigl(\om(\vec{q_f}
                                                 +\vec{k})\\
                                               & \ \om(\vec{q_i})+(\vec{q_f}+\vec{k})\vec{q_i}\bigr)\Bigr]\bigl(\om(\vec{q_f})\om(\vec{p_i})-\vec{q_f}\vec{p_i}\bigr)\Biggr]^2.\\
  \end{split}
\end{equation}
Then the expression for the cross section can be written as
\begin{equation}
  \begin{split}
    \sigma = & \ \frac{1}{4}\frac{(2\pi)^{-9}\kappa^6}{\sqrt{\bigl((p_iq_i)^2-m^4\bigr)}}\int d^4l_1d^4l_2\hat{g}(l_1)\hat{g}(l_2)^{\ast}\int\frac{d^3\vec{k}}
               {2\om(\vec{k})}\frac{\abs{q_f}^2d\abs{q_f}d\Omega}{2\om(\vec{q_f})}\frac{1}{2\om(\vec{q_f}+\vec{k})}\\
             & \ \frac{F_1(\abs{q_f},\abs{k},\abs{k}\cos\theta)}{(q_f-p_i)^22q_f(l_1-k)(q_f-p_i)^22q_f(l_2-k)}\sum_{j=1}^{2}\frac{1}{\lvert f^{\prime}(\abs{q_f})
               \rvert_{\abs{q_f}=\abs{q_f}^{(j)}}\rvert}\delta(\abs{q_f}-\abs{q_f}^{(j)}).\\
  \end{split}
\end{equation}
From this expression we can identify the differential cross section $\frac{d\sigma}{d\Omega}$. We do the $\abs{q_f}$-integration and obtain
\begin{equation}
  \begin{split}
    \frac{d\sigma}{d\Omega} = & \ \frac{1}{4}\frac{(2\pi)^{-9}\kappa^6}{\sqrt{\bigl((p_iq_i)^2-m^4\bigr)}}\int d^4l_1d^4l_2\hat{g}(l_1)\hat{g}(l_2)^{\ast}\int\frac{d^3\vec{k}}
                                {2\om(\vec{k})}\sum_{j=1}^{2}\frac{1}{\lvert f^{\prime}(\abs{q_f}^{(j)})\rvert}\\
                              & \ \frac{(\abs{q_f}^{(j)})^2}{2\sqrt{\bigl((\abs{q_f}^{(j)})^2+m^2\bigr)}}\frac{F_1(\abs{q_f}^{(j)},\abs{k},\abs{k}\cos\theta)}{2\sqrt{\bigl(
                                (\abs{q_f}^{(j)})^2+\abs{k}^2+\abs{k}\abs{q_f}^{(j)}\cos\theta+m^2\bigr)}}\\
                              & \ \Biggl(\frac{1}{m^2-2\sqrt{(\abs{q_f}^{(j)})^2+m^2}\om(\vec{p_i})+2\abs{q_f}^{(j)}\abs{p_i}\cos\al+p_i^2}\Biggr)^2\\
                              & \ \frac{1}{2\bigl[\sqrt{(\abs{q_f}^{(j)})^2+m^2}l_1^0-\abs{q_f}^{(j)}\abs{l_1}\cos\be-\sqrt{(\abs{q_f}^{(j)})^2+m^2}\abs{k}+\abs{q_f}^{(j)}
                                \abs{k}\cos\theta\bigr]}\\
                              &  \ \frac{1}{2\bigl[\sqrt{(\abs{q_f}^{(j)})^2+m^2}l_2^0-\abs{q_f}^{(j)}\abs{l_2}\cos\ga-\sqrt{(\abs{q_f}^{(j)})^2+m^2}\abs{k}+\abs{q_f}^{(j)}
                                \abs{k}\cos\theta\bigr]}\\
  \end{split}
\end{equation}
where we have introduced
\begin{equation}
  \begin{split}
    \al = & \ \sphericalangle(\vec{q_f},\vec{p_i})\\
    \be = & \ \sphericalangle(\vec{q_f},\vec{l_1})\\
    \ga = & \ \sphericalangle(\vec{q_f},\vec{l_2})\\
    \abs{l_i} = & \ |\vec{l_i}|,\ i=1,2.\\
  \end{split}
\end{equation}
In the next step we omit all small quantities in the numerator, and we set $\abs{k}=0$ in the denominators if there occur no singularities. Then the expression can be simplified to 
\begin{equation}
  \begin{split}
    \frac{d\sigma}{d\Omega} = & \ \frac{1}{4}\frac{(2\pi)^{-9}\kappa^6}{\sqrt{\bigl((p_iq_i)^2-m^4\bigr)}}\frac{\om(\vec{p_i})}{2\abs{p_i}}\int d^4l_1d^4l_2\hat{g}(l_1)
                                \hat{g}(l_2)^{\ast}\int\frac{d^3\vec{k}}{2\om(\vec{k})}\\
                              & \ \sum_{j=1}^{2}\frac{\abs{p_i}^2}{4\om(\vec{p_i})^2}F_1(\abs{q_f}^{(j)}|_{\abs{k}=0},0,0)\Biggl(\frac{1}{m^2-2\om(\vec{p_i})^2
                                +2\abs{q_f}^{(j)}|_{\abs{k}=0}\abs{p_i}\cos\al+p_i^2}\Biggr)^2\\
                              & \ \frac{1}{2\bigl[\om(\vec{p_i})l_1^0-\abs{q_f}^{(j)}|_{\abs{k}=0}\abs{l_1}\cos\be-\om(\vec{p_i})\abs{k}+\abs{q_f}^{(j)}|_{\abs{k}=0}\abs{k}
                                \cos\theta\bigr]}\\
                              & \ \frac{1}{2\bigl[\om(\vec{p_i})l_2^0-\abs{q_f}^{(j)}|_{\abs{k}=0}\abs{l_2}\cos\ga-\om(\vec{p_i})\abs{k}+\abs{q_f}^{(j)}|_{\abs{k}=0}\abs{k}
                                \cos\theta\bigr]}.\\
  \end{split}
\end{equation}
Now one observes that the function $F_1$ is independent of $\cos\theta$ and depends only on the initial momenta $p_i$ and $q_i$ so we can take it outside the integral. Then we get
\begin{equation}
  \begin{split}
    \frac{d\sigma}{d\Omega} = & \ \frac{1}{4}\frac{(2\pi)^{-9}\kappa^6}{\sqrt{\bigl((p_iq_i)^2-m^4\bigr)}}\sum_{j=1}^{2}F_2(\abs{q_f}^{(j)}|_{\abs{k}=0})\int d^4l_1d^4l_2
                                \hat{g}(l_1)\hat{g}(l_2)^{\ast}\\
                              & \ \int\frac{d^3\vec{k}}{2\om(\vec{k})}\frac{1}{2\bigl[\om(\vec{p_i})(l_1^0-\abs{k})-\abs{q_f}^{(j)}|_{\abs{k}=0}(\abs{l_1}\cos\be-\abs{k}
                                \cos\theta)\bigr]}\\
                              & \ \frac{1}{2\bigl[\om(\vec{p_i})(l_2^0-\abs{k})-\abs{q_f}^{(j)}|_{\abs{k}=0}(\abs{l_2}\cos\ga-\abs{k}
                                \cos\theta)\bigr]}\\
  \end{split}
\label{dsdO}
\end{equation}
where the function $F_2$ is given by
\begin{equation}
  F_2(\abs{q_f}^{(j)}|_{\abs{k}=0})=\frac{\abs{p_i}}{8\om(\vec{p_i})}F_1(\abs{q_f}^{(j)}|_{\abs{k}=0},0,0)\Biggl(\frac{1}{m^2-2\om(\vec{p_i})^2+2\abs{q_f}^{(j)}|_{\abs{k}=0}
                                    \abs{p_i}\cos\al+p_i^2}\Biggr)^2.
\end{equation}
If we use (\ref{limitqf}) we can write the differential cross section in the form
\begin{equation}
  \begin{split}
    \frac{d\sigma}{d\Omega} = & \ \frac{1}{16}\frac{(2\pi)^{-9}\kappa^6}{\sqrt{\bigl((p_iq_i)^2-m^4\bigr)}}\int d^4l_1d^4l_2\hat{g}(l_1)\hat{g}(l_2)^{\ast}\Bigg[F_2(+\abs{p_i})
                                \int\frac{d^3\vec{k}}{2\om(\vec{k})}\\
                              & \ \frac{1}{\bigl[\om(\vec{p_i})(l_1^0-\abs{k})-\abs{p_i}(\abs{l_1}\cos\be-\abs{k}\cos\theta)\bigr]\bigl[\om(\vec{p_i})(l_2^0
                                -\abs{k})-\abs{p_i}(\abs{l_2}\cos\ga-\abs{k}\cos\theta)\bigr]}\\
                              & +F_2(-\abs{p_i})\int\frac{d^3\vec{k}}{2\om(\vec{k})}\\ 
                              & \ \frac{1}{\bigl[\om(\vec{p_i})(l_1^0-\abs{k})+\abs{p_i}(\abs{l_1}\cos\be-\abs{k}\cos\theta)\bigr]\bigl[\om(\vec{p_i})(l_2^0
                                -\abs{k})+\abs{p_i}(\abs{l_2}\cos\ga-\abs{k}\cos\theta)\bigr]}\Biggr].\\
  \end{split}
\label{diffcrosssec}
\end{equation}
In order to do the integration w.r.t. $\vec{k}$ we use an identity due to Feynman
\begin{equation}
  \frac{1}{ab}=\int_0^1\frac{dx}{[a x+b(1-x)]^2},\quad a,b\in \mathbb{C}.
\end{equation}
We consider the first term in (\ref{diffcrosssec}). Then we define the functions $a$ and $b$ by 
\begin{equation}
  \begin{split}
    a = & \ \bigl[\om(\vec{p_i})(l_1^0-\abs{k})-\abs{p_i}(\abs{l_1}\cos\be-\abs{k}\cos\theta)\bigr] \\
    b = & \ \bigl[\om(\vec{p_i})(l_2^0-\abs{k})-\abs{p_i}(\abs{l_2}\cos\ga-\abs{k}\cos\theta)\bigr]. \\
  \end{split}
\label{ab}
\end{equation}
The integral becomes 
\begin{equation}
  \begin{split}
    \int\frac{d^3\vec{k}}{2\om(\vec{k})}\frac{1}{ab} = & \ \frac{1}{2}\int\abs{k}d\abs{k}d\Omega_{\abs{k}}\int_0^1\frac{dx}{\bigl[ax+b(1-x)\bigr]^2}\\
                                                     = & \ \frac{2\pi}{2}\int\abs{k}d\abs{k}\int_0^1dx\int_{-1}^{+1}\frac{d\cos\theta}{\bigl[ax+b(1-x)\bigr]^2}.\\
  \end{split}
\end{equation}
We insert $a$ and $b$ and obtain
\begin{equation}
  \begin{split}
    = & \ \pi\int\abs{k}d\abs{k}\int_0^1dx\int_{-1}^{+1}\\
      & \ \frac{d\cos\theta}{\bigl[-\om(\vec{p_i})\abs{k}+\abs{p_i}\abs{k}\cos\theta+(\om(\vec{p_i})l_1^0-\abs{p_i}\abs{l_1}\cos\be)x+(\om(\vec{p_i})l_2^0-\abs{p_i}
        \abs{l_2}\cos\ga)(1-x)\bigr]^2}.\\
  \end{split}
\label{theta-int}
\end{equation}
We set $\cos\theta=z$ and we observe that the integral w.r.t. $z$ is of the form 
\begin{equation}
  \int_{-1}^{+1}\frac{dz}{(cz+d)^2}=\frac{2}{d^2-c^2}
\label{bs1}
\end{equation}
see \cite{bs:tdm}, where we've set
\begin{equation}
  \begin{split}
    c = & \ \abs{p_i}\abs{k}\\
    d = & \ -\om(\vec{p_i})\abs{k}+\bigl(\om(\vec{p_i})l_1^0-\abs{p_i}\abs{l_1}\cos\be\bigr)x+\bigl(\om(\vec{p_i})l_2^0-\abs{p_i}\abs{l_2}\cos\ga\bigr)(1-x).\\
  \end{split}
\end{equation}
Then we obtain for (\ref{theta-int})
\begin{equation}
    = 2\pi\int\abs{k}d\abs{k}\int_0^1dx\frac{1}{p_i^2\abs{k}^2+A(l_1,l_2,x)\abs{k}+B(l_1,l_2,x)}    
\end{equation}
where $A(l_1,l_2,x)$ and $B(l_1,l_2,x)$ are given by
\begin{equation}
  \begin{split}
    A(l_1,l_2,x) = & -2\om(\vec{p_i})\Bigl[\bigl(\om(\vec{p_i})l_1^0-\abs{p_i}\abs{l_1}\cos\be\bigr)x+\bigl(\om(\vec{p_i})l_2^0-\abs{p_i}\abs{l_2}\cos\ga\bigr)(1-x)\Bigr]\\
    B(l_1,l_2,x) = & \ \Bigl[\bigl(\om(\vec{p_i})l_1^0-\abs{p_i}\abs{l_1}\cos\be\bigr)x+\bigl(\om(\vec{p_i})l_2^0-\abs{p_i}\abs{l_2}\cos\ga\bigr)(1-x)\Bigr]^2.\\
  \end{split}
\end{equation}
Then this integral w.r.t. $\abs{k}$ is of the form
\begin{equation}
  \int \frac{ydy}{c_1y^2+c_2y+c_3}=\frac{1}{2c_1}\ln\bigl(c_1y^2+c_2y+c_3\bigr)-\frac{c_2}{2c_1}\int\frac{dy}{c_1y^2+c_2y+c_3}
\label{bs44}
\end{equation}
where the constants $c_i,\ i=1,2,3$ are given by
\begin{equation}
  \begin{split}
    c_1 = & \ p_i^2=m^2\\
    c_2 = & \ A(l_1,l_2,x)\\
    c_3 = & \ B(l_1,l_2,x).\\
  \end{split}
\end{equation}
In our case the discriminant $\Delta:=4c_1c_3-c_2^2$ is given by
\begin{equation}
  \begin{split}
    \Delta = & +4p_i^2\Bigl[\bigl(\om(\vec{p_i})l_1^0-\abs{p_i}\abs{l_1}\cos\be\bigr)x+\bigl(\om(\vec{p_i})l_2^0-\abs{p_i}\abs{l_2}\cos\ga\bigr)(1-x)\Bigr]^2\\
             & -4\om(\vec{p_i})^2\Bigl[\bigl(\om(\vec{p_i})l_1^0-\abs{p_i}\abs{l_1}\cos\be\bigr)x+\bigl(\om(\vec{p_i})l_2^0-\abs{p_i}\abs{l_2}\cos\ga\bigr)(1-x)\Bigr]^2\\
           = & \ -4\abs{p_i}^2\Bigl[\bigl(\om(\vec{p_i})l_1^0-\abs{p_i}\abs{l_1}\cos\be\bigr)x+\bigl(\om(\vec{p_i})l_2^0-\abs{p_i}\abs{l_2}\cos\ga\bigr)(1-x)\Bigr]^2\\
  \end{split}
\end{equation}
so we see that $\Delta\leq 0$. Then the integral (\ref{bs44}) is given by
\begin{equation}
  \int \frac{ydy}{c_1y^2+c_2y+c_3} =  \frac{1}{2c_1}\ln(c_1y^2+c_2y+c_3)-\frac{c_2}{2c_1\sqrt{-\Delta}}\ln\Biggl[\frac{2c_1y+c_2-\sqrt{-\Delta}}{2c_1y+c_2
                                      +\sqrt{-\Delta}}\Biggr].
\end{equation}
Here we consider gravitons up to an energy $\om_0$ which is much smaller than the energy of the incident particles. Since every real detector has a finite energy resolution below which he cannot detect any particles we have to integrate over all these contributions up to the value $\om_0$. Then the final result of the $\abs{k}$-integration is given by the following expression
\begin{equation}
  \begin{split}
    \int_0^{\om_0}\frac{\abs{k}d\abs{k}}{p_i^2\abs{k}^2+A\abs{k}+B} = & \ \frac{1}{2c_1}\Biggl[\ln\bigl(c_1\om_0^2+c_2\om_0+c_3\bigr)\\
                                                                      & -\frac{c_2}{\sqrt{-\Delta}}\ln\Biggl(\frac{2c_1\om_0+c_2-\sqrt{-\Delta}}{2c_1\om_0+c_2
                                                                        +\sqrt{-\Delta}}\Biggr)\\
                                                                      & -\ln(c_3)+\frac{c_2}{\sqrt{-\Delta}}\ln\Biggl(\Bigg\lvert\frac{c_2-\sqrt{-\Delta}}{c_2+\sqrt{-\Delta}}
                                                                        \Bigg\rvert\Biggr)\Biggr].\\
  \end{split}
\end{equation}
The second term in (\ref{diffcrosssec}) can be treated in the same way. The only difference is the sign of $\abs{p_i}$ in the definition of the functions $a$ and $b$, see (\ref{ab}), which has no consequences for the calculation of the integrals (\ref{bs1}) and (\ref{bs44}). 
%
%
\subsection{Adiabatic Limit}
In the preceeding subsection we've found the differential cross section for brems\-strahlung in a scattering process of two massive scalar particles. Now we want to discuss the adiabatic limit. For the differential cross section we've found
\begin{equation}
  \begin{split}
    \frac{d\sigma}{d\Omega} = & \ \frac{1}{16}\frac{(2\pi)^{-9}\kappa^6}{\sqrt{\bigl((p_iq_i)^2-m^4\bigr)}}\int d^4l_1d^4l_2\hat{g}(l_1)\hat{g}(l_2)^{\ast}
                                F_2(+\abs{p_i})\\
                              & \ 2\pi\int_0^1dx\frac{1}{2c_1}\Biggl[\ln\bigl(c_1\om_0^2+c_2\om_0+c_3\bigr)\\
                              & -\frac{c_2}{\sqrt{-\Delta}}\ln\Biggl(\frac{2c_1\om_0+c_2-\sqrt{-\Delta}}{2c_1\om_0+c_2+\sqrt{-\Delta}}\Biggr)\\
                              & \ -\ln(c_3)+\frac{c_2}{\sqrt{-\Delta}}\ln\Biggl(\Bigg\lvert\frac{c_2-\sqrt{-\Delta}}{c_2+\sqrt{-\Delta}}\Bigg\rvert\Biggr)\Biggr]\\
  \end{split}
\label{diffcrosssec1}
\end{equation}  
where $c_1=m^2$ and $c_i=c_i(l_1,l_2,x),\ i=2,3$. This expression, as it stands, is well defined and finite due to the presence of the test functions. To obtain a physically relevant result we have to remove this cutoff and therefore test the infrared behaviour of (\ref{diffcrosssec1}). We will show below that part of this expression becomes singular in the adiabatic limit. In the adiabatic limit in momentum space we let the test functions tend to a delta distribution in their argument. This will be done in the same way as in subsection 9.1, see (\ref{adia}). We scale the argument of the test functions with a parameter $\vep$ in the integrals of (\ref{diffcrosssec1}) and consider the limit $\vep\rightarrow 0$. The functions $c_3$ and $\Delta$ are homogeneous of degree 2 in the variables $l_1$ and $l_2$, i.e. 
\begin{equation}
  \begin{split}
    c_3(\vep l_1,\vep l_2,x)=\vep^2 c_3(l_1,l_2,x)\\
    \Delta(\vep l_1,\vep l_2,x)=\vep^2\Delta(l_1,l_2,x)\\
  \end{split}
\end{equation}
while the function $c_2$ is homogeneous of degree 1, i.e. $c_2(\vep l_1,\vep l_2,x)=\vep c_2(l_1,l_2,x)$. With this in mind let us now look at the various logarithms in (\ref{diffcrosssec1}). The argument of the first logarithm contains the constant $m^2\om_0^2$ so it stays finite if $\vep$ goes to zero. The constant in front of the second and the fourth logarithm has equal powers of $\vep$ in the numerator as well as in the denominator so it tends to a constant. The argument of the second logarithm goes to one, so the logarithm itself goes to zero and the fourth goes to another constant. It remains to consider the third logarithm. This one becomes singular in the adiabatic limit. This can be seen as follows:
\begin{equation}
  \begin{split}
    \frac{d\sigma}{d\Omega} = & -\frac{\pi}{8}\frac{(2\pi)^{-9}\kappa^6}{\sqrt{\bigl((p_iq_i)^2-m^4\bigr)}}\int d^4l_1d^4l_2\hat{g}(l_1)\hat{g}(l_2)^{\ast}
                                F_2(+\abs{p_i})\frac{1}{2c_1}\int_0^1dx\\
                              & \ \ln\Biggl(\Bigl[\bigl(\om(\vec{p_i})l_1^0-\abs{p_i}\abs{l_1}\cos\be\bigr)x+\bigl(\om(\vec{p_i})l_2^0-\abs{p_i}\abs{l_2}\cos\ga\bigr)
                                (1-x)\Bigr]^2\Biggr)\\
                              & +\text{finite terms}.\\
  \end{split}
\end{equation}
If we concentrate on the singular part only we obtain by inserting the scaled test functions 
\begin{equation}
  \begin{split}
    & \ \int d^4l_1d^4l_2\hat{g}(l_1)\hat{g}(l_2)^{\ast}\ln\bigl\lvert\bigl(\om(\vec{p_i})l_1^0-\abs{p_i}\abs{l_1}\cos\be\bigr)x+\bigl(\om(\vec{p_i})l_2^0
      -\abs{p_i}\abs{l_2}\cos\ga\bigr)(1-x)\bigr\rvert\\
  = & \ \int d^4l_1d^4l_2\hat{g}_{\vep}(l_1)\hat{g}_{\vep}(l_2)^{\ast}\ln\bigl\lvert\bigl(\om(\vec{p_i})l_1^0-\abs{p_i}\abs{l_1}\cos\be\bigr)x+\bigl(\om(\vec{p_i})l_2^0
      -\abs{p_i}\abs{l_2}\cos\ga\bigr)(1-x)\bigr\rvert\\
  = & \ \frac{1}{\vep^8}\int d^4l_1d^3l_2\hat{g}_0\Bigl(\frac{l_1}{\vep}\Bigr)\hat{g}_0\Bigl(\frac{l_2}{\vep}\Bigr)^{\ast}\ln\bigl\lvert\bigl(\om(\vec{p_i})l_1^0-\abs{p_i}
      \abs{l_1}\cos\be\bigr)x+\bigl(\om(\vec{p_i})l_2^0-\abs{p_i}\abs{l_2}\cos\ga\bigr)\\
    & \ \times(1-x)\bigr\rvert\\
  = & \int d^4k_1d^4k_2\hat{g}_0(k_1)\hat{g}_0(k_2)^{\ast}\ln\bigl\lvert\bigl(\om(\vec{p_i})\vep k_1^0-\abs{p_i}\vep\abs{k_1}\cos\be\bigr)x+\bigl(\om(\vec{p_i})\vep k_2^0-
      \abs{p_i}\vep\abs{k_2}\cos\ga\bigr)\\
    & \ \times(1-x)\bigr\rvert\\
  = & \ (2\pi)^4\ln\lvert\vep\rvert+o(1)\\
  \end{split}
\end{equation}
where we've introduced $k_i=\frac{l_i}{\vep}$. So we get a logarithmic divergence if $\vep$ tends to zero. The origin of this singularity is the massless graviton propagator as can be clearly seen from the above calculation. Since all the other terms in (\ref{T3}) have the same structure of propagators we would get the same logarihmic divergence as in our example. Furthermore there can be no cancellation between these terms since they all have a positive sign. So we can omit the discussion of the other terms. 

This result is similar to the case of quantum electrodynamics \cite{sch:qed} where a logarithmic divergence in the differential cross section was obtained as well in the bremsstrahlung contribution to the scattering of an electron due to an external classical source. As is well known from QED, the infrared divergence in the bremsstrahlung must cancel against contributions from radiative corrections, otherwise scattering theory would break down.
%
%
\subsection{Graviton and Matter Self-Energy}
In this section we discuss the infrared behaviour of radiative corrections to two particle scattering coming form the graviton self-energy and matter self-energy. Let us start with the graviton self-energy tensor. This is a fourth-order contribution to the $S$-matrix and the time ordered product $T_4(x_1,\ldots, x_4)$ is given by
\begin{equation}
  \begin{split}
    T_4(x_1,\ldots, x_4) = & \ i\kappa^4\Bigl[:\vp(x_1)^{\tot}_{\pal}\vp(x_1)_{\pbe}\vp(x_4)^{\tot}_{\pro}\vp(x_4)_{\psig}:\\
                           & -m^2\eta^{\ro\si}:\vp(x_1)^{\tot}_{\pal}\vp(x_1)_{\pbe}\vp(x_4)^{\tot}\vp(x_4):\\
                           & -m^2\eta^{\al\be}:\vp(x_1)^{\tot}\vp(x_1)\vp(x_4)^{\tot}_{\pro}\vp(x_4)_{\psig}:\\
                           & +m^4\eta^{\al\be}\eta^{\ro\si}:\vp(x_1)^{\tot}\vp(x_1)\vp(x_4)^{\tot}\vp(x_4):\Bigr]\\
                           & \times D^F_0(x_1-x_2)b^{\al\be\la\ep}\Pi(x_2-x_3)_{\la\ep\mu\nu}b^{\mu\nu\ro\si}D^F_0(x_3-x_4)\\
  \end{split}
\label{T4}
\end{equation}
where the normalization terms are already included in the graviton self-energy tensor $\Pi$ \cite{gr:cqg3}. The $S$-matrix to fourth order is then given by
\begin{equation}
  S_4(g)=\int d^4x_1\ldots d^4x_4T_4(x_1,\ldots, x_4)g(x_1)\ldots g(x_4),\quad g\in\mathcal{S}(\mathbb{R}^4).
\end{equation}
We consider the following initial and final states
\begin{equation}
  \begin{split}
    \Phi_i = & \ \int d^3\vec{p_1}\,d^3\vec{q_1}\hat{\psi}_1(\vec{p_1})\hat{\psi}_2(\vec{q_1})a(\vec{p_1})^{\tot}a(\vec{q_1})^{\tot}\Omega\\
    \Phi_f = & \ \int d^3\vec{p_2}\,d^3\vec{q_2}\hat{\Psi}(\vec{p_2},\vec{q_2})a(\vec{p_2})^{\tot}a(\vec{q_2})^{\tot}\Omega.\\
  \end{split}
\end{equation}
We are interested in the matrix element of $S_4(g)$ between these states
\begin{equation}
  S_{fi}=\bigl(\Phi_f,S_4(g)\Phi_i\bigr).
\end{equation}
In the following we will consider the first term of $T_4$ only, because all the other terms have the same propagator structure and differ from the first one in the powers of external momenta or powers of $m$ only. We refer with an index $(1)$ to the first term of (\ref{T4}). Then the matrix element becomes
\begin{equation}
  \begin{split}
    S_{fi}^{(1)} = & \ i\kappa^4\int d^3\vec{p_1}\,d^3\vec{q_1}\,d^3\vec{p_2}\,d^3\vec{q_2}\,d^4x_1\ldots d^4x_4\hat{\psi}_1(\vec{p_1})\hat{\psi}_2(\vec{q_1})
                     \hat{\Psi}(\vec{p_2},\vec{q_2})\\
                   & \ g(x_1)\ldots g(x_4)D^F_0(x_1-x_2)b^{\al\be\la\ep}\Pi(x_2-x_3)_{\la\ep\mu\nu}b^{\mu\nu\ro\si}D^F_0(x_3-x_4)\\
                   & \ \bigl(\Omega,a(\vec{p_2})a(\vec{q_2}):\vp(x_1)^{\tot}_{\pal}\vp(x_1)_{\pbe}\vp(x_4)^{\tot}_{\pro}\vp(x_4)_{\psig}:a(\vec{p_1})^{\tot}a(\vec{q_1})^{\tot}
                     \Omega\bigr).\\
  \end{split}
\label{VPSfi2}
\end{equation}
We calculate the vacuum expectation value and write everything in momentum space. Then we obtain
\begin{multline}
    S_{fi}^{(1)} =  \ (2\pi)^{-4}i\kappa^4\int d^3\vec{p_1}\,d^3\vec{q_1}\,d^3\vec{p_2}\,d^3\vec{q_2}\,d^4l_1\ldots d^4l_4\hat{\psi}_1(\vec{p_1})\hat{\psi}_2(\vec{q_1})
                    \hat{\Psi}(\vec{p_2},\vec{q_2})\\
                    \hat{g}(l_1)\ldots\hat{g}(l_4)\bigl(16\om(\vec{p_1})\om(\vec{q_1})\om(\vec{p_2})\om(\vec{q_2})\bigr)^{-1/2}b^{\al\be\la\ep}b^{\mu\nu\ro\si}\\
                    \Bigl[\hat{D}^F_0(l_2+l_3+l_4-q_2+q_1)\hat{\Pi}(l_3+l_4-q_2+q_1)_{\la\ep\mu\nu}\hat{D}^F_0(l_4-q_2+q_1)p_2^{\al}p_1^{\be}q_2^{\ro}q_1^{\si}\\
                 +  \hat{D}^F_0(l_2+l_3+l_4-q_2+p_1)\hat{\Pi}(l_3+l_4-q_2+p_1)_{\la\ep\mu\nu}\hat{D}^F_0(l_4-q_2+p_1)p_2^{\al}q_1^{\be}q_2^{\ro}p_1^{\si}\\
                 +  \hat{D}^F_0(l_2+l_3+l_4-p_2+q_1)\hat{\Pi}(l_3+l_4-p_2+q_1)_{\la\ep\mu\nu}\hat{D}^F_0(l_4-p_2+q_1)q_2^{\al}p_1^{\be}p_2^{\ro}q_1^{\si}\\
                 +  \hat{D}^F_0(l_2+l_3+l_4-p_2+p_1)\hat{\Pi}(l_3+l_4-p_2+p_1)_{\la\ep\mu\nu}\hat{D}^F_0(l_4-p_2+p_1)q_2^{\al}q_1^{\be}p_2^{\ro}p_1^{\si}\Bigr]\\
                    \de^{(4)}(l_1+\ldots+l_4-q_2+q_1-p_2+p_1).\\
\label{VPSfi6}
\end{multline} 
The adiabatic limit in the variable $l_1$ is trivial, yielding just a factor $(2\pi)^2$. Then we have to investigate the adiabatic limit in the remaining variables $l_2,\ldots,l_4$. We do this by introducing test functions with scaled arguments, see section 9.1. We consider the part depending on the variables $l_2,\ldots,l_4$ in the first term of (\ref{VPSfi6}). 
\begin{equation}
  \begin{split}
    & \ \int d^4l_2\ldots d^4l_4\hat{g}_{\vep}(l_2)\ldots\hat{g}_{\vep}(l_4)\hat{D}^F_0(l_2+l_3+l_4-q_2+q_1)b^{\al\be\la\ep}\\
    & \ \hat{\Pi}(l_3+l_4-q_2+q_1)_{\la\ep\mu\nu}b^{\mu\nu\ro\si}\hat{D}^F_0(l_4-q_2+q_1)\de^{(4)}(l_2+l_3+l_4-p_2+p_1-q_2+q_1)\\
  = & \ \vep^{-12}\int d^4l_2\ldots d^3l_4\hat{g}_0(\vep^{-1}l_2)\ldots\hat{g}_0(\vep^{-1}l_4)\hat{D}^F_0(l_2+l_3+l_4-q_2+q_1)b^{\al\be\la\ep}\\
    & \ \hat{\Pi}(l_3+l_4-q_2+q_1)_{\la\ep\mu\nu}b^{\mu\nu\ro\si}\hat{D}^F_0(l_4-q_2+q_1)\de^{(4)}(l_2+l_3+l_4-p_2+p_1-q_2+q_1)\\
  = & \ \int d^4k_2\ldots d^4k_4\hat{g}_0(k_2)\ldots\hat{g}_0(k_4)\hat{D}^F_0\bigl(\vep(k_2+k_3+k_4)-q_2+q_1\bigr)b^{\al\be\la\ep}\\
    & \ \hat{\Pi}\bigl(\vep(k_3+k_4)-q_2+q_1\bigr)_{\la\ep\mu\nu}b^{\mu\nu\ro\si}\hat{D}^F_0(\vep k_4-q_2+q_1)\\
    & \ \de^{(4)}\bigl(\vep(k_2+k_3+k_4)-p_2+p_1-q_2+q_1\bigr).\\
  \end{split}
\label{AL}
\end{equation}
The graviton self energy tensor was explicitely calculated in \cite{gr:cqg3} and it is given in momentum space by
\begin{equation}
  \begin{split}
    \hat{\Pi}(p)_{\la\ep\mu\nu} = & \ \frac{\pi}{960(2\pi)^5}\Bigl[-656\,p_{\la}p_{\ep}p_{\mu}p_{\nu}-208\,p^2(p_{\la}p_{\ep}\eta_{\mu\nu}+p_{\mu}p_{\nu}\eta_{\la\ep})\\
                                  & +162\,p^2(p_{\la}p_{\mu}\eta_{\ep\nu}+p_{\la}p_{\nu}\eta_{\ep\mu}+p_{\ep}p_{\mu}\eta_{\la\nu}+p_{\ep}p_{\nu}\eta_{\la\mu})\\
                                  & -162\,p^4(\eta_{\la\mu}\eta_{\ep\nu}+\eta_{\la\nu}\eta_{\ep\mu})+118\,p^4\eta_{\la\ep}\eta_{\mu\nu}\Bigr]\ln\Bigl(\frac{-p^2-i0}{M^2}\Bigr)\\
                                  & + \text{normalization}.\\
  \end{split}
\label{Pi}
\end{equation}
where the normalization is a polynomial of degree four which has the same tensor structure as the one in (\ref{Pi}). $M>0$ is an arbitrary mass scale. Now, according to (\ref{AL}), the argument of the tensor $\hat{\Pi}$ is given by
\begin{equation}
  \begin{split}
    p^2 = & \ \bigl(\vep(k_3+k_4)-q_2+q_1\bigr)^2\\
        = & \ 2m^2-2q_2q_1+2\vep(k_3+k_4)(-q_2+q_1)+\vep^2(k_3+k_4)^2
  \end{split}
\end{equation}
since the momenta $q_1$ and $q_2$ have to be on the mass shell. We see that in the adiabatic limit $\vep\rightarrow 0$ this goes to $2(m^2-q_2q_1)$. The Feynman propagator in momentum space is given by
\begin{equation}
   \hat{D}^F_0(p)=-(2\pi)^{-2}\frac{1}{p^2+i0}.
\end{equation}
Again we observe that in the adiabatic limit the arguments of the two propagators in (\ref{AL}) approach $2(m^2-q_2q_1)$. Therefore, ignoring the tensor structure of $\hat{\Pi}$, we can argue that in the adiabatic limit we obtain
\begin{equation}
  \int d^4k_2\ldots d^4k_4\ldots=\ln\Bigl\lvert\frac{-2(m^2-q_2q_1)}{M^2}\Bigr\rvert+o(1)
\end{equation}
where we have used that
\begin{equation}
  \int d^4k\hat{g}_0(k)=(2\pi)^2.
\end{equation}
This result is finite due to the mass $m$ in the numerator of the logarithm. So we conclude, that the adiabatic limit for radiation corrections coming from the graviton self-energy tensor is infrared finite.

Now we come to the matter self-energy. As in the previous case we restrict ourself to a typical term of the time-ordered product which shows all essential features of the infrared behaviour. Furhtermore it is enough to consider the matter self-energy function in a scattering process where one massive scalar particle is scattered at a fixed external gravitational field. The time-ordered product to third order has the form
\begin{equation}
  T_3(x_1,x_2,x_3)\sim i\kappa^3:\vp(x_1)^{\tot}\vp(x_3):\Sigma(x_1-x_2)D^F_m(x_2-x_3)h^{ext}(x_3)
\end{equation}
where $h^{ext}$ is a classical external gravitational field and $\Sigma$ is the matter self-energy function. The $S$-matrix to third order is then given by
\begin{equation}
  S_3(g)=\int d^4x_1\ldots d^4x_3T_3(x_1,x_2,x_3)g(x_1)g(x_2)g(x_3),\quad g\in\mathcal{S}(\mathbb{R}^4).
\end{equation}
Then we want to calculate the $S$-matrix element
\begin{equation}
  S_{fi}=\bigl(\Phi_f,S_3(g)\Phi_i\bigr)
\end{equation}
where $\Phi_i$ and $\Phi_f$ are one-particle wave packets in momentum space. The $S$-matrix element in momentum space then takes the form
\begin{equation}
  \begin{split}
    S_{fi} = & \ \frac{i\kappa^3}{(2\pi)^3}\int\frac{d^3\vec{p}}{\sqrt{2\om(\vec{p})}}\frac{d^3\vec{q}}{\sqrt{2\om(\vec{q})}}\hat{\psi}(\vec{p})\hat{\Psi}(\vec{q})
               \int d^4l_1\ldots d^4l_3\hat{g}(l_1)\ldots\hat{g}(l_3)\\
             & \ \hat{\Sigma}(-l_1+q)\hat{D}^F_m(-l_1-l_2+q)\hat{h}^{ext}(-l_1-l_2-l_3+q-p).\\
  \end{split}
\end{equation}
In this expression we can perform the adiabatic limit in the variable $l_3$, since it appears in the argument of the external source only. We set this variable equal to zero in $h^{ext}$ and obtain a factor $(2\pi)^2$ from the integration. To study the singularity structure we look at the integrals w.r.t. $l_1$ and $l_2$. 
\begin{equation}
  \begin{split}
    \int d^4l_1d^4l_2\ldots = & \ \int d^4k_1d^4k_2\hat{g}_0(k_1)\hat{g}_0(k_2)\hat{\Sigma}(-\vep k_1+q)\\
                              & \ \hat{D}^F_m(-\vep k_1-\vep k_2+q)\hat{h}^{ext}(-\vep k_1-\vep k_2+q-p).\\
  \end{split}
\end{equation}
The matter self-energy function $\Sigma$ including it's normalization was calculated in \cite{gr:scmqg}, so we quote here the result in momentum space only. 
\begin{equation}
  \begin{split}
    \hat{\Sigma}(p) = & \ \frac{\Gamma}{2\pi}\Biggl\{\Bigl(p^2-\frac{3m^2}{2}+\frac{m^4}{2p^2}\Bigr)\Biggl[\ln\Bigl\lvert\frac{p^2-m^2}{m^2}\Bigr\rvert-i\pi\theta(p^2-m^2)
                        \Biggr]\\
                      & +(p^2-m^2)\Bigl(c_2-\frac{5}{4}\Bigr)\Biggr\}\\
  \end{split}
\end{equation}
where $c_2$ is a normalization constant and $\Gamma$ is given by
\begin{equation}
  \Gamma=\frac{-m^2\pi}{2(2\pi)^4}.
\end{equation}
Taking only the leading singularities in the Feynman propagator $\hat{D}^F_m(q-\vep(k_1+k_2))$ it becomes
\begin{equation}
  \hat{D}^F_m(q-\vep(k_1+k_2))=(2\pi)^{-2}\frac{1}{2\vep q(k_1+k_2)-i0}.
\end{equation}
The self energy function has singularity structure of the following form
\begin{equation}
  \hat{\Sigma}=-2\vep qk_1\ln\Bigl\lvert\frac{2\vep qk_1}{m^2}\Bigr\rvert+o(1).
\end{equation}
Then we obtain an overall logarithmic divergence
\begin{equation}
  \int d^4k_1d^4k_2\ldots=\ln|\vep|\int d^4k_1d^4k_2\hat{g}_0(k_1)\hat{g}_0(k_2)\frac{1}{2q(k_1+k_2)-i0}+o(1).
\end{equation}
since the factor $\vep$ infront of the logarithm is canceled by the Feynman propagator.

In the theory of massive scalar matter coupled to gravity there remains to discuss the so called vertex-function which is again a third-order process. Due to the self-coupling of the gravitational field, which was extensively studied in \cite{sw:s2qgi} there exists vertex functions of different type. This is not the case in QED. Since all infrared divergences calculated so far have the same form $\sim \ln\vep$, it is quite plausible that they exactly cancel as they do in QED. In this case Weinberg's short argument \cite{wei:iphgrav} would be right. For the full proof one needs the various vertex corrections. This is beyond the scope of the present work.
\vspace{1cm}
\frenchspacing
\begin{quotation}
  \glqq Alle Anf\"ange sind dunkel. Gerade dem Mathematiker, der in seiner ausgebildeten Wissenschaft in strenger und formaler Weise mit seinen Begriffen operiert, tut es not, 
  von Zeit zu Zeit daran erinnert zu werden, da\ss\, die Urspr\"unge in dunklere Tiefen zur\"uckweisen, als er mit seinen Methoden zu erfassen vermag. Jenseits alles 
  Einzelwissens bleibt die Aufgabe, zu \emph{begreifen}. Trotz des entmutigenden Hin- und Herschwankens der Philosophie von System zu System k\"onnen wir nicht darauf 
  verzichten, wenn sich nicht Erkenntnis in ein sinnloses Chaos verwandeln soll.\grqq
\end{quotation}
Hermann Weyl, 1918 in \emph{Raum $\cdot$ Zeit $\cdot$ Materie}
%
\newpage
\appendix
\addcontentsline{toc}{section}{\protect\numberline{}{Appendices}}
\noindent {\Large {\bf Appendices}}
\vspace{.75cm}
\nonfrenchspacing

\noindent
In the subsequent appendices A and B we give the explicit divergence forms for the various types in $d_QT_1$. 

\section{Divergences for Types $A,B,C,D,H$ and $K$}
Here we give the unique divergence expressions for $d_QT_1$.

\noindent 1. Type $A$:
We calculate the expression $\partial_{\mu}\wti{T}_{1/1}^{\mu,A}$ explicitly:
\begin{equation}
  \begin{split}
    \partial_{\mu}\wti{T}_{1/1}^{\mu,A} = & \ d_1\, u^{\mu}_{\pal\pmu}h^{\ro\si}_{\pal}h^{\ro\si}+
                                              d_2\, u^{\mu}_{\pal} h^{\ro\si}_{\pmu\pal}h^{\ro\si}+
                                              d_3\, u^{\mu}_{\pal}h^{\ro\si}_{\pmu}h^{\ro\si}_{\pal} \\
                                          & + d_4\, u^{\mu}_{\pmu}h^{\ro\si}_{\pal}h^{\ro\si}_{\pal}+
                                              d_5\, u^{\mu}h^{\ro\si}_{\pmu\pal}h^{\ro\si}_{\pal} \\
  \end{split}
  \label{eq:TypA-Monome}
\end{equation}
The constants $d_1,\ldots,d_5$ are given by
\begin{equation}
  \begin{split}
    d_1 := & \ \ti{c}_1+\ti{c}_5,\quad d_2 :=\ti{c}_1+\ti{c}_7,\quad d_3 :=\ti{c}_1+\ti{c}_6+\ti{c}_7, \\
    d_4 := & \ \ti{c}_2+\ti{c}_5,\quad d_5 :=2\,\ti{c}_2+\ti{c}_6 \\
  \end{split}
  \label{eq:d-TypA}
\end{equation}
From first order gauge invariance we obtain
\begin{equation}
  d_1=0,\quad d_2=0,\quad d_3=-ia_8,\quad d_4=\frac{i}{2}\,a_8,\quad d_5=0
  \label{eq:Zuordnung-TypA}
\end{equation}
The coefficient matrix $M_A$ of (\ref{eq:d-TypA}) is in $GL(5,\mathbb{Z})$. We invert these equations and obtain
\begin{align}
  \ti{c}_1 = & \ d_1+\frac{1}{2}\,\bigl(d_2-d_3+d_5-2\,d_4\bigr) \\
  \ti{c}_2 = & \ \frac{1}{2}\,\bigl(d_5-d_3+d_2\bigr) \\
  \ti{c}_5 = & \ \frac{1}{2}\,\bigl(d_3-d_2-d_5+2\,d_4\bigr) \\
  \ti{c}_6 = & \ d_3-d_2 \\
  \ti{c}_7 = & \ d_4-d_1+\frac{1}{2}\,\bigl(d_2+d_3-d_5\bigr)
\end{align}
These equations give, together with (\ref{eq:Zuordnung-TypA}), the desired divergence for $d_QT_1|_{Type A}$.

\noindent 2. Type $B$:
We calculate the expression $\partial_{\mu}\wti{T}_{1/1}^{\mu, B}$ explicitly:
\begin{equation}
  \begin{split}
    \partial_{\mu}\wti{T}_{1/1}^{\mu,B} = & \ d_6\, u^{\mu}_{\pmu\pal}h^{}_{\pal}h+ 
                                              d_7\, u^{\mu}_{\pal}h^{}_{\pmu\pal}h+ 
                                              d_8\, u^{\mu}_{\pal}h^{}_{\pmu}h_{\pal} \\
                                          & + d_9\, u^{\mu}_{\pmu}h^{}_{\pal}h^{}_{\pal}+ 
                                              d_{10}\, u^{\mu}h_{\pmu\pal}h_{\pal} \\
  \end{split}
  \label{eq:TypB-Monome}
\end{equation}
The constants $d_6,\ldots,d_{10}$ are given by
\begin{equation}
  \begin{split} 
    d_6 := & \ \ti{c}_8+\ti{c}_{12},\quad d_7:=\ti{c}_8+\ti{c}_{14},\quad d_8:=\ti{c}_8+\ti{c}_{13}+\ti{c}_{14}, \\
    d_9 := & \ \ti{c}_9+\ti{c}_{12},\quad d_{10}:=2\,\ti{c}_9+\ti{c}_{13} \\
  \end{split}
  \label{eq:d-TypB}
\end{equation}
From first order gauge invariance we obtain
\begin{equation}
  d_6=0,\quad d_7=0,\quad d_8=-ia_2,\quad d_9=\frac{i}{2}\,a_2,\quad d_{10}=0
  \label{eq:Zuordnung-TypB}
\end{equation}
The coefficient matrix $M_B$ of (\ref{eq:d-TypB}) is in $GL(5,\mathbb{Z})$. We invert these equations and obtain
\begin{align}
  \ti{c}_8 = & \ d_6+\frac{1}{2}\,\bigl(d_{10}-2\,d_9-d_8+d_7\bigr) \\
  \ti{c}_9 = & \ \frac{1}{2}\,\bigl(d_{10}-d_8+d_7\bigr) \\
  \ti{c}_{12} = & \ \frac{1}{2}\,\bigl(d_8+2\,d_9-d_{10}-d_7\bigr) \\
  \ti{c}_{13} = & \ d_8-d_7 \\
  \ti{c}_{14} = & \ d_9-d_6-\frac{1}{2}\,\bigl(d_{10}-d_8-d_7\bigr)
\end{align}
These equations give, together with (\ref{eq:Zuordnung-TypB}), the desired divergence for $d_QT_1|_{Type B}$.

\noindent 3. Type $C$:
We calculate the expression $\partial_{\mu}\wti{T}_{1/1}^{\mu, C}$ explicitly:
\begin{equation}
  \begin{split}
    \partial_{\mu}\wti{T}_{1/1}^{\mu,C} = & \ d_{11}\, u^{\al}_{\pnu\pmu}h^{\al\mu}_{\pnu}h+
                                              d_{12}\, u^{\al}_{\pnu}h^{\al\mu}_{\pnu\pmu}h+
                                              d_{13}\, u^{\al}_{\pnu}h^{\al\mu}_{\pnu}h^{}_{\pmu} \\
                                          & + d_{14}\, u^{\al}_{\pnu\pmu}h^{\al\mu}h^{}_{\pnu}+
                                              d_{15}\, u^{\al}_{\pnu}h^{\al\mu}_{\pmu}h^{}_{\pnu}+
                                              d_{16}\, u^{\al}_{\pnu}h^{\al\mu}h^{}_{\pmu\pnu} \\
                                          & + d_{17}\, u^{\al}_{\pmu}h^{\al\mu}_{\pnu}h^{}_{\pnu}+
                                              d_{18}\, u^{\al} h^{\al\mu}_{\pmu\pnu}h^{}_{\pnu}+
                                              d_{19}\, u^{\al} h^{\al\mu}_{\pnu}h^{}_{\pmu\pnu} \\
  \end{split}
  \label{eq:TypC-Monome}
\end{equation}
The constants $d_{11},\ldots,d_{19}$ are given by
\begin{equation}
  \begin{split} 
    d_{11} := & \ \ti{c}_{15}+\ti{c}_{21},\quad d_{12}:=\ti{c}_{15}+\ti{c}_{24},\quad d_{13}:=\ti{c}_{15}+
                \ti{c}_{25}+\ti{c}_{26}, \\
    d_{14} := & \ \ti{c}_{16}+\ti{c}_{22},\quad d_{15}:=\ti{c}_{16}+\ti{c}_{23}+\ti{c}_{24},\quad d_{16}:=
                \ti{c}_{16}+\ti{c}_{25} \\
    d_{17} := & \ \ti{c}_{17}+\ti{c}_{21}+\ti{c}_{22},\quad d_{18}:=\ti{c}_{17}+\ti{c}_{23},\quad d_{19}:=
                \ti{c}_{17}+\ti{c}_{26} \\
  \end{split}
  \label{eq:d-TypC}
\end{equation}
From first order gauge invariance we get
\begin{equation}
  \begin{split}
    d_{11} = & -ia_6,\quad d_{12}=0,\quad d_{13}=-\frac{i}{2}\,a_5,\quad d_{14}=-\frac{i}{2}\,a_5, \\
    d_{15} = & -\frac{i}{2}\,a_4,\quad d_{16}=0,\quad d_{17}=0,\quad d_{18}=0,\quad d_{19}=0 \\
  \end{split}
  \label{eq:Zuordnung-TypC}
\end{equation}
The coefficient matrix $M_C$ of (\ref{eq:d-TypC}) is in $GL(9,\mathbb{Z})$. We invert these equations and obtain
\begin{align}
  \ti{c}_{15} = & \ \frac{1}{2}\,\bigl(d_{11}+d_{12}+d_{14}-d_{15}-d_{17}+d_{18}\bigr) \\  
  \ti{c}_{16} = & \ \frac{1}{2}\,\bigl(d_{11}-d_{13}+d_{14}+d_{16}-d_{17}+d_{19}\bigr) \\
  \ti{c}_{17} = & \ \frac{1}{2}\,\bigl(d_{12}-d_{13}-d_{15}+d_{16}+d_{18}+d_{19}\bigr) \\
  \ti{c}_{21} = & \ \frac{1}{2}\,\bigl(d_{11}-d_{12}-d_{14}+d_{15}+d_{17}-d_{18}\bigr) \\
  \ti{c}_{22} = & \ \frac{1}{2}\,\bigl(-d_{11}+d_{13}+d_{14}-d_{16}+d_{17}-d_{19}\bigr) \\
  \ti{c}_{23} = & \ \frac{1}{2}\,\bigl(-d_{12}+d_{13}+d_{15}-d_{16}+d_{18}-d_{19}\bigr) \\
  \ti{c}_{24} = & \ \frac{1}{2}\,\bigl(-d_{11}+d_{12}-d_{14}+d_{15}+d_{17}-d_{18}\bigr) \\
  \ti{c}_{25} = & \ \frac{1}{2}\,\bigl(-d_{11}+d_{13}-d_{14}+d_{16}+d_{17}-d_{19}\bigr) \\
  \ti{c}_{26} = & \ \frac{1}{2}\,\bigl(-d_{12}+d_{13}+d_{15}-d_{16}-d_{18}+d_{19}\bigr)
\end{align}
These equations give, together with (\ref{eq:Zuordnung-TypC}), the desired divergence for $d_QT_1|_{Type C}$.

\noindent 4. Type $D$:
We calculate the expression $\partial_{\mu}\wti{T}_{1/1}^{\mu, D}$ explicitly:
\begin{equation}
  \begin{split}
    \partial_{\mu}\wti{T}_{1/1}^{\mu,D} = & \ d_{20}\, u^{\al}_{\pmu\pnu}h^{\al\si}_{\pnu}h^{\si\mu}+
                                              d_{21}\, u^{\al}_{\pnu}h^{\al\si}_{\pmu\pnu}h^{\si\mu}+
                                              d_{22}\, u^{\al}_{\pnu}h^{\al\si}_{\pnu}h^{\si\mu}_{\pmu} \\
                                          & + d_{23}\, u^{\al}_{\pmu\pnu}h^{\al\si}h^{\si\mu}_{\pnu}+
                                              d_{24}\, u^{\al}_{\pnu}h^{\al\si}_{\pmu}h^{\si\mu}_{\pnu}+
                                              d_{25}\, u^{\al}_{\pnu}h^{\al\si}h^{\si\mu}_{\pmu\pnu} \\
                                          & + d_{26}\, u^{\al}_{\pmu}h^{\al\si}_{\pnu}h^{\si\mu}_{\pnu}+
                                              d_{27}\, u^{\al}h^{\al\si}_{\pmu\pnu}h^{\si\mu}_{\pnu}+
                                              d_{28}\, u^{\al}h^{\al\si}_{\pnu}h^{\si\mu}_{\pmu\pnu} \\
  \end{split}
  \label{eq:TypD-Monome}
\end{equation}
The constants $d_{20},\ldots,d_{28}$ are given by
\begin{equation}
  \begin{split}
    d_{20} := & \ \ti{c}_{27}+\ti{c}_{33},\quad d_{21}:=\ti{c}_{27}+\ti{c}_{36},\quad d_{22}:=\ti{c}_{27}+
                \ti{c}_{37}+\ti{c}_{38} \\
    d_{23} := & \ \ti{c}_{28}+\ti{c}_{34},\quad d_{24}:=\ti{c}_{28}+\ti{c}_{35}+\ti{c}_{36},\quad d_{25}:=
                \ti{c}_{28}+\ti{c}_{37}, \\
    d_{26} := & \ \ti{c}_{29}+\ti{c}_{33}+\ti{c}_{34},\quad d_{27}:=\ti{c}_{29}+\ti{c}_{35},\quad d_{28}:=
                \ti{c}_{29}+\ti{c}_{38} \\
  \end{split}
  \label{eq:d-TypD}
\end{equation}
From first order gauge invariance we obtain
\begin{equation}
  \begin{split}
    d_{20} = & -\frac{i}{2}\,a_9,\quad d_{21}=0,\quad d_{22}=-\frac{i}{2}\,a_{10},\quad d_{23}=-ia_{12} \\
    d_{24} = & -\frac{i}{2}\,a_9,\quad d_{25}=0,\quad d_{26}=0,\quad d_{27}=0,\quad d_{28}=0 \\
  \end{split}
  \label{eq:Zuordnung-TypD}
\end{equation}
The coefficient matrix $M_D$ of (\ref{eq:d-TypD}) is in $GL(9,\mathbb{Z})$. We invert these equations and obtain
\begin{align}
  \ti{c}_{27} = & \ \frac{1}{2}\,\bigl(d_{20}+d_{21}+d_{23}-d_{24}-d_{26}+d_{27}\bigr) \\
  \ti{c}_{28} = & \ \frac{1}{2}\,\bigl(d_{20}-d_{22}+d_{23}+d_{25}-d_{26}+d_{28}\bigr) \\
  \ti{c}_{29} = & \ \frac{1}{2}\,\bigl(d_{21}-d_{22}-d_{24}+d_{25}+d_{27}+d_{28}\bigr) \\
  \ti{c}_{33} = & \ \frac{1}{2}\,\bigl(d_{20}-d_{21}-d_{23}+d_{24}+d_{26}-d_{27}\bigr) \\
  \ti{c}_{34} = & \ \frac{1}{2}\,\bigl(-d_{20}+d_{22}+d_{23}-d_{25}+d_{26}-d_{28}\bigr) \\
  \ti{c}_{35} = & \ \frac{1}{2}\,\bigl(-d_{21}+d_{22}+d_{24}-d_{25}+d_{27}-d_{28}\bigr) \\
  \ti{c}_{36} = & \ \frac{1}{2}\,\bigl(-d_{20}+d_{21}-d_{23}+d_{24}+d_{26}-d_{27}\bigr) \\
  \ti{c}_{37} = & \ \frac{1}{2}\,\bigl(-d_{20}+d_{22}-d_{23}+d_{25}+d_{26}-d_{28}\bigr) \\
  \ti{c}_{38} = & \ \frac{1}{2}\,\bigl(-d_{21}+d_{22}+d_{24}-d_{25}-d_{27}+d_{28}\bigr)
\end{align}                   
These equations give, together with (\ref{eq:Zuordnung-TypD}), the desired divergence for $d_QT_1|_{Type D}$.

\noindent 5. Type $H$:
We calculate the expression $\partial_{\mu}\wti{T}_{1/1}^{\mu, H}$ explicitly:
\begin{equation}
  \begin{split}
    \partial_{\mu}\wti{T}_{1/1}^{\mu,H} = & \ d_{80}\, u^{\mu}_{\pnu\pmu}\ti{u}^{\al}_{\pnu}u^{\al}+
                                              d_{81}\, u^{\mu}_{\pnu}\ti{u}^{\al}_{\pmu\pnu}u^{\al}+
                                              d_{82}\, u^{\mu}_{\pnu}\ti{u}^{\al}_{\pnu}u^{\al}_{\pmu} \\
                                          & + d_{83}\, u^{\mu}_{\pnu\pmu}\ti{u}^{\al}u^{\al}_{\pnu}+
                                              d_{84}\, u^{\mu}_{\pnu}\ti{u}^{\al}_{\pmu}u^{\al}_{\pnu}+
                                              d_{85}\, u^{\mu}_{\pnu}\ti{u}^{\al}u^{\al}_{\pmu\pnu} \\
                                          & + d_{86}\, u^{\mu}_{\pmu}\ti{u}^{\al}_{\pnu}u^{\al}_{\pnu}+
                                              d_{87}\, u^{\mu}\ti{u}^{\al}_{\pnu\pmu}u^{\al}_{\pnu}+
                                              d_{88}\, u^{\mu}\ti{u}^{\al}_{\pnu}u^{\al}_{\pmu\pnu} \\
  \end{split}
  \label{eq:TypH-Monome}
\end{equation}
The constants $d_{80},\ldots,d_{88}$ are given by
\begin{equation}
  \begin{split}
    d_{80} := & \ \ti{c}_{84}+\ti{c}_{87},\quad d_{81}:=\ti{c}_{84}+\ti{c}_{90},\quad d_{82}:=\ti{c}_{84}+
                \ti{c}_{91}+\ti{c}_{92}, \\
    d_{83} := & \ \ti{c}_{85}+\ti{c}_{88},\quad d_{84}:=\ti{c}_{85}+\ti{c}_{89}+\ti{c}_{90},\quad d_{85}:=
                \ti{c}_{85}+
                \ti{c}_{91}, \\
    d_{86} := & \ \ti{c}_{86}+\ti{c}_{87}+\ti{c}_{88},\quad d_{87}:=\ti{c}_{86}+\ti{c}_{89},\quad d_{88}:=
                \ti{c}_{86}+\ti{c}_{92} \\
  \end{split}
  \label{eq:d-TypH}
\end{equation}
From first order gauge invariance we obtain
\begin{equation}
  \begin{split}
    d_{80} = & 0,\quad d_{81}=0,\quad d_{82}=\frac{i}{2}\,b_{19},\quad d_{83}=0,\quad  d_{84}=
               -\frac{i}{2}\,\bigl(b_1-b_{19}\bigr), \\ 
    d_{85} = & -\frac{i}{2}\,b_2,\quad d_{86}=-\frac{i}{2}\,\bigl(b_4+b_{19}\bigr),\quad d_{87}=0,\quad 
               d_{88}=-\frac{i}{2}\,b_3 \\
  \end{split}
  \label{eq:Zuordnung-TypH}
\end{equation}
The coefficient matrix $M_H$ of (\ref{eq:d-TypH}) is in $GL(9,\mathbb{Z})$. We invert these equations and obtain
\begin{align}
  \ti{c}_{84} = & \ \frac{1}{2}\,\bigl(d_{80}+d_{81}+d_{83}-d_{84}-d_{86}+d_{87}\bigr) \\
  \ti{c}_{85} = & \ \frac{1}{2}\,\bigl(d_{80}-d_{82}+d_{83}+d_{85}-d_{86}+d_{88}\bigr) \\
  \ti{c}_{86} = & \ \frac{1}{2}\,\bigl(d_{81}-d_{82}-d_{84}+d_{85}+d_{87}+d_{88}\bigr) \\
  \ti{c}_{87} = & \ \frac{1}{2}\,\bigl(d_{80}-d_{81}-d_{83}+d_{84}+d_{86}-d_{87}\bigr) \\
  \ti{c}_{88} = & \ \frac{1}{2}\,\bigl(-d_{80}+d_{82}+d_{83}-d_{85}+d_{86}-d_{88}\bigr) \\
  \ti{c}_{89} = & \ \frac{1}{2}\,\bigl(-d_{81}+d_{82}+d_{84}-d_{85}+d_{87}-d_{88}\bigr) \\
  \ti{c}_{90} = & \ \frac{1}{2}\,\bigl(-d_{80}+d_{81}-d_{83}+d_{84}+d_{86}-d_{87}\bigr) \\
  \ti{c}_{91} = & \ \frac{1}{2}\,\bigl(-d_{80}+d_{82}-d_{83}+d_{85}+d_{86}-d_{88}\bigr) \\
  \ti{c}_{92} = & \ \frac{1}{2}\,\bigl(-d_{81}+d_{82}+d_{84}-d_{85}-d_{87}+d_{88}\bigr)
\end{align}
These equations give, together with (\ref{eq:Zuordnung-TypH}), the desired divergence for $d_QT_1|_{Type H}$.
 
\noindent 6. Type $K$:
We calculate the expression $\partial_{\mu}\wti{T}_{1/1}^{\mu, K}$ explicitly:
\begin{equation}
  \partial_{\mu}\wti{T}_{1/1}^{\mu,K}=d_{102}\, u^{\si}_{\pmu\pnu}\ti{u}^{\mu}_{\pnu}u^{\si}+
                                      d_{103}\, u^{\si}_{\pnu}\ti{u}^{\mu}_{\pnu\pmu}u^{\si}+
                                      d_{104}\, u^{\si}_{\pmu}\ti{u}^{\mu}_{\pnu}u^{\si}_{\pnu}+
                                      d_{105}\, u^{\si}_{\pal\pmu}\ti{u}^{\al}u^{\si}_{\pmu}
  \label{eq:TypK-Monome}
\end{equation}
The constants $d_{102},\ldots,d_{105}$ are given by
\begin{equation}
  d_{102}:=\ti{c}_{108}+c_{110},\ d_{103}:=\ti{c}_{108}+\ti{c}_{112},\ d_{104}:=-\ti{c}_{108}+c_{110}+
           \ti{c}_{111},\ d_{105}:=\ti{c}_{111}
  \label{eq:d-TypK}
\end{equation}
From first order gauge invariance we obtain
\begin{equation}
  d_{102}=\frac{i}{2}\,b_{18},\quad d_{103}=0,\quad d_{104}=-\frac{i}{2}\,b_{13},\quad d_{105}=\frac{i}{2}\,
  b_{17}
  \label{eq:Zuordnung-TypK}
\end{equation}
The coefficient matrix $M_K$ of (\ref{eq:d-TypK}) is in $GL(4,\mathbb{Z})$. We invert these equations and obtain
\begin{align}
  \ti{c}_{110} = & \ \frac{1}{2}\,\bigl(d_{102}-d_{104}+d_{105}\bigr) \\
  c_{112}      = & \ \frac{1}{2}\,\bigl(d_{102}+d_{104}-d_{105}\bigr) \\
  \ti{c}_{113} = & \ d_{105} \\
  \ti{c}_{114} = & \ d_{103}+\frac{1}{2}\,\bigl(-d_{102}+d_{104}-d_{105}\bigr)
\end{align}
These equations give, together with (\ref{eq:Zuordnung-TypK}), the desired divergence for $d_QT_1|_{Type K}$.

\section{Divergences for Types $E,F,G$ and $J$}
Here we calculate the explicit divergence forms in terms of the coupling parameters $a_1,\ldots,a_{12},b_1,\ldots,b_{21}$ for the types $E,F,G$ and $J$. In contrast to the other types the system of equations between the $c_i$ and $d_j$ are no longer invertible in a unique way. There are some ambiguities, if we express the $c_i$ in terms of the $d_j$. Let us begin with type $E$.

\noindent 1. Type $E$:
Let $M_E\in Mat(18\times 15,\mathbb{Z})$ the coefficient matrix of the system (\ref{eq:d-TypE}). We have to consider the equation 
\begin{equation}
  M_E\cdot \mathbf{c}^E=\mathbf{d}^E
  \label{eq:d-TypEMatrix}
\end{equation}
The general solution of this equation is the sum of an arbitrary solution and the general solution of the corresponding homogeneous equation
\begin{equation}
  M_E\cdot \mathbf{c}^E=0
  \label{eq:homogene TypE-Matrix}
\end{equation}
The matrix $M_E$ has $rank(M_E)=12$. So there are three free parameters $\lambda_1,\lambda_2,\lambda_3\in\mathbb{C}$ in the solution of (\ref{eq:homogene TypE-Matrix}). We obtain
\begin{equation}
  \begin{split}
    \mathbf{c}^E_0(\la_1,\la_2,\la_3) = & \ \Bigl(-\la_1-\la_2+\la_3,\,-\la_1+\la_2-\la_3,\,\la_1
                                             -\la_2-\la_3,\,-\la_1-\la_2+\la_3,\, \\
                                           & -\la_1-\la_2+\la_3,\,-\la_1+\la_2-\la_3,\,-\la_1+\la_2-\la_3,\,\la_1
                                             -\la_2-\la_3,\, \\
                                           & \ \la_1-\la_2-\la_3,\,\la_1+\la_2-\la_3,\,\la_1-\la_2+\la_3,\,-\la_1
                                             +\la_2+\la_3,\,2\,\la_1,\, \\
                                           & \ 2\,\la_2,\,2\,\la_3\Bigr) \\
  \end{split}
  \label{eq:allgemeine homogene TypE-Loesung}
\end{equation}
A special solution of equation~(\ref{eq:d-TypEMatrix}) is given by
\begin{equation}
  \begin{split}
    \mathbf{c}^E_s = & \ \Bigl(\frac{1}{2}\,\bigl(d_{37}+d_{40}-d_{39}\bigr),\,d_{29}-d_{31}-d_{44}+d_{42}
                          -\frac{1}{2}\,\bigl(d_{39}-d_{37}-d_{40}\bigr),\, \\
                        & \ d_{32}+d_{36}-d_{33}-d_{44}-\frac{1}{2}\,\bigl(d_{39}-d_{37}-d_{40}\bigr),\,d_{29}
                          -d_{44}-\frac{1}{2}\,\bigl(d_{39}-d_{37}-d_{40}\bigr),\, \\
                        & \ d_{32}-d_{44}-\frac{1}{2}\,\bigl(d_{39}-d_{37}-d_{40}\bigr),\,-d_{32}+d_{33}+d_{44}
                          +\frac{1}{2}\,\bigl(d_{39}-d_{37}-d_{40}\bigr),\, \\
                        & \ \frac{1}{2}\,\bigl(d_{39}+d_{37}-d_{40}\bigr),\,\frac{1}{2}\,\bigl(d_{40}+d_{39}
                          -d_{37}\bigr),\,d_{31}-d_{29}+d_{44}+\frac{1}{2}\,\bigl(d_{39}-d_{37} \\
                        & -d_{40}\bigr),\,d_{44}+\frac{1}{2}\,\bigl(d_{39}-d_{37}-d_{40}\bigr),\,d_{44}+d_{31}
                          -d_{29}-d_{42}+d_{45}+\frac{1}{2}\,\bigl(d_{39} \\
                        & -d_{37}-d_{40}\bigr),\,-d_{32}-d_{36}+d_{33}+d_{46}+d_{44}+\frac{1}{2}\,\bigl(d_{39}-d_{37}
                          -d_{40}\bigr),\,0,\,0,\,0\Bigr) \\
  \end{split}
  \label{eq:spezielle inhomogene TypE-Loesung}
\end{equation}
The general solution of (\ref{eq:d-TypEMatrix}) is then given by
\begin{equation}
  \mathbf{c}^E=\mathbf{c}^E_s+\mathbf{c}^E_0
  \label{eq:allgemeine TypE-Loesung}
\end{equation}
With the equations (\ref{eq:Zuordnung-TypE}) we can write the expression $d_QT_1|_{Type E}$ as a divergence.
The two parts of the solution to (\ref{eq:d-TypEMatrix}), $\mathbf{c}^E_s$ and $\mathbf{c}^E_0$, correspond to a $Q$-vertex of the form 
\begin{equation} 
  \wti{T}_{1/1}^{\mu, E}=\wti{T}_{1/1}^{\mu, E, s}+\wti{T}_{1/1}^{\mu, E, 0}
  \label{eq:TypE-Q-vertex}
\end{equation}
where the distinguished part $\wti{T}_{1/1}^{\mu, E, s}$ is given by (\ref{eq:TypE}) with coefficients (\ref{eq:spezielle inhomogene TypE-Loesung}). They are uniquely determined by the parameters of the theory, see (\ref{eq:Zuordnung-TypE}). The homogeneous part $\wti{T}_{1/1}^{\mu, E, 0}$ is given by (\ref{eq:TypE}) with coefficients (\ref{eq:allgemeine homogene TypE-Loesung}) and one observes that it can be written in the form
\begin{equation}
  \begin{split}
    \wti{T}_{1/1}^{\mu, E, 0} = & \ \bigl(\la_1+\la_2-\la_3\bigr)\partial_{\ro}T_1^{\mu\ro, E} \\
                                  & +\bigl(\la_1-\la_2+\la_3\bigr)\partial_{\ro}T_2^{\mu\ro, E} \\
                                  & +\bigl(\la_1-\la_2-\la_3\bigr)\partial_{\ro}T_3^{\mu\ro, E} \\
  \end{split}
  \label{eq:homogener TypE-Q-vertex}
\end{equation}
with
\begin{equation}
  \begin{split}
    T_1^{\mu\ro, E} = & \ u^{\al}_{\psig}h^{\al\mu}h^{\ro\si}-u^{\al}_{\psig}h^{\al\ro}h^{\mu\si} \\
    T_2^{\mu\ro, E} = & \ u^{\al}h^{\al\mu}_{\psig}h^{\ro\si}-u^{\al}h^{\al\ro}_{\psig}h^{\mu\si} \\
    T_3^{\mu\ro, E} = & \ u^{\al}h^{\al\ro}h^{\mu\si}_{\psig}-u^{\al}h^{\al\mu}h^{\ro\si}_{\psig} \\
  \end{split}
  \label{eq:antisymmetrische TypE-Anteile}
\end{equation}
they are antisymmetric in their indices, i.e.
\begin{equation}
  T_i^{\mu\ro, E}=-T_i^{\ro\mu, E}\quad i=1,2,3
  \label{eq:Antisymmetrie der TypE-Anteile}
\end{equation}
The homogeneous part $\wti{T}_{1/1}^{\mu, E, 0}$ can be written as the divergence of an antisymmetric tensor which is independent of the parameters of the theory.

\noindent 2. Type $F$:
In analogy to type $E$ we first calculate the expression $\partial_{\mu}\wti{T}_{1/1}^{\mu, F}$. This expression is of the form
\begin{equation}
  \partial_{\mu}\wti{T}_{1/1}^{\mu, F}=\sum_{i=47}^{61}d_i\, \partial_{\mu}\partial_{\ro}\partial_{\si}|
                                       u^{\mu}h^{\ro\nu}h^{\nu\si}
  \label{eq:TypF-Monome}
\end{equation}
Here the three derivatives are distributed among fields in all possible combinations. The new constants $d_i$ are defined as follows
\begin{equation} 
  \begin{split}
    d_{47}:= & \ c_{57}+c_{58}+2\,c_{63},\quad d_{48}:=c_{56}+c_{57}+c_{67},\quad d_{49}:=c_{57}+c_{59}+c_{66}, \\
    d_{50}:= & \ c_{58}+c_{62}+c_{67},\quad d_{51}:=c_{55}+c_{58}+c_{66},\quad d_{52}:=c_{59}+c_{62}+c_{64}, \\
    d_{53}:= & \ c_{55}+c_{59}+2\,c_{65},\quad d_{54}:=c_{54}+c_{66}+c_{60},\quad d_{55}:=c_{60}+c_{61}+c_{64}, \\
    d_{56}:= & \ c_{60}+c_{65},\quad d_{57}:=c_{54}+c_{61}+c_{67},\quad d_{58}:=c_{61}+c_{68}, \\
    d_{59}:= & \ c_{56}+c_{62}+2\,c_{68},\quad d_{60}:=c_{54}+c_{63},\quad d_{61}:=c_{55}+c_{64}+c_{56} \\
  \end{split}
  \label{eq:d-TypF}
\end{equation}
From first order gauge invariance we obtain
\begin{equation}
  \begin{split}
    d_{47} = & -i\Bigl(\frac{1}{2}\,a_9+a_{12}\Bigr),\quad d_{48}=-ib_1,\quad d_{49}=-\frac{i}{2}\,a_9, \\
    d_{50} = & -i\Bigl(\frac{1}{2}\,a_{10}+b_2\Bigr),\quad d_{51}=0,\quad d_{52}=-ib_3,\quad d_{53}=0, \\
    d_{54} = & \ i\Bigl(a_5+\frac{1}{2}\,a_9+a_{12}\Bigr),\quad d_{55}=-ib_4,\quad d_{56}=\frac{i}{2}\,\bigl(2\,a_6
               +a_9+a_{12}\bigr), \\
    d_{57} = & \ \frac{i}{2}\,\bigl(2\,a_4+a_9+a_{10}\bigr),\quad d_{58}=\frac{i}{2}\,\bigl(2\,a_3+a_{10}+a_{11}
               -2\,b_5\bigr), \\
    d_{59} = & -ib_6,\quad d_{60}=0,\quad d_{61}=0 \\
  \end{split}
  \label{eq:Zuordnung-TypF}
\end{equation}
Let $M_F\in Mat(15\times 15,\mathbb{Z})$ the coefficient matrix of (\ref{eq:d-TypF}). Then we determine the general solution of 
\begin{equation}
  M_F\cdot \mathbf{c}^F=\mathbf{d}^F
  \label{eq:d-TypF-MatrixForm}
\end{equation}
where $\mathbf{c}^F\in \mathbb{C}^{15}$ and $\mathbf{d}^F\in \mathbb{C}^{15}$ are the column vectors with components $(c_{54},\ldots,c_{68})$ and $(d_{47},\ldots,d_{61})$ respectively. The matrix $M_F$ has $rank(M_F)=11$. The general solution of the corresponding homogeneous system  
\begin{equation}
  M_F\cdot \mathbf{c}^F=0
  \label{eq:homogene TypF-Matrix}
\end{equation}
is labeled by $4$ independent parameters $\la_1,\ldots,\la_4\in\mathbb{C}$ and is given by
\begin{equation}
  \begin{split}
    \mathbf{c}^F_0(\la_1,\la_2,\la_3,\la_4) = & \ \Bigl(-\la_2+\la_3,\,\la_1+\la_3-\la_4,\,-\la_1-2\,\la_3,\,
                                                   \la_1-\la_2+2\,\la_3,\, \\
                                                 & -\la_1-\la_2,\,-\la_1-\la_3-\la_4,\,-\la_4,\,-\la_3,\,\la_1,\,
                                                   \la_2-\la_3,\,\la_3+\la_4,\, \\
                                                 & \ \la_4,\,\la_2-\la_3+\la_4,\,\la_2,\,\la_3\Bigr) \\
  \end{split}
  \label{eq:allgemeine homogene TypF-Loesung}
\end{equation}
A special solution to (\ref{eq:d-TypF-MatrixForm}) is given by
\begin{equation}
  \begin{split}
    \mathbf{c}^F_s = & \ \Bigl(d_{56}+d_{57}-d_{55}-\frac{1}{2}\,\bigl(d_{53}+d_{59}-d_{52}-d_{61}\bigr),\,
                          \frac{1}{2}\,\bigl(d_{53}-d_{59}-d_{52}+d_{61}\bigr),\, \\
                        & \ d_{59},\,d_{49}+d_{57}-d_{55}-d_{54}-d_{53}-d_{59}+d_{61}+2\,d_{56},\,d_{47}-d_{49}
                          +d_{52} \\
                        & +d_{57}-d_{55}-2\,d_{60}+d_{54},\,\frac{1}{2}\,\bigl(d_{53}+d_{59}+d_{52}-d_{61}\bigr),\,
                          d_{56},\,-d_{56}+d_{55} \\
                        & +\frac{1}{2}\,\bigl(d_{53}+d_{59}-d_{52}-d_{61}\bigr),\,0,\,-d_{56}-d_{57}+d_{55}-d_{60}
                          +\frac{1}{2}\,\bigl(d_{53}+d_{59} \\
                        & -d_{52}-d_{61}\bigr),\,-\frac{1}{2}\,\bigl(d_{53}+d_{59}-d_{52}-d_{61}\bigr),\,0,\,d_{54}
                          -2\,d_{56}-d_{57}+d_{55} \\
                        & +\frac{1}{2}\,\bigl(d_{53}+d_{59}-d_{52}-d_{61}\bigr),\,0,\,0\Bigr) \\ 
  \end{split}
  \label{eq:spezielle inhomogene TypF-Loesung}
\end{equation}
The general solution to (\ref{eq:d-TypF-MatrixForm}) is then given by
\begin{equation}
  \mathbf{c}^F=\mathbf{c}^F_s+\mathbf{c}^F_0
  \label{eq:allgemeine TypF-Loesung}
\end{equation}
With the equations (\ref{eq:Zuordnung-TypF}) we can write the expression $d_QT_1|_{Type F}$ as a divergence.
According to the two parts of the solution to (\ref{eq:d-TypF-MatrixForm}), $\mathbf{c}^F_s$ and $\mathbf{c}^F_0$ we can represent the $Q$-vertex as a sum 
\begin{equation}
  \wti{T}_{1/1}^{\mu, F}=\wti{T}_{1/1}^{\mu, F, s}+\wti{T}_{1/1}^{\mu, F, 0}
  \label{eq:TypF-Q-vertex}
\end{equation}
The distinguished part $\wti{T}_{1/1}^{\mu, F, s}$ is given by (\ref{eq:TypF}) with coefficients (\ref{eq:spezielle inhomogene TypF-Loesung}). They are uniquely determined by the parameters of the theory, see (\ref{eq:Zuordnung-TypF}). The homogeneous part $\wti{T}_{1/1}^{\mu, F, 0}$ is given by (\ref{eq:TypF}) with coefficients (\ref{eq:allgemeine homogene TypF-Loesung}) and it has the form
\begin{equation}
  \begin{split}
    \wti{T}_{1/1}^{\mu, F, 0} = & \ \bigl(\la_1+\la_3\bigr)\partial_{\si}T_1^{\mu\si, F}+\la_4\partial_{\ro}T_2^{\mu\ro, F} \\
                                & +\bigl(\la_2-\la_3\bigr)\partial_{\ro}T_3^{\mu\ro, F}+\la_3\partial_{\ro}T_4^{\mu\ro, F} \\
  \end{split}
  \label{eq:homogener TypF-Q-vertex}
\end{equation}
with
\begin{equation}
  \begin{split}
    T_1^{\mu\si, F} = & \ u^{\ro}h^{\mu\nu}_{\pro}h^{\nu\si}-u^{\ro}h^{\mu\nu}h^{\nu\si}_{\pro} \\
    T_2^{\mu\ro, F} = & \ u^{\mu}h^{\ro\nu}_{\psig}h^{\nu\si}-u^{\ro}h^{\mu\nu}_{\psig}h^{\nu\si} \\
    T_3^{\mu\ro, F} = & \ u^{\mu}_{\psig}h^{\si\nu}h^{\nu\ro}-u^{\ro}_{\psig}h^{\mu\nu}h^{\nu\si} \\
    T_4^{\mu\ro, F} = & \ u^{\mu}h^{\ro\nu}h^{\nu\si}_{\psig}-u^{\ro}h^{\mu\nu}h^{\nu\si}_{\psig} \\
  \end{split}  
\label{eq:antisymmetrische TypF-Anteile}
\end{equation}
They are antisymmetric in their indices, i.e
\begin{equation}
  T_i^{\mu\si, F}=-T_i^{\si\mu, F}, \quad i=1,\ldots,4
 \label{eq:Antisymmetrie der TypF-Anteile}
\end{equation}
The homogeneous part $\wti{T}_{1/1}^{\mu, F, 0}$ can be written as the divergence of an antisymmetric tensor which is independent of the parameters of the theory. 

\noindent 3. Type $G$:
We calculate $\partial_{\mu}\wti{T}_{1/1}^{\mu, G}$. This has the form
\begin{equation}
  \partial_{\mu}\wti{T}_{1/1}^{\mu, G}=\sum_{i=62}^{79}d_i\, \partial_{\mu}\partial_{\ro}\partial_{\si}|u^{\mu}h^{\ro\si}h
  \label{eq:TypG-Monome}
\end{equation}
The new constants $d_i$ are defined by
\begin{equation}
  \begin{split}
    d_{62}:= & \ c_{72}+c_{78},\quad d_{63}:=c_{70}+c_{72}+c_{81},\quad d_{64}:=c_{72}+c_{77}+c_{82}, \\
    d_{65}:= & \ c_{73}+c_{78},\quad d_{66}:=c_{73}+c_{74}+c_{81},\quad d_{67}:=c_{71}+c_{73}+c_{82}, \\
    d_{68}:= & \ c_{74}+c_{79},\quad d_{69}:=c_{71}+c_{74}+c_{83},\quad d_{70}:=c_{69}+c_{75}+c_{81}, \\
    d_{71}:= & \ c_{75}+c_{79},\quad d_{72}:=c_{75}+c_{76}+c_{83},\quad d_{73}:=c_{69}+c_{76}+c_{82}, \\
    d_{74}:= & \ c_{76}+c_{80},\quad d_{75}:=c_{70}+c_{77}+c_{83},\quad d_{76}:=c_{77}+c_{80}, \\
    d_{77}:= & \ c_{71}+c_{80},\quad d_{78}:=c_{70}+c_{79},\quad d_{79}:=c_{69}+c_{78} \\
  \end{split}
  \label{eq:d-TypG}
\end{equation}
From first order gauge invariance we obtain
\begin{equation}
  \begin{split}
    d_{62} = & -ia_6,\quad d_{63}=-ib_7,\quad d_{64}=-\frac{i}{2}\,a_5,\quad d_{65}=-\frac{i}{2}\,a_5, \\
    d_{66} = & -i\Bigl(\frac{1}{2}\,a_4+b_8\Bigr),\quad d_{67}=0,\quad d_{68}=-ib_9,\quad d_{69}=0,\\
    d_{70} = & \ i\Bigl(a_1+a_6+\frac{1}{2}\,a_7\Bigr),\quad d_{71}=-ib_{10},\quad \\
    d_{72} = & \ i\bigl(a_1-b_{11}\bigr)+\frac{i}{2}\,\bigl(a_4+a_5+a_7\bigr),\quad d_{73}=i\Bigl(2\,a_2
               +\frac{1}{2}\,a_5+a_8\Bigr), \\
    d_{74} = & \ 0,\quad d_{75}=-ib_{12},\quad d_{76}=0,\quad d_{77}=0,\quad d_{78}=0,\quad d_{79}=0 \\
  \end{split}
  \label{eq:Zuordnung-TypG}
\end{equation}
Let $M_G\in Mat(18\times 15,\mathbb{Z})$ be the coefficient matrix of (\ref{eq:d-TypG}). We determine the general solution of 
\begin{equation}
  M_G\cdot \mathbf{c}^G=\mathbf{d}^G
  \label{eq:d-TypG-MatrixForm}
\end{equation}
where $\mathbf{c}^G\in \mathbb{C}^{15}$ and $\mathbf{d}^G\in \mathbb{C}^{18}$ are the column vectors with components $(c_{69},\ldots,c_{83})$ and $(d_{62},\ldots,d_{79})$ respectively. The matrix $M_G$ has $rank(M_G)=12$. The general solution of the corresponding homogeneous system
\begin{equation}
  M_G\cdot \mathbf{c}^G=0
  \label{eq:homogene TypG-Matrix}
\end{equation}
is labeled by three independent parameters $\la_1,\la_2,\la_3\in\mathbb{C}$ and is given by
\begin{equation}
  \begin{split} 
    \mathbf{c}^G_0(\la_1,\la_2,\la_3) = & \ \Bigl(\la_3-\la_1-\la_2,\,\la_2-\la_1-\la_3,\,\la_1-\la_2
                                             -\la_3,\,\la_3-\la_1-\la_2,\,\la_3 \\
                                           & -\la_1-\la_2,\,\la_2-\la_1-\la_3,\,\la_2-\la_1-\la_3,\,\la_1-\la_2
                                             -\la_3,\,\la_1-\la_2 \\
                                           & -\la_3,\,\la_1+\la_2-\la_3,\,\la_1-\la_2+\la_3,\,\la_2+\la_3-\la_1,\,
                                             2\,\la_1,\,2\,\la_2,\,2\,\la_3\Bigr) \\
  \end{split}
  \label{eq:allgemeine homogene TypG-Loesung}
\end{equation}
A special solution to (\ref{eq:d-TypG-MatrixForm}) is given by
\begin{equation}
  \begin{split}
    \mathbf{c}^G_s = & \ \Bigl(\frac{1}{2}\,\bigl(d_{73}-d_{72}+d_{70}\bigr),\,\frac{1}{2}\,\bigl(d_{66}
                          +d_{67}-d_{69}\bigr)-d_{65}-d_{64}+d_{62}+d_{75},\, \\
                        & \ \frac{1}{2}\,\bigl(d_{67}-d_{66}+d_{69}\bigr),\,d_{62}-d_{65}-\frac{1}{2}\,\bigl(d_{69}
                          -d_{66}-d_{67}\bigr),\,\frac{1}{2}\,\bigl(d_{66}+d_{67}-d_{69}\bigr),\, \\
                        & \ \frac{1}{2}\,\bigl(d_{69}+d_{66}-d_{67}\bigr),\,\frac{1}{2}\,\bigl(d_{72}-d_{73}
                          +d_{70}\bigr),\,\frac{1}{2}\,\bigl(d_{72}+d_{73}-d_{70}\bigr),\,d_{65}+d_{64} \\
                        & -d_{62}+\frac{1}{2}\,\bigl(d_{69}-d_{66}-d_{67}\bigr),\,d_{65}+\frac{1}{2}\,\bigl(d_{69}
                          -d_{66}-d_{67}\bigr),\,d_{65}+d_{64}-d_{62} \\
                        & -d_{75}+d_{78}+\frac{1}{2}\,\bigl(d_{69}-d_{67}-d_{66}\bigr),\,d_{74}+\frac{1}{2}\,\bigl(
                          d_{66}-d_{67}-d_{69}\bigr),\,0,\,0,\,0\Bigr) \\
  \end{split}
  \label{eq:spezielle inhomogene TypG-Loesung}
\end{equation}
The general solution to (\ref{eq:d-TypG-MatrixForm}) then reads
\begin{equation}
  \mathbf{c}^G=\mathbf{c}^G_S+\mathbf{c}^G_0
  \label{eq:allgemeine TypG-Loesung}
\end{equation}
With the equations (\ref{eq:Zuordnung-TypG}) we can write the expression $d_QT_1|_{Type G}$ as a divergence.
According to the two parts of the solution to (\ref{eq:d-TypG-MatrixForm}), $\mathbf{c}^G_s$ and $\mathbf{c}^G_0$, we can represent the $Q$-vertex as a sum
\begin{equation}
  \wti{T}_{1/1}^{\mu, G}=\wti{T}_{1/1}^{\mu, G, s}+\wti{T}_{1/1}^{\mu, G, 0}
  \label{eq:TypG-Q-vertex}
\end{equation}
where the distinguished part $\wti{T}_{1/1}^{\mu, G, s}$ is given by (\ref{eq:TypG}) with coefficients (\ref{eq:spezielle inhomogene TypG-Loesung}). They are uniquely determined by the parameters of the theory, see (\ref{eq:Zuordnung-TypG}).The homogeneous part $\wti{T}_{1/1}^{\mu, G, 0}$ is given by (\ref{eq:TypG}) with coefficients (\ref{eq:allgemeine homogene TypG-Loesung}) and one observes that it can be written in the form
\begin{equation}
  \begin{split}
    \wti{T}_{1/1}^{\mu, G, 0} = & \ \bigl(\la_1+\la_2-\la_3\bigr)\partial_{\ro}T_1^{\mu\ro, G} \\
                                & +\bigl(\la_1-\la_2+\la_3\bigr)\partial_{\ro}T_2^{\mu\ro, G} \\
                                & +\bigl(\la_1-\la_2-\la_3\bigr)\partial_{\ro}T_3^{\mu\ro, G} \\
  \end{split}
  \label{eq:homogener TypG-Q-vertex}
\end{equation}
with
\begin{equation}
  \begin{split}
    T_1^{\mu\ro, G} = & \ u^{\mu}_{\psig}h^{\ro\si}h-u^{\ro}_{\psig}h^{\mu\si}h \\
    T_2^{\mu\ro, G} = & \ u^{\mu}h^{\ro\si}_{\psig}h-u^{\ro}h^{\mu\si}_{\psig}h \\
    T_3^{\mu\ro, G} = & \ u^{\ro}h^{\mu\si}h^{}_{\psig}-u^{\mu}h^{\ro\si}h^{}_{\psig} \\
  \end{split}
  \label{eq:antisymmetrische TypG-Anteile}
\end{equation}
They are antisymmetric in their indices, i.e.
\begin{equation}
  T_i^{\mu\ro, G}=-T_i^{\ro\mu, G}, \quad i=1,2,3
  \label{eq:Antisymmetrie der TypG-Anteile}
\end{equation}
The homogeneous part $\wti{T}_{1/1}^{\mu, G, 0}$ can be written as a divergence of an antisymmetric tensor which is independent of the parameters of the theory.

\noindent 4. Type $J$:
We calculate the expression $\partial_{\mu}\wti{T}_{1/1}^{\mu, J}$. This has the form
\begin{equation}
  \partial_{\mu}\wti{T}_{1/1}^{\mu, J}=\sum_{i=89}^{101}d_i\, \partial_{\mu}\partial_{\al}\partial_{\ro}|
                                       u^{\mu}\ti{u}^{\al}u^{\ro}
  \label{eq:TypJ-Monome}
\end{equation}
The new constants $d_i$ are defined by
\begin{equation}
  \begin{split}
    d_{89}:= & -c_{96}+c_{101}-c_{102},\quad d_{90}:=-c_{96}+c_{104}-c_{100}+c_{108}, \\
    d_{91}:= & \ c_{96}-c_{103}+c_{107},\quad d_{92}=c_{103}-c_{101},\quad d_{93}:=-c_{97}+c_{107}-c_{99}, \\
    d_{94}:= & \ c_{97}-c_{100}+c_{105}+c_{108},\quad d_{95}:=c_{97}-c_{102}+c_{106}, \\
    d_{96}:= & \ c_{107}-c_{102},\quad d_{97}:=c_{103}-c_{98}-c_{106},\quad d_{98}:=c_{98}-c_{104}+c_{105}, \\
    d_{99}:= & \ c_{106}-c_{99},\quad d_{100}:=c_{98}+c_{99}-c_{101},\quad d_{101}:=c_{100}+c_{108} \\
  \end{split}
  \label{eq:d-TypJ}
\end{equation}
From first order gauge invariance we obtain
\begin{equation}
  \begin{split}
    d_{89} = & -\frac{i}{2}\,\bigl(b_3+b_{18}\bigr),\quad d_{90}=0,\quad d_{91}=-\frac{i}{2}\,\bigl(b_1-b_{13}\bigr), \\
    d_{92} = & \ \frac{i}{2}\,\bigl(b_2+b_{17}\bigr),\quad d_{93}=\frac{i}{2}\,b_{18}+ib_{12},\quad d_{94}=0, \\
    d_{95} = & -\frac{i}{2}\,\bigl(b_1+b_4+2\,b_7+b_{13}\bigr),\quad d_{96}=0,\quad d_{97}=\frac{i}{2}\,b_2+ib_8, \\
    d_{98} = & -\frac{i}{2}\,b_3-ib_9,\quad d_{99}=\frac{i}{2}\,b_{17}+ib_{11},\quad d_{100}=0,\quad d_{101}=0 \\
  \end{split}
  \label{eq:Zuordnung-TypJ}
\end{equation}    
Let $M_J\in Mat(13\times 12,\mathbb{Z})$ be the coefficient matrix of (\ref{eq:d-TypJ}). Then we determine the general solution of
\begin{equation}
  M_J\cdot \mathbf{c}^J=\mathbf{d}^J
  \label{eq:d-TypJ-MatrixForm}
\end{equation}
where $\mathbf{c}^J\in \mathbb{C}^{12}$ and $\mathbf{d}^J\in \mathbb{C}^{13}$ are the column vectors with components $(c_{96},\ldots,c_{107})$ and $(d_{89},\ldots,d_{101})$ respectively. The matrix $M_J$ has $rank(M_J)=9$. The general solution of the corresponding homogeneous system
\begin{equation}
  M_J\cdot \mathbf{c}^J=0
  \label{eq:homogene TypJ-Matrix}
\end{equation}
is labeled by three independent parameters $\la_1,\ldots,\la_3\in\mathbb{C}$ and is given by
\begin{equation}
  \begin{split}
     \mathbf{c}^J_0(\la_1,\ldots,\la_3) = & \ \Bigl(\la_1,\,\la_3-\la_2,\,\la_1-\la_2+\la_3,\,
                                              \la_2,\,0,\,\\
                                          & \ \la_1+\la_3,\,\la_3,\,\la_1+\la_3,\,\la_1,\,
                                              \la_2-\la_3,\\
                                          & \ \la_2,\,\la_3\Bigr) \\       
  \end{split}
  \label{eq:allgemeine homogene TypJ-Loesung}
\end{equation}
A special solution to (\ref{eq:d-TypJ-MatrixForm}) is given by
\begin{equation}
  \begin{split}
     \mathbf{c}^J_s = & \ \Bigl(d_{101}-d_{90}-2\,(d_{98}-d_{100}-d_{99}-d_{89}),\,-d_{91}-d_{97}-d_{100}-d_{99}
                          -d_{89}+d_{95},\, \\
                      & -d_{101}-d_{90}-d_{91}-d_{97},\,-2\,d_{98}+2\,d_{100}+d_{99},\,d_{101},\, \\
                      & -d_{101}-d_{90}-d_{91}-d_{97}-2\,d_{98}+d_{100}+d_{99},\,-d_{91}-d_{97}-d_{100}-d_{99}
                        -d_{89},\, \\
                      & -d_{101}-d_{90}-d_{91},\,0,\,d_{101}+d_{90}+d_{91}+d_{97}+d_{98},\,0,\,0\Bigr) \\
  \end{split}
  \label{eq:spezielle inhomogene TypJ-Loesung}
\end{equation}
The general solution to (\ref{eq:d-TypJ-MatrixForm}) is then given by
\begin{equation}
  \mathbf{c}^J=\mathbf{c}^J_s+\mathbf{c}^J_0
  \label{eq:allgemeine TypJ-Loesung}
\end{equation}
With the equations (\ref{eq:Zuordnung-TypJ}) we can write the expression $d_QT_1|_{Type J}$ as a divergence.
According to the two parts of the solution to (\ref{eq:d-TypJ-MatrixForm}), $\mathbf{c}^J_s$ and $\mathbf{c}^J_0$, we can represent the $Q$-vertex as a sum
\begin{equation}
  \wti{T}_{1/1}^{\mu, J}=\wti{T}_{1/1}^{\mu, J, s}+\wti{T}_{1/1}^{\mu, J, 0}
  \label{eq:TypJ-Q-vertex}
\end{equation}
where the distinguished part $\wti{T}_{1/1}^{\mu, J, s}$ is given by (\ref{eq:TypJ}) with coefficients (\ref{eq:spezielle inhomogene TypJ-Loesung}). They are uniquely determined by the parameters of the theory, see (\ref{eq:Zuordnung-TypJ}). The homogeneous part $\wti{T}_{1/1}^{\mu, J, 0}$ is given by (\ref{eq:TypJ}) with coefficients (\ref{eq:allgemeine homogene TypJ-Loesung}) and one observes that it can be written in the form
\begin{equation}
    \wti{T}_{1/1}^{\mu, J, 0}=\bigl(\la_2-\la_3\bigr)\partial_{\ro}T_1^{\mu\ro, J}+\la_1\partial_{\al}T_2^{\mu\al, J}
                              +\la_3\partial_{\ro}T_3^{\mu\ro, J}
  \label{eq:homogener TypJ-Q-vertex}
\end{equation}
with
\begin{equation}
  \begin{split}
    T_1^{\mu\ro, J} = & \ u^{\al}_{\pal}\ti{u}^{\ro}u^{\mu}-u^{\al}_{\pal}\ti{u}^{\mu}u^{\ro} \\
    T_2^{\mu\al, J} = & \ u^{\al}_{\pro}\ti{u}^{\mu}u^{\ro}-u^{\mu}_{\pro}\ti{u}^{\al}u^{\ro} \\
    T_3^{\mu\ro, J} = & \ u^{\ro}_{\pal}\ti{u}^{\al}u^{\mu}-u^{\mu}_{\pal}\ti{u}^{\al}u^{\ro} \\
  \end{split}
  \label{eq:antisymmetrische TypJ-Anteile}
\end{equation}
They are antisymmetric in their indices, i.e.
\begin{equation}
  T_i^{\mu\ro, J}=-T_i^{\ro\mu, J}, \quad i=1,2,3
  \label{eq:Antisymmetrie der TypJ-Anteile}
\end{equation}
The homogeneous part $\wti{T}_{1/1}^{\mu, J, 0}$ can be written as a divergence of an antisymmetric tensor which is independent of the parameters of the theory.

\newpage
\addcontentsline{toc}{section}{\protect\numberline{}{References}}
\input{diss.bbl}
\end{document}

%% file: faust.tex
\begin{verse}
  {\it Geschrieben steht: \glqq im Anfang war das Wort!\grqq \\
       Hier stock' ich schon! Wer hilft mir weiter fort? \\
       Ich kann das Wort so hoch unm"oglich sch"atzen, \\
       Ich mu\ss\, es anders "ubersetzen, \\
       Wenn ich vom Geiste recht erleuchtet bin. \\
       Geschrieben steht: im Anfang war der Sinn. \\
       Bedenke wohl die erste Zeile, \\
       Da\ss\, deine Feder sich nicht "ubereile! \\
       Ist es der Sinn, der alles wirkt und schafft? \\
       Es sollte stehn: im Anfang war die Kraft! \\
       Doch, auch indem ich dieses niederschreibe, \\
       Schon warnt mich was, da\ss\, ich dabei nicht bleibe. \\
       Mir hilft der Geist! Auf einmal seh' ich Rat \\
       Und schreibe getrost: im Anfang war die Tat!}

\end{verse}
Goethe, {\it Faust I}

%% file: divclass.tex

%
%

%

\section{Classification of Divergences}
In the previous section we have defined the trilinear coupling of the graviton field $T_1^h$ as well as the coupling to ghost and anti-ghost fields $T_1^u$. In this section we try to write the gauge variation $d_QT_1$ as a divergence $\partial_{\mu}T_{1/1}^{\mu}$. We proceed in a systematic way: Because of the great variety of different terms in $d_QT_1$ it is most convenient to use a separate ansatz for $T_{1/1}^{\mu}$. Since the operator $d_Q$ applied to our $T_1$ increases the ghost number of the result by one we have to make an ansatz with $n_g(T_{1/1}^{\mu})=1$. Furthermore the application of $d_Q$ increases the number of partial derivatives by one. Due to this, every term in $T_{1/1}^{\mu}$ must have two derivatives. The terms appearing in this ansatz can be classified according to their field content. In the so called graviton sector we have $T_{1/1}^{\mu}\sim uhh$ and the ghost sector contains $T_{1/1}^{\mu}\sim u\ti{u}u$. Inside each sector there is a further classification w.r.t. the tensor indices: There are seven different Lorentz types in the graviton sector, namely
\begin{align*}
  \text{type}\, A: & \ u^{\al}h^{\ro\si}h^{\ro\si} \\
  \text{type}\, B: & \ u^{\al}hh \\
  \text{type}\, C: & \ u^{\al}h^{\al\mu}h \\
  \text{type}\, D: & \ u^{\al}h^{\al\si}h^{\si\ro} \\
  \text{type}\, E: & \ u^{\al}h^{\al\nu}h^{\ro\si} \\
  \text{type}\, F: & \ u^{\nu}h^{\ro\mu}h^{\mu\si} \\
  \text{type}\, G: & \ u^{\nu}h^{\ro\si}h. 
\end{align*}
In the ghost sector we've three different Lorentz types, namely
\begin{align*}
  \text{type}\, H: & \ u^{\mu}\ti{u}^{\al}u^{\al} \\
  \text{type}\, J: & \ u^{\al}\ti{u}^{\mu}u^{\ro} \\
  \text{type}\, K: & \ u^{\si}\ti{u}^{\mu}u^{\si}. \\
\end{align*}
In the following subsections we explain in detail the way in which the divergences for these different Lorentz types can be found.

\subsection{Graviton Sector} 
Divergences of type $A$ have the structure $\partial_{\al}\partial_{\nu}\partial_{\nu}|u^{\al}h^{\ro\si}h^{\ro\si}$, where we have to distribute the three derivatives in all possible ways among the fields. Taking into account that all fields obey the wave equation we find the so called  basis elements from which all divergences of type $A$ can be constructed:
\begin{align*}
  e^A_1 & = u^{\al}_{\pal\pnu}h^{\ro\si}_{\pnu}h^{\ro\si} \\
  e^A_2 & = u^{\al}_{\pal}h^{\ro\si}_{\pnu}h^{\ro\si}_{\pnu} \\
  e^A_3 & = u^{\al}_{\pnu}h^{\ro\si}_{\pal\pnu}h^{\ro\si} \\
  e^A_4 & = u^{\al}_{\pnu}h^{\ro\si}_{\pal}h^{\ro\si}_{\pnu} \\
  e^A_5 & = u^{\al}h^{\ro\si}_{\pal\pnu}h^{\ro\si}_{\pnu}. \\
\end{align*}
For the construction of the divergences we have two different partial derivatives with indices $\al$ and $\nu$ which can be taken out, i.e. we can form
\begin{align*}
  div^A_1 & = \partial_{\al}\bigl(\partial_{\nu}\partial_{\nu}|u^{\al}h^{\ro\si}h^{\ro\si}\bigr) \\
  div^A_2 & = \partial_{\nu}\bigl(\partial_{\al}\partial_{\nu}|u^{\al}h^{\ro\si}h^{\ro\si}\bigr) \\
\end{align*}
where the remaining derivatives inside the bracket must be distributed in all possible ways among the fields. Then we find the following divergences of the form $div^A_1$
\begin{align*}
  \partial_{\al}\bigl(u^{\al}_{\pnu}h^{\ro\si}_{\pnu}h^{\ro\si}\bigr) & = e^A_1+e^A_3+e^A_4 \\
  \partial_{\al}\bigl(u^{\al}h^{\ro\si}_{\pnu}h^{\ro\si}_{\pnu}\bigr) & = e^A_2+2e^A_5. \\
\end{align*}
Divergences of the form $div^A_2$ are
\begin{align*}
  \partial_{\nu}\bigl(u^{\al}_{\pal\pnu}h^{\ro\si}h^{\ro\si}\bigr) & = 2e^A_1 \\
  \partial_{\nu}\bigl(u^{\al}_{\pal}h^{\ro\si}_{\pnu}h^{\ro\si}\bigr) & = e^A_1+e^A_2 \\
  \partial_{\nu}\bigl(u^{\al}_{\pnu}h^{\ro\si}_{\pal}h^{\ro\si}\bigr) & = e^A_3+e^A_4 \\
  \partial_{\nu}\bigl(u^{\al}h^{\ro\si}_{\pal\pnu}h^{\ro\si}\bigr) & = e^A_3+e^A_5 \\
  \partial_{\nu}\bigl(u^{\al}h^{\ro\si}_{\pal}h^{\ro\si}_{\pnu}\bigr) & = e^A_4+e^A_5.
\end{align*}
In this way we've found the set of different divergences of Lorentz type $A$. Collecting all terms in the brackets we can make the following ansatz for $T_{1/1}^{\mu,A}$: 
\begin{equation}
   \begin{split}
     T_{1/1}^{\mu,A} = & \ c_1\, u^{\mu}_{\pal} h^{\ro\si}_{\pal} h^{\ro\si}+
                         c_2\, u^{\mu} h^{\ro\si}_{\pal} h^{\ro\si}_{\pal}+
                         c_3\, u^{\al}_{\pal\pmu} h^{\ro\si} h^{\ro\si}+
                         c_4\, u^{\al} h^{\ro\si}_{\pal\pmu} h^{\ro\si} \\
                       & +c_5\, u^{\al}_{\pal} h^{\ro\si}_{\pmu} h^{\ro\si}+
                         c_6\, u^{\al} h^{\ro\si}_{\pal} h^{\ro\si}_{\pmu}+
                         c_7\, u^{\al}_{\pmu} h^{\ro\si}_{\pal} h^{\ro\si} \\
   \end{split}
  \label{eq:TypA}
\end{equation}
where the constants $c_1,\ldots,c_7$ are for the moment free constants to be determined by gauge invariance.

Divergences of type $B$ have the structure $\partial_{\al}\partial_{\nu}\partial_{\nu}|u^{\al}hh$. The basis elements of this type are
\begin{align*}
  e^B_1 & = u^{\al}_{\pal\pnu}h^{}_{\pnu}h \\
  e^B_2 & = u^{\al}_{\pal}h^{}_{\pnu}h^{}_{\pnu} \\
  e^B_3 & = u^{\al}_{\pnu}h^{}_{\pal\pnu}h \\
  e^B_4 & = u^{\al}_{\pnu}h^{}_{\pal}h^{}_{\pnu} \\
  e^B_5 & = u^{\al}h^{}_{\pal\pnu}h^{}_{\pnu}.
\end{align*}
Again we can form two different types of divergences corresponding to the two partial derivatives with indices $\al$ and $\nu$.
\begin{align*}
  div^B_1 & = \partial_{\al}\bigl(\partial_{\nu}\partial_{\nu}|u^{\al}hh\bigr) \\
  div^B_2 & = \partial_{\nu}\bigl(\partial_{\al}\partial_{\nu}|u^{\al}hh\bigr).
\end{align*}
We find the following divergences of the form $div^B_1$
\begin{align*}
  \partial_{\al}\bigl(u^{\al}_{\pnu}h^{}_{\pnu}h\bigr) & = e^B_1+e^B_3+e^B_4 \\
  \partial_{\al}\bigl(u^{\al}h^{}_{\pnu}h^{}_{\pnu}\bigr) & = e^B_2+2e^B_5.
\end{align*}
For divergences of the form $div^B_2$ we find
\begin{align*}
  \partial_{\nu}\bigl(u^{\al}_{\pal\pnu}hh\bigr) & = 2e^B_1 \\
  \partial_{\nu}\bigl(u^{\al}_{\pal}h^{}_{\pnu}h\bigr) & = e^B_1+e^B_2 \\
  \partial_{\nu}\bigl(u^{\al}_{\pnu}h^{}_{\pal}h\bigr) & = e^B_3+e^B_4 \\
  \partial_{\nu}\bigl(u^{\al}h^{}_{\pal\pnu}h\bigr) & = e^B_3+e^B_5 \\
  \partial_{\nu}\bigl(u^{\al}h^{}_{\pal}h^{}_{\pnu}\bigr) & = e^B_4+e^B_5.
\end{align*}
For $T_{1/1}^{\mu,B}$ we can therefore make the ansatz
\begin{equation}
   \begin{split}
     T_{1/1}^{\mu,B} = & \ c_8\, u^{\mu}_{\pal} h^{}_{\pal} h+
                           c_9\, u^{\mu} h^{}_{\pal} h^{}_{\pal}+
                           c_{10}\, u^{\al}_{\pal\pmu} h h+ 
                           c_{11}\, u^{\al} h^{}_{\pal\pmu} h \\ 
                       & + c_{12}\, u^{\al}_{\pal} h^{}_{\pmu} h+ 
                           c_{13}\, u^{\al} h^{}_{\pal} h^{}_{\pmu}+
                           c_{14}\, u^{\al}_{\pmu} h_{\pal} h. \\
   \end{split}
  \label{eq:TypB}
\end{equation}

Divergences of type $C$ have the structure $\partial_{\mu}\partial_{\nu}\partial_{\nu}|u^{\al}h^{\al\mu}h$. The basis elements of this type are
\begin{align*}
  e^C_1 & = u^{\al}_{\pmu\pnu}h^{\al\mu}_{\pnu}h \\
  e^C_2 & = u^{\al}_{\pmu\pnu}h^{\al\mu}h^{}_{\pnu} \\
  e^C_3 & = u^{\al}_{\pmu}h^{\al\mu}_{\pnu}h^{}_{\pnu} \\
  e^C_4 & = u^{\al}_{\pnu}h^{\al\mu}_{\pmu\pnu}h \\
  e^C_5 & = u^{\al}_{\pnu}h^{\al\mu}h^{}_{\pmu\pnu} \\
  e^C_6 & = u^{\al}_{\pnu}h^{\al\mu}_{\pmu}h^{}_{\pnu} \\
  e^C_7 & = u^{\al}_{\pnu}h^{\al\mu}_{\pnu}h^{}_{\pmu} \\
  e^C_8 & = u^{\al}h^{\al\mu}_{\pmu\pnu}h^{}_{\pnu} \\
  e^C_9 & = u^{\al}h^{\al\mu}_{\pnu}h^{}_{\pmu\pnu}.
\end{align*}
We find two types of divergences
\begin{align*}
  div^C_1 & = \partial_{\mu}\bigl(\partial_{\nu}\partial_{\nu}|u^{\al}h^{\al\mu}h\bigr) \\
  div^C_2 & = \partial_{\nu}\bigl(\partial_{\mu}\partial_{\nu}|u^{\al}h^{\al\mu}h\bigr). \\
\end{align*}
We find the following divergences of the form $div^C_1$
\begin{align*}
  \partial_{\mu}\bigl(u^{\al}_{\pnu}h^{\al\mu}_{\pnu}h\bigr) & = e^C_1+e^C_4+e^C_7 \\
  \partial_{\mu}\bigl(u^{\al}_{\pnu}h^{\al\mu}h^{}_{\pnu}\bigr) & = e^C_2+e^C_6+e^C_5 \\
  \partial_{\mu}\bigl(u^{\al}h^{\al\mu}_{\pnu}h^{}_{\pnu}\bigr) & = e^C_3+e^C_8+e^C_9.
\end{align*}
For divergences of the form $div^C_2$ we find
\begin{align*}
  \partial_{\nu}\bigl(u^{\al}_{\pmu\pnu}h^{\al\mu}h\bigr) & = e^C_1+e^C_2 \\
  \partial_{\nu}\bigl(u^{\al}_{\pmu}h^{\al\mu}_{\pnu}h\bigr) & = e^C_1+e^C_3 \\
  \partial_{\nu}\bigl(u^{\al}_{\pmu}h^{\al\mu}h^{}_{\pnu}\bigr) & = e^C_2+e^C_3 \\
  \partial_{\nu}\bigl(u^{\al}_{\pnu}h^{\al\mu}_{\pmu}h\bigr) & = e^C_4+e^C_6 \\
  \partial_{\nu}\bigl(u^{\al}_{\pnu}h^{\al\mu}h^{}_{\pmu}\bigr) & = e^C_5+e^C_7 \\
  \partial_{\nu}\bigl(u^{\al}h^{\al\mu}_{\pmu\pnu}h\bigr) & = e^C_4+e^C_8 \\
  \partial_{\nu}\bigl(u^{\al}h^{\al\mu}_{\pmu}h^{}_{\pnu}\bigr) & = e^C_6+e^C_8 \\
  \partial_{\nu}\bigl(u^{\al}h^{\al\mu}_{\pnu}h^{}_{\pmu}\bigr) & = e^C_7+e^C_9 \\
  \partial_{\nu}\bigl(u^{\al}h^{\al\mu}h^{}_{\pmu\pnu}\bigr) & = e^C_5+e^C_9.
\end{align*}
For $T_{1/1}^{\mu,C}$ we can make the ansatz
\begin{equation}
   \begin{split}
     T_{1/1}^{\mu,C} = & \ c_{15}\, u^{\al}_{\pnu} h^{\al\mu}_{\pnu} h+
                           c_{16}\, u^{\al}_{\pnu} h^{\al\mu} h^{}_{\pnu}+
                           c_{17}\, u^{\al} h^{\al\mu}_{\pnu} h^{}_{\pnu}+
                           c_{18}\, u^{\al}_{\pnu\pmu} h^{\al\nu} h \\
                       & + c_{19}\, u^{\al} h^{\al\nu}_{\pnu\pmu} h+
                           c_{20}\, u^{\al} h^{\al\nu} h^{}_{\pnu\pmu}+ 
                           c_{21}\, u^{\al}_{\pnu} h^{\al\nu}_{\pmu} h+
                           c_{22}\, u^{\al}_{\pnu} h^{\al\nu} h^{}_{\pmu} \\
                       & + c_{23}\, u^{\al} h^{\al\nu}_{\pnu} h^{}_{\pmu}+ 
                           c_{24}\, u^{\al}_{\pmu} h^{\al\nu}_{\pnu} h+
                           c_{25}\, u^{\al}_{\pmu} h^{\al\nu} h^{}_{\pnu}+
                           c_{26}\, u^{\al} h^{\al\nu}_{\pmu} h^{}_{\pnu}. \\
   \end{split}
  \label{eq:TypC}
\end{equation} 

Divergences of type $D$ have the structure $\partial_{\ro}\partial_{\nu}\partial_{\nu}|u^{\al}h^{\al\si}h^{\si\ro}$. The basis elements of this type are
\begin{align*}
  e^D_1 & = u^{\al}_{\pro\pnu}h^{\al\si}_{\pnu}h^{\si\ro} \\
  e^D_2 & = u^{\al}_{\pro\pnu}h^{\al\si}h^{\si\ro}_{\pnu} \\
  e^D_3 & = u^{\al}_{\pro}h^{\al\si}_{\pnu}h^{\si\ro}_{\pnu} \\
  e^D_4 & = u^{\al}_{\pnu}h^{\al\si}_{\pro\pnu}h^{\si\ro} \\
  e^D_5 & = u^{\al}_{\pnu}h^{\al\si}h^{\si\ro}_{\pro\pnu} \\
  e^D_6 & = u^{\al}_{\pnu}h^{\al\si}_{\pro}h^{\si\ro}_{\pnu} \\
  e^D_7 & = u^{\al}_{\pnu}h^{\al\si}_{\pnu}h^{\si\ro}_{\pro} \\
  e^D_8 & = u^{\al}h^{\al\si}_{\pro\pnu}h^{\si\ro}_{\pnu} \\
  e^D_9 & = u^{\al}h^{\al\si}_{\pnu}h^{\si\ro}_{\pro\pnu}.
\end{align*}
We find two types of divergences
\begin{align*}
  div^D_1 & = \partial_{\ro}\bigl(\partial_{\nu}\partial_{\nu}|u^{\al}h^{\al\si}h^{\si\ro}\bigr) \\
  div^D_2 & = \partial_{\nu}\bigl(\partial_{\ro}\partial_{\nu}|u^{\al}h^{\al\si}h^{\si\ro}\bigr). 
\end{align*}
We find the following divergences of type $div^D_1$
\begin{align*}
  \partial_{\ro}\bigl(u^{\al}_{\pnu}h^{\al\si}_{\pnu}h^{\si\ro}\bigr) & = e^D_1+e^D_4+e^D_7 \\
  \partial_{\ro}\bigl(u^{\al}_{\pnu}h^{\al\si}h^{\si\ro}_{\pnu}\bigr) & = e^D_2+e^D_6+e^D_5 \\
  \partial_{\ro}\bigl(u^{\al}h^{\al\si}_{\pnu}h^{\si\ro}_{\pnu}\bigr) & = e^D_3+e^D_8+e^D_9. 
\end{align*}
Divergences of type $div^D_2$ are
\begin{align*}
  \partial_{\nu}\bigl(u^{\al}_{\pro\pnu}h^{\al\si}h^{\si\ro}\bigr) & = e^D_1+e^D_2 \\
  \partial_{\nu}\bigl(u^{\al}_{\pro}h^{\al\si}_{\pnu}h^{\si\ro}\bigr) & = e^D_1+e^D_3 \\
  \partial_{\nu}\bigl(u^{\al}_{\pro}h^{\al\si}h^{\si\ro}_{\pnu}\bigr) & = e^D_2+e^D_3 \\
  \partial_{\nu}\bigl(u^{\al}_{\pnu}h^{\al\si}_{\pro}h^{\si\ro}\bigr) & = e^D_4+e^D_6 \\
  \partial_{\nu}\bigl(u^{\al}_{\pnu}h^{\al\si}h^{\si\ro}_{\pro}\bigr) & = e^D_7+e^D_5 \\
  \partial_{\nu}\bigl(u^{\al}h^{\al\si}_{\pro\pnu}h^{\si\ro}\bigr) & = e^D_4+e^D_8 \\
  \partial_{\nu}\bigl(u^{\al}h^{\al\si}_{\pro}h^{\si\ro}_{\pnu}\bigr) & = e^D_6+e^D_8 \\
  \partial_{\nu}\bigl(u^{\al}h^{\al\si}_{\pnu}h^{\si\ro}_{\pro}\bigr) & = e^D_7+e^D_9 \\
  \partial_{\nu}\bigl(u^{\al}h^{\al\si}h^{\si\ro}_{\pro\pnu}\bigr) & = e^D_5+e^D_9.
\end{align*}
For $T_{1/1}^{\mu,D}$ we can make the ansatz 
\begin{equation}
   \begin{split}
     T_{1/1}^{\mu,D} = & \ c_{27}\, u^{\al}_{\pro} h^{\al\si}_{\pro} h^{\si\mu}+ 
                           c_{28}\, u^{\al}_{\pro} h^{\al\si} h^{\si\mu}_{\pro}+
                           c_{29}\, u^{\al} h^{\al\si}_{\pro} h^{\si\mu}_{\pro}+
                           c_{30}\, u^{\al}_{\pro\pmu} h^{\al\si} h^{\si\ro} \\
                       & + c_{31}\, u^{\al} h^{\al\si}_{\pro\pmu} h^{\si\ro}+
                           c_{32}\, u^{\al} h^{\al\si} h^{\si\ro}_{\pro\pmu}+
                           c_{33}\, u^{\al}_{\pro} h^{\al\si}_{\pmu} h^{\si\ro}+
                           c_{34}\, u^{\al}_{\pro} h^{\al\si} h^{\si\ro}_{\pmu} \\ 
                       & + c_{35}\, u^{\al} h^{\al\si}_{\pro} h^{\si\ro}_{\pmu}+
                           c_{36}\, u^{\al}_{\pmu} h^{\al\si}_{\pro} h^{\si\ro}+
                           c_{37}\, u^{\al}_{\pmu} h^{\al\si} h^{\si\ro}_{\pro}+
                           c_{38}\, u^{\al} h^{\al\si}_{\pmu} h^{\si\ro}_{\pro}. \\
   \end{split}
  \label{eq:TypD}
\end{equation}       

Divergences of type $E$ have the structure $\partial_{\ro}\partial_{\si}\partial_{\nu}|u^{\al}h^{\al\nu}h^{\ro\si}$. The basis elements are
\begin{align*}
  e^E_1 & = u^{\al}_{\pro\psig\pnu}h^{\al\nu}h^{\ro\si} \\
  e^E_2 & = u^{\al}_{\pro\psig}h^{\al\nu}_{\pnu}h^{\ro\si} \\
  e^E_3 & = u^{\al}_{\pro\psig}h^{\al\nu}h^{\ro\si}_{\pnu} \\
  e^E_4 & = u^{\al}_{\pro\pnu}h^{\al\nu}_{\psig}h^{\ro\si} \\
  e^E_5 & = u^{\al}_{\pro\pnu}h^{\al\nu}h^{\ro\si}_{\psig} \\
  e^E_6 & = u^{\al}_{\pro}h^{\al\nu}_{\psig\pnu}h^{\ro\si} \\
  e^E_7 & = u^{\al}_{\pro}h^{\al\nu}_{\psig}h^{\ro\si}_{\pnu} \\
  e^E_8 & = u^{\al}_{\pro}h^{\al\nu}_{\pnu}h^{\ro\si}_{\psig} \\
  e^E_9 & = u^{\al}_{\pro}h^{\al\nu}h^{\ro\si}_{\psig\pnu} \\
  e^E_{10} & = u^{\al}_{\pnu}h^{\al\nu}_{\pro\psig}h^{\ro\si} \\
  e^E_{11} & = u^{\al}_{\pnu}h^{\al\nu}_{\pro}h^{\ro\si}_{\psig} \\
  e^E_{12} & = u^{\al}_{\pnu}h^{\al\nu}h^{\ro\si}_{\pro\psig} \\
  e^E_{13} & = u^{\al}h^{\al\nu}_{\pro\psig\pnu}h^{\ro\si} \\
  e^E_{14} & = u^{\al}h^{\al\nu}_{\pro\psig}h^{\ro\si}_{\pnu} \\
  e^E_{15} & = u^{\al}h^{\al\nu}_{\pro\pnu}h^{\ro\si}_{\psig} \\
  e^E_{16} & = u^{\al}h^{\al\nu}_{\pro}h^{\ro\si}_{\psig\pnu} \\
  e^E_{17} & = u^{\al}h^{\al\nu}_{\pnu}h^{\ro\si}_{\pro\psig} \\
  e^E_{18} & = u^{\al}h^{\al\nu}h^{\ro\si}_{\pro\psig\pnu}.
\end{align*}
Although there are three different partial derivatives for type $E$ we have only two different divergences because of the symmetry of the tensor field $h^{\ro\si}$
\begin{align*}
  div^E_1 & = \partial_{\ro}\bigl(\partial_{\si}\partial_{\nu}|u^{\al}h^{\al\nu}h^{\ro\si}\bigr) \\
  div^E_2 & = \partial_{\nu}\bigl(\partial_{\ro}\partial_{\si}|u^{\al}h^{\al\nu}h^{\ro\si}\bigr). 
\end{align*}
We find the following divergences of the form $div^E_1$
\begin{align*}
  \partial_{\ro}\bigl(u^{\al}_{\psig\pnu}h^{\al\nu}h^{\ro\si}\bigr) & = e^E_1+e^E_4+e^E_5 \\
  \partial_{\ro}\bigl(u^{\al}_{\psig}h^{\al\nu}_{\pnu}h^{\ro\si}\bigr) & = e^E_2+e^E_6+e^E_8 \\
  \partial_{\ro}\bigl(u^{\al}_{\psig}h^{\al\nu}h^{\ro\si}_{\pnu}\bigr) & = e^E_3+e^E_7+e^E_9 \\
  \partial_{\ro}\bigl(u^{\al}h^{\al\nu}_{\psig\pnu}h^{\ro\si}\bigr) & = e^E_6+e^E_{13}+e^E_{15} \\
  \partial_{\ro}\bigl(u^{\al}h^{\al\nu}_{\psig}h^{\ro\si}_{\pnu}\bigr) & = e^E_7+e^E_{14}+e^E_{16} \\
  \partial_{\ro}\bigl(u^{\al}_{\pnu}h^{\al\nu}_{\psig}h^{\ro\si}\bigr) & = e^E_4+e^E_{10}+e^E_{11} \\
  \partial_{\ro}\bigl(u^{\al}h^{\al\nu}h^{\ro\si}_{\psig\pnu}\bigr) & = e^E_9+e^E_{16}+e^E_{18} \\
  \partial_{\ro}\bigl(u^{\al}h^{\al\nu}_{\pnu}h^{\ro\si}_{\psig}\bigr) & = e^E_8+e^E_{15}+e^E_{17} \\
  \partial_{\ro}\bigl(u^{\al}_{\pnu}h^{\al\nu}h^{\ro\si}_{\psig}\bigr) & = e^E_5+e^E_{11}+e^E_{12}.
\end{align*}
Divergences of the form $div^E_2$ are
\begin{align*}
  \partial_{\nu}\bigl(u^{\al}_{\pro\psig}h^{\al\nu}h^{\ro\si}\bigr) & = e^E_1+e^E_2+e^E_3 \\
  \partial_{\nu}\bigl(u^{\al}_{\pro}h^{\al\nu}_{\psig}h^{\ro\si}\bigr) & = e^E_4+e^E_6+e^E_7 \\
  \partial_{\nu}\bigl(u^{\al}_{\pro}h^{\al\nu}h^{\ro\si}_{\psig}\bigr) & = e^E_5+e^E_8+e^E_9 \\
  \partial_{\nu}\bigl(u^{\al}h^{\al\nu}_{\pro\psig}h^{\ro\si}\bigr) & = e^E_{10}+e^E_{13}+e^E_{14} \\
  \partial_{\nu}\bigl(u^{\al}h^{\al\nu}_{\pro}h^{\ro\si}_{\psig}\bigr) & = e^E_{11}+e^E_{15}+e^E_{16} \\
  \partial_{\nu}\bigl(u^{\al}h^{\al\nu}h^{\ro\si}_{\pro\psig}\bigr) & = e^E_{12}+e^E_{17}+e^E_{18}.
\end{align*}
For $T_{1/1}^{\mu,E}$ we can make the ansatz
\begin{equation}
   \begin{split}
     T_{1/1}^{\mu,E} = & \ c_{39}\, u^{\al}_{\psig\pro} h^{\al\ro} h^{\mu\si}+
                           c_{40}\, u^{\al} h^{\al\ro}_{\psig\pro} h^{\mu\si}+
                           c_{41}\, u^{\al} h^{\al\ro} h^{\mu\si}_{\psig\pro}+
                           c_{42}\, u^{\al}_{\psig} h^{\al\ro}_{\pro} h^{\mu\si} \\
                       & + c_{43}\, u^{\al}_{\psig} h^{\al\ro} h^{\mu\si}_{\pro}+
                           c_{44}\, u^{\al} h^{\al\ro}_{\psig} h^{\mu\si}_{\pro}+
                           c_{45}\, u^{\al}_{\pro} h^{\al\ro}_{\psig} h^{\mu\si}+
                           c_{46}\, u^{\al}_{\pro} h^{\al\ro} h^{\mu\si}_{\psig} \\
                       & + c_{47}\, u^{\al} h^{\al\ro}_{\pro} h^{\mu\si}_{\psig}+
                           c_{48}\, u^{\al}_{\pro\psig} h^{\al\mu} h^{\si\ro}+
                           c_{49}\, u^{\al} h^{\al\mu}_{\pro\psig} h^{\ro\si}+
                           c_{50}\, u^{\al} h^{\al\mu} h^{\ro\si}_{\pro\psig} \\
                       & + c_{51}\, u^{\al}_{\pro} h^{\al\mu}_{\psig} h^{\ro\si}+
                           c_{52}\, u^{\al}_{\pro} h^{\al\mu} h^{\ro\si}_{\psig}+
                           c_{53}\, u^{\al} h^{\al\mu}_{\pro} h^{\ro\si}_{\psig}.\\
   \end{split}
  \label{eq:TypE}
\end{equation}

Divergences of type $F$ are of the form $\partial_{\ro}\partial_{\si}\partial_{\nu}|u^{\nu}h^{\ro\mu}h^{\mu\si}$. The basis elements are
\begin{align*}
  e^F_1 & = u^{\nu}_{\pro\psig\pnu}h^{\ro\mu}h^{\mu\si} \\
  e^F_2 & = u^{\nu}_{\pro\psig}h^{\ro\mu}_{\pnu}h^{\mu\si} \\
  e^F_3 & = u^{\nu}_{\pro\pnu}h^{\ro\mu}_{\psig}h^{\mu\si} \\
  e^F_4 & = u^{\nu}_{\pro\pnu}h^{\ro\mu}h^{\mu\si}_{\psig} \\
  e^F_5 & = u^{\nu}_{\pro}h^{\ro\mu}_{\psig\pnu}h^{\mu\si} \\
  e^F_6 & = u^{\nu}_{\pro}h^{\ro\mu}h^{\mu\si}_{\psig\pnu} \\
  e^F_7 & = u^{\nu}_{\pro}h^{\ro\mu}_{\psig}h^{\mu\si}_{\pnu} \\
  e^F_8 & = u^{\nu}_{\pro}h^{\ro\mu}_{\pnu}h^{\mu\si}_{\psig} \\
  e^F_9 & = u^{\nu}_{\pnu}h^{\ro\mu}_{\pro\psig}h^{\mu\si} \\
  e^F_{10} & = u^{\nu}_{\pnu}h^{\ro\mu}_{\pro}h^{\mu\si}_{\psig} \\
  e^F_{11} & = u^{\nu}_{\pnu}h^{\ro\mu}_{\psig}h^{\mu\si}_{\pro} \\
  e^F_{12} & = u^{\nu}h^{\ro\mu}_{\pro\psig\pnu}h^{\mu\si} \\
  e^F_{13} & = u^{\nu}h^{\ro\mu}_{\pro\psig}h^{\mu\si}_{\pnu} \\
  e^F_{14} & = u^{\nu}h^{\ro\mu}_{\pro\pnu}h^{\mu\si}_{\psig} \\
  e^F_{15} & = u^{\nu}h^{\ro\mu}_{\psig\pnu}h^{\mu\si}_{\pro}.
\end{align*}
Then we have two different types divergences, namely
\begin{align*}
  div^F_1 & = \partial_{\ro}\bigl(\partial_{\si}\partial_{\nu}|u^{\nu}h^{\ro\mu}h^{\mu\si}\bigr) \\
  div^F_2 & = \partial_{\nu}\bigl(\partial_{\ro}\partial_{\si}|u^{\nu}h^{\ro\mu}h^{\mu\si}\bigr).
\end{align*}
We find the following divergences of the form $div^F_1$
\begin{align*}
  \partial_{\ro}\bigl(u^{\nu}_{\psig\pnu}h^{\ro\mu}h^{\mu\si}\bigr) & = e^F_1+e^F_4+e^F_3 \\
  \partial_{\ro}\bigl(u^{\nu}_{\psig}h^{\ro\mu}_{\pnu}h^{\mu\si}\bigr) & = e^F_2+e^F_6+e^F_7 \\
  \partial_{\ro}\bigl(u^{\nu}_{\psig}h^{\ro\mu}h^{\mu\si}_{\pnu}\bigr) & = e^F_2+e^F_8+e^F_5 \\
  \partial_{\ro}\bigl(u^{\nu}h^{\ro\mu}_{\psig\pnu}h^{\mu\si}\bigr) & = e^F_5+e^F_{12}+e^F_{15} \\
  \partial_{\ro}\bigl(u^{\nu}h^{\ro\mu}_{\psig}h^{\mu\si}_{\pnu}\bigr) & = e^F_7+e^F_{13}+e^F_{15} \\
  \partial_{\ro}\bigl(u^{\nu}_{\pnu}h^{\ro\mu}_{\psig}h^{\mu\si}\bigr) & = e^F_3+e^F_9+e^F_{11} \\
  \partial_{\ro}\bigl(u^{\nu}h^{\ro\mu}h^{\mu\si}_{\psig\pnu}\bigr) & = e^F_6+e^F_{14}+e^F_{12} \\
  \partial_{\ro}\bigl(u^{\nu}h^{\ro\mu}_{\pnu}h^{\mu\si}_{\psig}\bigr) & = e^F_8+e^F_{14}+e^F_{13} \\
  \partial_{\ro}\bigl(u^{\nu}_{\pnu}h^{\ro\mu}h^{\mu\si}_{\psig}\bigr) & = e^F_4+e^F_{10}+e^F_9.
\end{align*}
Divergences of the form $div^F_2$ are
\begin{align*}
  \partial_{\nu}\bigl(u^{\nu}_{\pro\psig}h^{\ro\mu}h^{\mu\si}\bigr) & = e^F_1+2e^F_2 \\
  \partial_{\nu}\bigl(u^{\nu}_{\pro}h^{\ro\mu}_{\psig}h^{\mu\si}\bigr) & = e^F_3+e^F_5+e^F_7 \\
  \partial_{\nu}\bigl(u^{\nu}_{\pro}h^{\ro\mu}h^{\mu\si}_{\psig}\bigr) & = e^F_4+e^F_8+e^F_6 \\
  \partial_{\nu}\bigl(u^{\nu}h^{\ro\mu}_{\pro\psig}h^{\mu\si}\bigr) & = e^F_9+e^F_{12}+e^F_{13} \\ 
  \partial_{\nu}\bigl(u^{\nu}h^{\ro\mu}_{\pro}h^{\mu\si}_{\psig}\bigr) & = e^F_{10}+2e^F_{14} \\
  \partial_{\nu}\bigl(u^{\nu}h^{\ro\mu}_{\psig}h^{\mu\si}_{\pro}\bigr) & = e^F_{11}+2e^F_{15}.
\end{align*}
For $T_{1/1}^{\mu,F}$ we can make the ansatz
\begin{equation}
   \begin{split}
     T_{1/1}^{\mu,F} = & \ c_{54}\, u^{\ro}_{\psig\pro} h^{\mu\nu} h^{\nu\si}+ 
                           c_{55}\, u^{\ro} h^{\mu\nu}_{\psig\pro} h^{\nu\si}+
                           c_{56}\, u^{\ro} h^{\mu\nu} h^{\nu\si}_{\psig\pro}+
                           c_{57}\, u^{\ro}_{\psig} h^{\mu\nu}_{\pro} h^{\nu\si} \\
                       & + c_{58}\, u^{\ro}_{\psig} h^{\mu\nu} h^{\nu\si}_{\pro}+ 
                           c_{59}\, u^{\ro} h^{\mu\nu}_{\psig} h^{\nu\si}_{\pro}+
                           c_{60}\, u^{\ro}_{\pro} h^{\mu\nu}_{\psig} h^{\nu\si}+
                           c_{61}\, u^{\ro}_{\pro} h^{\mu\nu} h^{\nu\si}_{\psig} \\
                       & + c_{62}\, u^{\ro} h^{\mu\nu}_{\pro} h^{\nu\si}_{\psig}+ 
                           c_{63}\, u^{\mu}_{\pro\psig} h^{\ro\nu} h^{\nu\si}+
                           c_{64}\, u^{\mu} h^{\ro\nu}_{\pro\psig} h^{\nu\si}+
                           c_{65}\, u^{\mu} h^{\ro\nu}_{\psig} h^{\nu\si}_{\pro} \\
                       & + c_{66}\, u^{\mu}_{\pro} h^{\ro\nu}_{\psig} h^{\nu\si}+ 
                           c_{67}\, u^{\mu}_{\pro} h^{\ro\nu} h^{\nu\si}_{\psig}+
                           c_{68}\, u^{\mu} h^{\ro\nu}_{\pro} h^{\nu\si}_{\psig}. \\
   \end{split}
  \label{eq:TypF}
\end{equation}

Divergences of type $G$ are of the form $\partial_{\ro}\partial_{\si}\partial_{\nu}|u^{\nu}h^{\ro\si}h$. The basis elements are
\begin{align*}
  e^G_1 & = u^{\nu}_{\pro\psig\pnu}h^{\ro\si}h \\
  e^G_2 & = u^{\nu}_{\pro\psig}h^{\ro\si}_{\pnu}h \\
  e^G_3 & = u^{\nu}_{\pro\psig}h^{\ro\si}h^{}_{\pnu} \\
  e^G_4 & = u^{\nu}_{\pro\pnu}h^{\ro\si}_{\psig}h \\
  e^G_5 & = u^{\nu}_{\pro\pnu}h^{\ro\si}h^{}_{\psig} \\
  e^G_6 & = u^{\nu}_{\pro}h^{\ro\si}_{\psig\pnu}h \\
  e^G_7 & = u^{\nu}_{\pro}h^{\ro\si}h^{}_{\pro\pnu} \\
  e^G_8 & = u^{\nu}_{\pro}h^{\ro\si}_{\psig}h^{}_{\pnu} \\
  e^G_9 & = u^{\nu}_{\pro}h^{\ro\si}_{\pnu}h^{}_{\psig} \\
  e^G_{10} & = u^{\nu}_{\pnu}h^{\ro\si}_{\pro\psig}h \\
  e^G_{11} & = u^{\nu}_{\pnu}h^{\ro\si}h^{}_{\pro\psig} \\
  e^G_{12} & = u^{\nu}_{\pnu}h^{\ro\si}_{\pro}h^{}_{\psig} \\
  e^G_{13} & = u^{\nu}h^{\ro\si}_{\pro\psig\pnu}h \\
  e^G_{14} & = u^{\nu}h^{\ro\si}_{\pro\psig}h^{}_{\pnu} \\
  e^G_{15} & = u^{\nu}h^{\ro\si}_{\pro\pnu}h^{}_{\psig} \\
  e^G_{16} & = u^{\nu}h^{\ro\si}_{\pro}h^{}_{\psig\pnu} \\
  e^G_{17} & = u^{\nu}h^{\ro\si}_{\pnu}h^{}_{\pro\psig} \\
  e^G_{18} & = u^{\nu}h^{\ro\si}h^{}_{\pro\psig\pnu}.
\end{align*}
We have two types of divergences, namely
\begin{align*}
  div^G_1 & = \partial_{\ro}\bigl(\partial_{\si}\partial_{\nu}|u^{\nu}h^{\ro\si}h\bigr) \\
  div^G_2 & = \partial_{\nu}\bigl(\partial_{\ro}\partial_{\si}|u^{\nu}h^{\ro\si}h\bigr).
\end{align*}
We find the following divergences of type $div^G_1$
\begin{align*}
  \partial_{\ro}\bigl(u^{\nu}_{\psig\pnu}h^{\ro\si}h\bigr) & = e^G_1+e^G_4+e^G_5 \\
  \partial_{\ro}\bigl(u^{\nu}_{\psig}h^{\ro\si}_{\pnu}h\bigr) & = e^G_2+e^G_6+e^G_9 \\
  \partial_{\ro}\bigl(u^{\nu}_{\psig}h^{\ro\si}h^{}_{\pnu}\bigr) & = e^G_3+e^G_8+e^G_7 \\
  \partial_{\ro}\bigl(u^{\nu}h^{\ro\si}_{\psig\pnu}h\bigr) & = e^G_6+e^G_{13}+e^G_{15} \\
  \partial_{\ro}\bigl(u^{\nu}h^{\ro\si}_{\psig}h^{}_{\pnu}\bigr) & = e^G_8+e^G_{14}+e^G_{16} \\
  \partial_{\ro}\bigl(u^{\nu}_{\pnu}h^{\ro\si}_{\psig}h\bigr) & = e^G_4+e^G_{10}+e^G_{12} \\
  \partial_{\ro}\bigl(u^{\nu}h^{\ro\si}h^{}_{\psig\pnu}\bigr) & = e^G_7+e^G_{16}+e^G_{18} \\
  \partial_{\ro}\bigl(u^{\nu}h^{\ro\si}_{\pnu}h^{}_{\psig}\bigr) & = e^G_9+e^G_{15}+e^G_{17} \\
  \partial_{\ro}\bigl(u^{\nu}_{\pnu}h^{\ro\si}h^{}_{\psig}\bigr) & = e^G_5+e^G_{12}+e^G_{11}. 
\end{align*}
Divergences of the form $div^G_2$ are 
\begin{align*}
  \partial_{\nu}\bigl(u^{\nu}_{\pro\psig}h^{\ro\si}h\bigr) & = e^G_1+e^G_2+e^G_3 \\
  \partial_{\nu}\bigl(u^{\nu}_{\pro}h^{\ro\si}_{\psig}h\bigr) & = e^G_4+e^G_6+e^G_8 \\
  \partial_{\nu}\bigl(u^{\nu}_{\pro}h^{\ro\si}h^{}_{\psig}\bigr) & = e^G_5+e^G_9+e^G_7 \\
  \partial_{\nu}\bigl(u^{\nu}h^{\ro\si}_{\pro\psig}h\bigr) & = e^G_{10}+e^G_{13}+e^G_{14} \\
  \partial_{\nu}\bigl(u^{\nu}h^{\ro\si}_{\pro}h^{}_{\psig}\bigr) & = e^G_{12}+e^G_{15}+e^G_{16} \\
  \partial_{\nu}\bigl(u^{\nu}h^{\ro\si}h^{}_{\pro\psig}\bigr) & = e^G_{11}+e^G_{17}+e^G_{18}.
\end{align*}
For $T_{1/1}^{\mu,G}$ we can make the ansatz
\begin{equation}
   \begin{split}
     T_{1/1}^{\mu,G} = & \ c_{69}\, u^{\ro}_{\psig\pro} h^{\mu\si} h+
                           c_{70}\, u^{\ro} h^{\mu\si}_{\psig\pro} h+ 
                           c_{71}\, u^{\ro} h^{\mu\si} h^{}_{\psig\pro}+
                           c_{72}\, u^{\ro}_{\psig} h^{\mu\si}_{\pro} h \\
                       & + c_{73}\, u^{\ro}_{\psig} h^{\mu\si} h^{}_{\pro}+
                           c_{74}\, u^{\ro} h^{\mu\si}_{\psig} h^{}_{\pro}+
                           c_{75}\, u^{\ro}_{\pro} h^{\mu\si}_{\psig} h+
                           c_{76}\, u^{\ro}_{\pro} h^{\mu\si} h^{}_{\psig} \\
                       & + c_{77}\, u^{\ro} h^{\mu\si}_{\pro} h^{}_{\psig}+
                           c_{78}\, u^{\mu}_{\pro\psig} h^{\ro\si} h+ 
                           c_{79}\, u^{\mu} h^{\ro\si}_{\pro\psig} h+
                           c_{80}\, u^{\mu} h^{\ro\si} h^{}_{\pro\psig} \\
                       & + c_{81}\, u^{\mu}_{\pro} h^{\ro\si}_{\psig} h+
                           c_{82}\, u^{\mu}_{\pro} h^{\ro\si} h^{}_{\psig}+
                           c_{83}\, u^{\mu} h^{\ro\si}_{\pro} h^{}_{\psig}. \\
   \end{split}
  \label{eq:TypG}
\end{equation}
This completes the discussion of the divergences in the graviton sector. 

\subsection{Ghost Sector}
Now we come to the ghost sector where we've three different Lorentz types. Divergences of type $H$ are of the form $\partial_{\si}\partial_{\nu}\partial_{\nu}|\ti{u}^{\al}u^{\si}u^{\al}$. The basis elements are
\begin{align*}
  e^H_1 & = \ti{u}^{\al}_{\psig\pnu}u^{\si}_{\pnu}u^{\al} \\
  e^H_2 & = \ti{u}^{\al}_{\psig\pnu}u^{\si}u^{\al}_{\pnu} \\
  e^H_3 & = \ti{u}^{\al}_{\psig}u^{\si}_{\pnu}u^{\al}_{\pnu} \\
  e^H_4 & = \ti{u}^{\al}_{\pnu}u^{\si}_{\psig\pnu}u^{\al} \\
  e^H_5 & = \ti{u}^{\al}_{\pnu}u^{\si}u^{\al}_{\psig\pnu} \\
  e^H_6 & = \ti{u}^{\al}_{\pnu}u^{\si}_{\psig}u^{\al}_{\pnu} \\
  e^H_7 & = \ti{u}^{\al}_{\pnu}u^{\si}_{\pnu}u^{\al}_{\psig} \\
  e^H_8 & = \ti{u}^{\al}u^{\si}_{\psig\pnu}u^{\al}_{\pnu} \\
  e^H_9 & = \ti{u}^{\al}u^{\si}_{\pnu}u^{\al}_{\psig\pnu}. 
\end{align*}
We have two types of divergences corresponding to the two different partial derivatives
\begin{align*}
  div^H_1 & = \partial_{\si}\bigl(\partial_{\nu}\partial_{\nu}|\ti{u}^{\al}u^{\si}u^{\al}\bigr) \\
  div^H_2 & = \partial_{\nu}\bigl(\partial_{\si}\partial_{\nu}|\ti{u}^{\al}u^{\si}u^{\al}\bigr). \\
\end{align*}
We find the following divergences of the form $div^H_1$
\begin{align*}
  \partial_{\si}\bigl(\ti{u}^{\al}_{\pnu}u^{\si}_{\pnu}u^{\al}\bigr) & = e^H_1+e^H_4+e^H_7 \\
  \partial_{\si}\bigl(\ti{u}^{\al}_{\pnu}u^{\si}u^{\al}_{\pnu}\bigr) & = e^H_2+e^H_6+e^H_5 \\
  \partial_{\si}\bigl(\ti{u}^{\al}u^{\si}_{\pnu}u^{\al}_{\pnu}\bigr) & = e^H_3+e^H_8+e^H_9. 
\end{align*}
Divergences of the form $div^H_2$ are
\begin{align*}
  \partial_{\nu}\bigl(\ti{u}^{\al}_{\psig\pnu}u^{\si}u^{\al}\bigr) & = e^H_1+e^H_2 \\
  \partial_{\nu}\bigl(\ti{u}^{\al}_{\psig}u^{\si}_{\pnu}u^{\al}\bigr) & = e^H_1+e^H_3 \\
  \partial_{\nu}\bigl(\ti{u}^{\al}_{\psig}u^{\si}u^{\al}_{\pnu}\bigr) & = e^H_2+e^H_3 \\
  \partial_{\nu}\bigl(\ti{u}^{\al}_{\pnu}u^{\si}_{\psig}u^{\al}\bigr) & = e^H_4+e^H_6 \\
  \partial_{\nu}\bigl(\ti{u}^{\al}_{\pnu}u^{\si}u^{\al}_{\psig}\bigr) & = e^H_7+e^H_5 \\
  \partial_{\nu}\bigl(\ti{u}^{\al}u^{\si}_{\psig\pnu}u^{\al}\bigr) & = e^H_4+e^H_8 \\
  \partial_{\nu}\bigl(\ti{u}^{\al}u^{\si}_{\psig}u^{\al}_{\pnu}\bigr) & = e^H_6+e^H_8 \\
  \partial_{\nu}\bigl(\ti{u}^{\al}u^{\si}_{\pnu}u^{\al}_{\psig}\bigr) & = e^H_7+e^H_9 \\
  \partial_{\nu}\bigl(\ti{u}^{\al}u^{\si}u^{\al}_{\psig\pnu}\bigr) & = e^H_5+e^H_9.
\end{align*}
Then we can make the ansatz for $T_{1/1}^{\mu,H}$
\begin{equation}
   \begin{split}
     T_{1/1}^{\mu,H} = & \ c_{84}\, u^{\mu}_{\psig} \ti{u}^{\al}_{\psig} u^{\al}+
                           c_{85}\, u^{\mu}_{\psig} \ti{u}^{\al} u^{\al}_{\psig}+
                           c_{86}\, u^{\mu} \ti{u}^{\al}_{\psig} u^{\al}_{\psig}+ 
                           c_{87}\, u^{\si}_{\psig} \ti{u}^{\al}_{\pmu} u^{\al} \\
                       & + c_{88}\, u^{\si}_{\psig} \ti{u}^{\al} u^{\al}_{\pmu}+
                           c_{89}\, u^{\si} \ti{u}^{\al}_{\psig} u^{\al}_{\pmu}+
                           c_{90}\, u^{\si}_{\pmu} \ti{u}^{\al}_{\psig} u^{\al}+ 
                           c_{91}\, u^{\si}_{\pmu} \ti{u}^{\al} u^{\al}_{\psig} \\
                       & + c_{92}\, u^{\si} \ti{u}^{\al}_{\pmu} u^{\al}_{\psig}+
                           c_{93}\, u^{\si}_{\psig\pmu} \ti{u}^{\al} u^{\al}+
                           c_{94}\, u^{\si} \ti{u}^{\al}_{\psig\pmu} u^{\al}+
                           c_{95}\, u^{\si} \ti{u}^{\al} u^{\al}_{\psig\pmu}. \\
   \end{split}
  \label{eq:TypH}
\end{equation}

Divergences of type $J$ are of the form $\partial_{\al}\partial_{\ro}\partial_{\nu}|\ti{u}^{\al}u^{\ro}u^{\nu}$. The fact that the ghost fields anticommute reduces the number of independent basis elements considerably, e.g terms of the form $\ti{u}^{\al}_{\pal\pro\pnu}u^{\ro}u^{\nu},\,\ti{u}^{\al}_{\pal}u^{\ro}_{\pro}u^{\nu}_{\pnu}$ and $\ti{u}^{\al}_{\pal}u^{\ro}_{\pnu}u^{\nu}_{\pro}$ must vanish. Then we have the following basis elements
\begin{align*}
  e^J_1 & = \ti{u}^{\al}_{\pal\pro}u^{\ro}_{\pnu}u^{\nu} \\
  e^J_2 & = \ti{u}^{\al}_{\pal\pro}u^{\ro}u^{\nu}_{\pnu} \\
  e^J_3 & = \ti{u}^{\al}_{\pro\pnu}u^{\ro}_{\pal}u^{\nu} \\
  e^J_4 & = \ti{u}^{\al}_{\pal}u^{\ro}_{\pro\pnu}u^{\nu} \\
  e^J_5 & = \ti{u}^{\al}_{\pro}u^{\ro}_{\pal\pnu}u^{\nu} \\
  e^J_6 & = \ti{u}^{\al}_{\pro}u^{\ro}u^{\nu}_{\pal\pnu} \\
  e^J_7 & = \ti{u}^{\al}_{\pro}u^{\ro}_{\pal}u^{\nu}_{\pnu} \\
  e^J_8 & = \ti{u}^{\al}_{\pro}u^{\ro}_{\pnu}u^{\nu}_{\pal} \\
  e^J_9 & = \ti{u}^{\al}u^{\ro}_{\pal\pro\pnu}u^{\nu} \\
  e^J_{10} & = \ti{u}^{\al}u^{\ro}_{\pal\pro}u^{\nu}_{\pnu} \\
  e^J_{11} & = \ti{u}^{\al}u^{\ro}_{\pal\pnu}u^{\nu}_{\pro} \\
  e^J_{12} & = \ti{u}^{\al}u^{\ro}_{\pro\pnu}u^{\nu}_{\pal}. 
\end{align*}
We have two types of divergences, namely
\begin{align*}
  div^J_1 & = \partial_{\al}\bigl(\partial_{\ro}\partial_{\nu}|\ti{u}^{\al}u^{\ro}u^{\nu}\bigr) \\
  div^J_2 & = \partial_{\ro}\bigl(\partial_{\al}\partial_{\nu}|\ti{u}^{\al}u^{\ro}u^{\nu}\bigr).
\end{align*}
We find the following divergences of type $div^J_1$
\begin{align*}
  \partial_{\al}\bigl(\ti{u}^{\al}_{\pro}u^{\ro}_{\pnu}u^{\nu}\bigr) & = e^J_1+e^J_5+e^J_8 \\
  \partial_{\al}\bigl(\ti{u}^{\al}_{\pro}u^{\ro}u^{\nu}_{\pnu}\bigr) & = e^J_2+e^J_7+e^J_6 \\
  \partial_{\al}\bigl(\ti{u}^{\al}u^{\ro}_{\pro\pnu}u^{\nu}\bigr) & = e^J_4+e^J_9+e^J_{12}. 
\end{align*}
Divergences of type $div^J_2$ are
\begin{align*}
  \partial_{\ro}\bigl(\ti{u}^{\al}_{\pal\pnu}u^{\ro}u^{\nu}\bigr) & = -e^J_1-e^J_2 \\
  \partial_{\ro}\bigl(\ti{u}^{\al}_{\pal}u^{\ro}_{\pnu}u^{\nu}\bigr) & = e^J_1+e^J_5 \\
  \partial_{\ro}\bigl(\ti{u}^{\al}_{\pal}u^{\ro}u^{\nu}_{\pnu}\bigr) & = e^J_2-e^J_4 \\
  \partial_{\ro}\bigl(\ti{u}^{\al}u^{\ro}_{\pal\pnu}u^{\nu}\bigr) & = e^J_5+e^J_9+e^J_{11} \\
  \partial_{\ro}\bigl(\ti{u}^{\al}u^{\ro}_{\pal}u^{\nu}_{\pnu}\bigr) & = e^J_7+e^J_{10}-e^J_{12} \\
  \partial_{\ro}\bigl(\ti{u}^{\al}_{\pnu}u^{\ro}_{\pal}u^{\nu}\bigr) & = e^J_3-e^J_6-e^J_8 \\
  \partial_{\ro}\bigl(\ti{u}^{\al}u^{\ro}u^{\nu}_{\pal\pnu}\bigr) & = e^J_6-e^J_{10}-e^J_9 \\
  \partial_{\ro}\bigl(\ti{u}^{\al}u^{\ro}_{\pnu}u^{\nu}_{\pal}\bigr) & = e^J_8+e^J_{12}-e^J_{11} \\
  \partial_{\ro}\bigl(\ti{u}^{\al}_{\pnu}u^{\ro}u^{\nu}_{\pal}\bigr) & = -e^J_3-e^J_7-e^J_5.
\end{align*}
For $T_{1/1}^{\mu,J}$ we can make the ansatz
\begin{equation}
   \begin{split}
     T_{1/1}^{\mu,J} = & \ c_{96}\, u^{\al}_{\pro} \ti{u}^{\mu}_{\pal} u^{\ro}+
                           c_{97}\, u^{\al}_{\pal} \ti{u}^{\mu}_{\pro} u^{\ro}+
                           c_{98}\, u^{\al}_{\pro\pal} \ti{u}^{\mu} u^{\ro}+
                           c_{99}\, u^{\ro}_{\pal\pro} \ti{u}^{\al} u^{\mu} \\
                       & + c_{100}\, u^{\ro} \ti{u}^{\al}_{\pal\pro} u^{\mu}+ 
                           c_{101}\, u^{\ro} \ti{u}^{\al} u^{\mu}_{\pal\pro}+
                           c_{102}\, u^{\ro}_{\pal} \ti{u}^{\al}_{\pro} u^{\mu}+ 
                           c_{103}\, u^{\ro}_{\pal} \ti{u}^{\al} u^{\mu}_{\pro} \\
                       & + c_{104}\, u^{\ro} \ti{u}^{\al}_{\pal} u^{\mu}_{\pro}+ 
                           c_{105}\, u^{\ro}_{\pro} \ti{u}^{\al}_{\pal} u^{\mu}+
                           c_{106}\, u^{\ro}_{\pro} \ti{u}^{\al} u^{\mu}_{\pal}+
                           c_{107}\, u^{\ro} \ti{u}^{\al}_{\pro} u^{\mu}_{\pal}. \\
   \end{split}
  \label{eq:TypJ}
\end{equation}

Divergences of type $K$ are of the form $\partial_{\al}\partial_{\nu}\partial_{\nu}|\ti{u}^{\al}u^{\si}u^{\si}$. The basis elements are
\begin{align*}
  e^K_1 & = \ti{u}^{\al}_{\pal\pnu}u^{\si}_{\pnu}u^{\si} \\
  e^K_2 & = \ti{u}^{\al}_{\pnu}u^{\si}_{\pal\pnu}u^{\si} \\
  e^K_3 & = \ti{u}^{\al}_{\pnu}u^{\si}_{\pal}u^{\si}_{\pnu} \\
  e^K_4 & = \ti{u}^{\al}u^{\si}_{\pal\pnu}u^{\si}_{\pnu}. 
\end{align*}
Then we have two types of divergences
\begin{align*}
  div^K_1 & = \partial_{\al}\bigl(\partial_{\nu}\partial_{\nu}|\ti{u}^{\al}u^{\si}u^{\si}\bigr) \\
  div^K_2 & = \partial_{\nu}\bigl(\partial_{\al}\partial_{\nu}|\ti{u}^{\al}u^{\si}u^{\si}\bigr). 
\end{align*}
We find the following divergence of type $div^K_1$
\begin{equation*}
  \partial_{\al}\bigl(\ti{u}^{\al}_{\pnu}u^{\si}_{\pnu}u^{\si}\bigr)=e^K_1+e^K_2-e^K_3.
\end{equation*}
The following divergences are of type $div^K_2$
\begin{align*}
  \partial_{\nu}\bigl(\ti{u}^{\al}_{\pal}u^{\si}_{\pnu}u^{\si}\bigr) & = e^K_1 \\
  \partial_{\nu}\bigl(\ti{u}^{\al}_{\pnu}u^{\si}_{\pal}u^{\si}\bigr) & = e^K_2+e^K_3 \\
  \partial_{\nu}\bigl(\ti{u}^{\al}u^{\si}_{\pal\pnu}u^{\si}\bigr) & = e^K_2+e^K_4 \\
  \partial_{\nu}\bigl(\ti{u}^{\al}u^{\si}_{\pal}u^{\si}_{\pnu}\bigr) & = e^K_3+e^K_4.
\end{align*}
We can make the ansatz for $T_{1/1}^{\mu,K}$ 
\begin{equation}
   \begin{split}
     T_{1/1}^{\mu,K} = & \ c_{108}\, u^{\si}_{\pal} \ti{u}^{\mu}_{\pal} u^{\si}+
                           c_{109}\, u^{\si}_{\pal\pmu} \ti{u}^{\al} u^{\si}+
                           c_{110}\, u^{\si}_{\pal} \ti{u}^{\al}_{\pmu} u^{\si} \\
                       & + c_{111}\, u^{\si}_{\pal} \ti{u}^{\al} u^{\si}_{\pmu}+
                           c_{112}\, u^{\si}_{\pmu} \ti{u}^{\al}_{\pal} u^{\si}. \\ 
   \end{split}
  \label{eq:TypK}
\end{equation}  

This completes the discussion of the ghost sector. Collecting all divergences we obtain the total divergence as the sum of these 10 different Lorentz types
\begin{equation}
  \partial_{\mu} T_{1/1}^{\mu} = \partial_{\mu}\sum_{i\in \{A,\ldots, K\}}T_{1/1}^{\mu,i}.
  \label{eq:T1/1}
\end{equation}
The parameters $c_1,\ldots,c_{112}\in\mathbb{C}$ are for the moment free constants, to be determined by gauge invariance. This expression\footnote{$T_{1/1}^{\mu}$ is called $Q$-vertex in the sequel because it is obtained from the usual vertex $T_1$ if one replaces a quantum field with the gauge variation of that field.} for $T_{1/1}^{\mu}$ contains all possible combinations of fields appearing after gauge variation of $T_1$. Without losing generality one can now eliminate a few terms in the types $A,\ldots,D,H$ and $K$\footnote{This relies on an idea of M. D\"utsch, see \cite{due:nqym}.}. For that purpose we consider a new $Q$-vertex $\wti{T}_{1/1}^{\mu}(x)$ for which the following relation holds
\begin{equation}
  T_{1/1}^{\mu}(x) = \wti{T}_{1/1}^{\mu}(x) + B^{\mu}(x)
  \label{eq:tiT1/1}
\end{equation}
where $B^{\mu}$ has the special form $B^{\mu}(x)=\partial_{\nu}^xA^{\nu\mu}(x)$ and $A^{\nu\mu}(x)$ is an anti-symmetrical tensor of rank $2$. Then we have
\begin{equation}
  \partial_{\mu} T_{1/1}^{\mu}(x) = \partial_{\mu} \wti{T}_{1/1}^{\mu}(x),
  \label{eq:T1/1=tiT1/1}
\end{equation}
because partial derivatives are commuting. Let us now construct such a tensor $A^{\nu\mu}$. We consider the type-$A$ term $c_3\,u^\al_{\pal\pmu}h^{\ro\si}h^{\ro\si}$. This can be written as
\begin{equation}
  c_3\,u^\al_{\pal\pmu}h^{\ro\si}h^{\ro\si} =
  c_3\,\bigl[\partial_{\al}\bigl(u^{\al}_{\pmu}h^{\ro\si}h^{\ro\si}\bigr) -
  2\,u^{\al}_{\pmu}h^{\ro\si}_{\pal}h^{\ro\si}\bigr].
  \label{eq:c3-Term}
\end{equation}
In an analogous way and using the wave equation we can write 
\begin{equation}
  0 = c_3\,u^{\mu}_{\pal\pal}h^{\ro\si}h^{\ro\si} =
  c_3\,\bigl[\partial_{\al}\bigl(u^{\mu}_{\pal}h^{\ro\si}h^{\ro\si}\bigr) -
  2\,u^{\mu}_{\pal}h^{\ro\si}_{\pal}h^{\ro\si}\bigr].
  \label{eq:c3-Term=0}
\end{equation}
Now we add $-c_3\,u^{\mu}_{\pal\pal}h^{\ro\si}h^{\ro\si}$ to $T_{1/1}^{\mu}$ and obtain
\begin{equation}
  T_{1/1}^{\mu} = \wti{T}_{1/1}^{\mu} +
  c_3\,\partial_{\nu}\bigl(u^{\nu}_{\pmu} - u^{\mu}_{\pnu}\bigr)h^{\ro\si}h^{\ro\si}.
  \label{eq:tiT1/1-Beispiel}
\end{equation}
The expression in brackets is anti-symmetric in $\nu,\mu$ and we get $\wti{T}_{1/1}^{\mu}$ if we replace the constants $c_1$ with $c_1+2\,c_3$ and $c_7$ with $c_7-2\,c_3$ in $T_{1/1}^{\mu}$. In this way we can eliminate the monomials with constants $c_3,c_4$ in type $A$, $c_{10},c_{11}$ in type $B$, $c_{18},c_{19},c_{20}$ in type $C$, $c_{30},c_{31},c_{32}$ in type $D$, $c_{93},c_{94},c_{95}$ in type $H$ and $c_{109}$ in type $K$. Then we obtain a smaller $Q$-vertex $\wti{T}_{1/1}^{\mu}$ from $T_{1/1}^{\mu}$ if we replace
\begin{equation*}
  \begin{split}
    c_i,\ i & \in\{1,2,5,6,7,8,9,12,13,14,15,16,17,21,22,23,24,25,26,27,28,29, \\
            & 33,34,35,36,37,84,85,86,87,88,89,90,91,92,108,111,112\} \\
  \end{split}
\end{equation*}
by
\begin{equation*}
  \begin{split}
    \ti{c}_1= & \ c_1+2\,c_3+c_4,\ \ti{c}_2=c_2+c_4,\ \ti{c}_5=c_5-c_4,
                \ \ti{c}_6=c_6-c_4, \\
    \ti{c}_7= & \ c_7-2\,c_3,\ \ti{c}_8=c_8+2\,c_{10}+c_{11},
                \ \ti{c}_9=c_9+c_{11},\ \ti{c}_{12}=c_{12}-c_{11}, \\
    \ti{c}_{13}= & \ c_{13}-c_{11},\ \ti{c}_{14}=c_{14}-2\,c_{10},
                   \ \ti{c}_{15}=c_{15}+c_{18}+c_{19},\ \ti{c}_{16}=
                   c_{16}+c_{18}+c_{20}, \\
    \ti{c}_{17}= & \ c_{17}+c_{19}+c_{20},\ \ti{c}_{21}=c_{21}-c_{19},
                   \ \ti{c}_{22}=c_{22}-c_{20},\ \ti{c}_{23}=c_{23}-c_{20}, \\
    \ti{c}_{24}= & \ c_{24}-c_{18},\ \ti{c}_{25}=c_{25}-c_{18},\ \ti{c}_{26}=
                   c_{26}-c_{19},\ \ti{c}_{27}=c_{27}+c_{30}+c_{31}, \\
    \ti{c}_{28}= & \ c_{28}+c_{30}+c_{32},\ \ti{c}_{29}=c_{29}+c_{31}+c_{32},
                   \ \ti{c}_{33}=c_{33}-c_{31},\ \ti{c}_{34}=c_{34}-c_{32}, \\
    \ti{c}_{35}= & \ c_{35}-c_{32},\ \ti{c}_{36}=c_{36}-c_{30},\ \ti{c}_{37}=
                   c_{37}-c_{30},\ \ti{c}_{38}=c_{38}-c_{31}, \\
    \ti{c}_{84}= & \ c_{84}+c_{93}+c_{94},\ \ti{c}_{85}=c_{85}+c_{93}+c_{95},
                   \ \ti{c}_{86}=c_{86}+c_{94}+c_{95},\ \ti{c}_{87}=c_{87}
                   -c_{94}, \\
    \ti{c}_{88}= & \ c_{88}-c_{95},\ \ti{c}_{89}=c_{89}-c_{95},\ \ti{c}_{90}
                   =c_{90}-c_{93},\ \ti{c}_{91}=c_{91}-c_{93}, \\
    \ti{c}_{92}= & \ c_{92}-c_{94},\ \ti{c}_{108}=c_{108}+c_{109},
                   \ \ti{c}_{111}=c_{111}+c_{109},\ \ti{c}_{112}=c_{112}
                   -c_{109}. \\
  \end{split}
\end{equation*} 
In the following we will always use this new $Q$-vertex $\wti{T}_{1/1}^{\mu}$. After elimination of these redundant terms in the types $A,\ldots,D,H$ and $K$ one can express the corresponding terms of $d_QT_1$ in an unique way as a divergence in the sense of vector analysis. This is done in appendix A. For the types $E,F,G$ and $J$ the situation is different. Here we have only monomials without derivatives acting with respect to $x^{\mu}$. For these types we obtain a $Q$-vertex which contains free constants. But it is shown in appendix B that this indeterminancy can be put into a form which drops out when we build the divergence.  
